\documentclass[a4paper, 12pt]{article}
\usepackage{latexsym,amsmath,amsfonts,amssymb}
\usepackage{tikz}
\usetikzlibrary{decorations.pathmorphing,cd,decorations.markings}
\usepackage[latin1]{inputenc}
\usepackage[american]{babel}
\usepackage{graphicx}
\usepackage{bbm}
\usepackage{cite}
\usepackage[colorlinks=true, citecolor=blue, linkcolor=blue, linktocpage=true]{hyperref}
\usepackage{tcolorbox}

\renewcommand{\baselinestretch}{1.2}
\newcommand{\ds}{\displaystyle}
\newcommand{\wt}{\widetilde}

\newcommand{\wh}{\widehat}

\newcommand{\matht}[1]{{\ensuremath{\boldsymbol{#1}}}}

\newcommand{\fP}{\mathfrak{P}}
\newcommand{\GW}{\text{GW}}

\newcommand{\eg}{\textit{e.g.}}

\newcommand{\ie}{\textit{i.e.}}

\newcommand{\rhs}{r.h.s.}

\numberwithin{equation}{section}

\newcommand{\nn}{\nonumber}
\newcommand{\mat}[1]{\begin{pmatrix} #1 \end{pmatrix}}
\newcommand{\smat}[1]{\big( \begin{smallmatrix} #1 \end{smallmatrix} \big)}
\newcommand{\be}{\begin{equation}} \newcommand{\ee}{\end{equation}}
\newcommand{\bea}{\begin{equation} \begin{aligned}} \newcommand{\eea}{\end{aligned} \end{equation}}

\newcommand{\cA}{\mathcal{A}}
\newcommand{\cB}{\mathcal{B}}
\newcommand{\cC}{\mathcal{C}}
\newcommand{\cD}{\mathcal{D}}

\newcommand{\cH}{\mathcal{H}}

\newcommand{\cK}{\mathcal{K}}
\newcommand{\cL}{\mathcal{L}}
\newcommand{\cM}{\mathcal{M}}
\newcommand{\cN}{\mathcal{N}}
\newcommand{\cO}{\mathcal{O}}
\newcommand{\cP}{\mathcal{P}}
\newcommand{\cQ}{\mathcal{Q}}
\newcommand{\cR}{\mathcal{R}}
\newcommand{\cS}{\mathcal{S}}
\newcommand{\cT}{\mathcal{T}}
\newcommand{\cU}{\mathcal{U}}

\newcommand{\cW}{\mathcal{W}}

\newcommand{\cZ}{\mathcal{Z}}

\newcommand{\bQ}{\mathbb{Q}}
\newcommand{\bR}{\mathbb{R}}

\newcommand{\bZ}{\mathbb{Z}}
\newcommand{\fD}{\mathfrak{D}}

\newcommand{\fg}{\mathfrak{g}}

\newcommand{\fM}{\mathfrak{M}}

\newcommand{\rme}{\mathrm{e}}
\newcommand{\rmm}{\mathrm{m}}
\newcommand{\sT}{\mathsf{T}}

\newcommand{\unit}{\mathbbm{1}}

\def\su{\mathfrak{su}}

\def\repa{\raise4pt\hbox{$\square$}\mkern-14mu\raise-4pt\hbox{$\square$}}
\def\repab{\overline{\raise4pt\hbox{$\square$}\mkern-14mu\raise-4pt\hbox{$\square$}\mkern-1mu}}

\DeclareMathOperator{\Tr}{Tr}

\DeclareMathOperator{\im}{\mathbb{I}m}

\DeclareMathOperator{\diag}{diag}

\DeclareMathOperator{\Aut}{Aut}

\begin{document}
\thispagestyle{empty}
\fontsize{12pt}{20pt}
\begin{flushright}
	SISSA  16/2022/FISI
\end{flushright}
\vspace{13mm}  
\begin{center}
	{\huge  The Holography of Non-Invertible \\[.5em] Self-Duality Symmetries}
	\\[13mm]
    	{\large Andrea Antinucci$^{a,\, b}$, \, Francesco Benini$^{a,\, b,\, c}$, \, Christian Copetti$^{a,\, b}$,
	\\[6mm]
	Giovanni Galati$^{a,\, b}$, and Giovanni Rizi$^{a,\, b}$}
	
	\bigskip
	{\it
		$^a$ SISSA, Via Bonomea 265, 34136 Trieste, Italy \\[.0em]
		$^b$ INFN, Sezione di Trieste, Via Valerio 2, 34127 Trieste, Italy \\[.6em]
		$^c$ ICTP, Strada Costiera 11, 34151 Trieste, Italy
	}
\end{center}

\bigskip

\begin{abstract}
\noindent
We study how non-invertible self-duality defects arise in theories with a holographic dual. We focus on the paradigmatic example of $\mathfrak{su}(N)$ $\cN = 4$ SYM. The theory is known to have non-invertible duality and triality defects at $\tau =i$ and $\tau = e^{2 \pi i /3}$, respectively. At these points in the gravitational moduli space, the gauged $SL(2,\bZ)$ duality symmetry of type IIB string theory is spontaneously broken to a finite subgroup $G$, giving rise to a discrete emergent $G$ gauge field. After reduction on the internal manifold, the low-energy physics is dominated by an interesting 5d Chern-Simons theory, further gauged by $G$, that we analyze and which gives rise to the self-duality defects in the boundary theory. Using the five-dimensional bulk theory, we compute the fusion rules of those defects in detail. The methods presented here are general and may be used to investigate such symmetries in other theories with a gravity dual.
\end{abstract}

\newpage
\pagenumbering{arabic}
\setcounter{page}{1}
\setcounter{footnote}{0}
\renewcommand{\thefootnote}{\arabic{footnote}}

{\renewcommand{\baselinestretch}{.88} \parskip=0pt
\setcounter{tocdepth}{2}
\tableofcontents}

%%%%%%%%%%%%%%%%%%%%%%%%%%%%%%%%%%%%%%%%%%
%%%%%%%%%%%%%%%%%%%%%%%%%%%%%%%%%%%%%%%%%%

\section{Introduction}
\label{sec: intro}

In recent years a great deal of effort has been devoted to the generalization of the concept of symmetry in quantum field theory, starting from \cite{Gaiotto:2014kfa}. According to the modern perspective, a $p$-form symmetry is implemented by topological operators $\fD_a(\gamma)$ supported on codimension-$(p+1)$ closed submanifolds $\gamma$, and its charged objects are $p$-dimensional operators.
An increasing amount of attention has been dedicated to symmetries whose defects do not fuse following the multiplication rules of a group, but rather form a more general categorical fusion algebra
\be
\label{categorical fusion algebra}
\fD_a \times \fD_b = \sum_{c} \cN_{a b}^c \; \fD_c \;.
\ee
This expression generalizes the group law in many ways: first, there can be more than one operator $\fD_c$ on the right-hand-side; second, the coefficients $\cN_{ab}^c$ are in general topological quantum field theories (rather than $c$-numbers) of the same dimensionality of the operators.
Such symmetries are called \emph{non-invertible} or \emph{categorical}. This structure is well understood for line operators in 2d quantum field theories (QFTs) and 3d topological quantum field theories (TQFTs) \cite{Verlinde:1988sn, Petkova:2000ip, Fuchs:2002cm, Frohlich:2003hm, Frohlich:2004ef, Frohlich:2006ch, Frohlich:2009gb, Carqueville:2012dk, Bhardwaj:2017xup, Chang:2018iay, Thorngren:2019iar, Komargodski:2020mxz, Thorngren:2021yso, Huang:2021zvu, Burbano:2021loy, Benini:2022hzx, Inamura:2022lun, Lin:2022dhv} using the formalism of fusion categories and modular tensor categories. The generalization to higher dimensional defects and $d>2$ QFTs started more recently with \cite{Choi:2021kmx,Kaidi:2021xfk} and is now a very active field of research \cite{Nguyen:2021yld, Wang:2021vki, Roumpedakis:2022aik, Bhardwaj:2022yxj, Arias-Tamargo:2022nlf, Choi:2022zal, Kaidi:2022uux, Choi:2022jqy, Cordova:2022ieu, Antinucci:2022eat, Bashmakov:2022jtl, Damia:2022bcd, Choi:2022rfe, Bhardwaj:2022lsg, Lin:2022xod, Bartsch:2022mpm, Apruzzi:2022rei, GarciaEtxebarria:2022vzq, Heckman:2022muc, Niro:2022ctq}. At the moment we know of, roughly speaking, three types of constructions of non-invertible symmetry defects.

In the first construction, that we refer to as of Kaidi-Ohmori-Zheng (KOZ) type \cite{Kaidi:2021xfk} (see also \cite{Bhardwaj:2022yxj}), one starts with a four-dimensional theory with a 0-form and a 1-form discrete symmetry, linked by a mixed 't~Hooft anomaly. If one gauges the 1-form symmetry, the 0-form symmetry defects become ill-defined because of the anomaly, but can be made well-defined by stacking them with suitable 3d TQFTs coupled to the dynamical 2-form gauge field for the 1-form symmetry. The resulting defects have non-invertible fusion laws due to the stacking rules of the 3d TQFTs. This type of categorical symmetries are ubiquitous, and many variations have been proposed. For example, classical Abelian symmetries suffering from an ABJ anomaly have been discovered to be realized at the quantum level as KOZ-type non-invertible symmetries \cite{Choi:2022jqy, Cordova:2022ieu, Choi:2022rfe}.

\begin{figure}[t]
\centering
\begin{tikzpicture}
	\filldraw[color=white!100!red, left color= white!85!red,  right color= white!100!red] (0,-1) -- (2,-1) -- (2,1) -- (0,1) --cycle;
	\draw[color=black, line width=2] (0,-1) node[below] {$I_g$} -- (0,1);
	\node at (-0.8,0) {$\mathsf{T}_\rho(x)$};
	\node at (1.2,0) {$\mathsf{T}_{\rho_g}(g \cdot x)$};
\begin{scope}[shift={(6.5,0)}]
	\filldraw[fill=white!90!red, color=white!90!red] (0,-1) -- (2,-1) -- (2,1) -- (0,1) --cycle;
	\draw[color=black, line width=2] (0,-1) node[yshift=-0.4cm] {$I_g$} -- (0,1);
	\draw[color=black, line width=2] (2,-1) node[yshift=-0.35cm] {$\varphi_{g}^\dagger$} -- (2,1);
	\node at (-0.8,0) {$\mathsf{T}_\rho(x)$};
	\node at (1,0) {$\mathsf{T}_{\rho_g}(x)$};
	\node at (2.8,0) {$\mathsf{T}_\rho(x)$};
	\node at (4.25,0) {$=$};
\begin{scope}[shift={(6.5,0)}]
	\node at (-0.7,0) {$\mathsf{T}_\rho(x)$};
	\node at (0.8,0) {$\mathsf{T}_\rho(x)$};
	\draw[color=red, line width=2] (0,-1) node[below, black] {$\fD_g$}-- (0,1);
\end{scope}
\end{scope}
\end{tikzpicture}
\caption{Left: a topological duality interface $I_g$. Right: the definition of a local topological defect $\fD_g$ at fixed points in the conformal manifold.%
\label{fig:selfdualitydef}}
\end{figure}
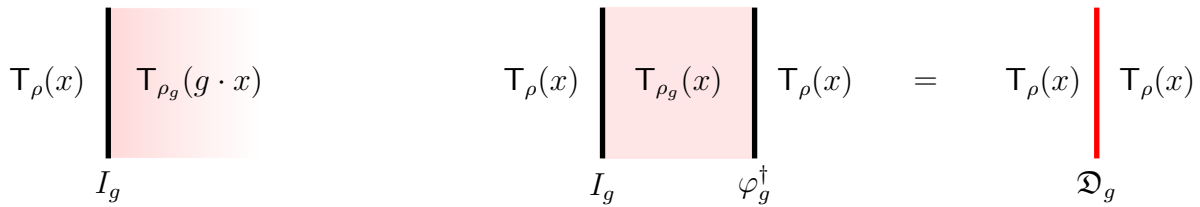

The second construction \cite{Choi:2021kmx, Choi:2022zal, Kaidi:2022uux} involves self-duality defects, and will be the main focus of this paper. One considers a family of gauge theories $\mathsf{T}_\rho(x)$, where $x\in\fM$ parametrizes a conformal manifold and $\rho$ labels possible different gauge groups with the same gauge algebra, with the action on both of a nontrivial duality group $\Gamma$. Its action is implemented by a topological interface $I_{g \in \Gamma}: \mathsf{T}_\rho(x) \to \mathsf{T}_{\rho_g}(g \cdot x)$ which is depicted in Figure~\ref{fig:selfdualitydef} left. One also assumes that there exists a second topological interface $\varphi_g : \mathsf{T}_\rho(x) \to \mathsf{T}_{\rho_g}(x)$ which acts on the global structure alone --- in the known examples, $\varphi _g$ is implemented by gauging some discrete 1-form symmetry of the theory on a half space. At generic points on $\fM$ where it acts faithfully, $\Gamma$ is not a symmetry because it is a map between two different descriptions of the same quantum theory. At points $x$ with a nontrivial stabilizer $G \subset \Gamma$, one might expect $G$ to become a symmetry. It does become an ordinary 0-form symmetry if it does not act on the global structure $\rho$. Consider, instead, the case that $G$ (that we take Abelian) acts faithfully on $\rho$. Then, for $g\in G$, it is still possible to construct a topological defect $\fD_g$ from the theory to itself, but it requires a composition $\fD_g = \varphi^\dagger_g \circ I_g: \mathsf{T}_\rho(x) \to \mathsf{T}_\rho(x)$ (Figure~\ref{fig:selfdualitydef} right). As a consequence it satisfies, schematically:
\bea
\label{composition law selfduality defects}
\fD_g \times \fD_h &= \cN_{g,h} \; \fD_{gh} \qquad\qquad\text{for $gh \neq \unit$} \;, \\
\fD_g \times \overline{\fD}_g &= \cC \;.
\eea
Here $\cN_{g,h}$ are decoupled 3d TQFTs, so the composition law is group-like but with TQFTs in place of coefficients. In the second line $\overline{\fD}_g \equiv \fD_{g^{-1}}$ is the orientation reversal of $\fD_g$. On the other hand, $\cC$ is a peculiar topological operator: it is a 3d ``condensate'', or ``higher gauging'' \cite{Gaiotto:2019xmp, Roumpedakis:2022aik}, of the 1-form symmetry used to change the global structure of the group. It is constructed by gauging that symmetry on a 3d submanifold, or alternatively, but summing over the surface operators that implement that symmetry.%
\footnote{The composition law (\ref{composition law selfduality defects}) is not a group for two reasons: the coefficients are TQFTs, and $\cC$ is different from $\fD_\unit$ which is trivial. However, if one places (\ref{composition law selfduality defects}) on $S^3$ then both $\cN_{g,h}$ and $\cC$ reduce to 1 and one recovers the group law of $G$.%
\label{foo: fusion to group}}
The construction of these defects in $\cN=4$ super-Yang-Mills (SYM) was carried out in \cite{Choi:2021kmx, Choi:2022zal, Kaidi:2022uux}.

The third construction \cite{Bhardwaj:2022yxj, Arias-Tamargo:2022nlf, Antinucci:2022eat} is of orbifold type. One starts with a theory with a discrete 0-form symmetry $G$ that acts nontrivially on higher-codimension symmetry defects $U$. After gauging $G$, the gauge-invariant symmetry defects are obtained by summing over the orbit of the action of $G$ on some $U$, and support topological line defects labeled by $\text{Rep} \bigl( \text{Stab}(U) \bigr)$. This gives rise to fusion rules (\ref{categorical fusion algebra}) where multiple operators appear in the sum on the \rhs, and is a generalization of the theory of orbifolds in 2d conformal field theories (CFTs) \cite{Dijkgraaf:1989hb} and of 3d $G$-crossed modular tensor categories (MTCs) \cite{Barkeshli:2014cna}.

Our interest in this paper is in understanding how categorical symmetries appear in holography. The common lore is that a global symmetry of the boundary theory appears as a gauge symmetry, accompanied by a gauge field, in the bulk. This is confirmed and well understood in the case of invertible symmetries --- both ordinary 0-form, continuous and discrete, as well as higher form. What happens for a non-invertible symmetry? What plays the role of a ``non-invertible'' gauge field?

We investigate this question in the specific case of \emph{self-duality defects}.%
\footnote{The holographic description of non-invertible defects of the KOZ and orbifold type has recently been investigated in \cite{Apruzzi:2022rei, GarciaEtxebarria:2022vzq, Heckman:2022muc}.}
The idea is simple.%
\footnote{We are grateful to Davide Gaiotto for pointing out the emergence of discrete gauge symmetries in type IIB string theory and their possible relevance to duality defects.}
The conformal manifold $\fM$ of the boundary theory is dual to a moduli space of bulk solutions in the gravitational description, while the choice of a global structure on the boundary corresponds to a certain boundary condition in gravity. The duality group $\Gamma$ is a discrete gauge symmetry of string theory, which however is completely Higgsed at generic points $x$ at which $\Gamma$ acts faithfully on $\fM$. At a special point $x$ which is stabilized by $G \subset \Gamma$, the duality symmetry $\Gamma$ is only Higgsed to $G$, and in the low-energy description appears an emergent $G$ gauge field that acts on the supergravity fields. In particular, it also acts on a low-energy topological sector of string theory whose topological (or conformal) boundary conditions encode the possible global structures of the boundary theory. It is this structure that plays the role of a ``non-invertible gauge field'', at least in this class of examples.
The derivation and explanation of how the supergravity theory with extra gauge field gives rise to the non-invertible fusion rules (\ref{composition law selfduality defects}) is the subject of this paper. We focus on the example of 4d $\cN=4$ SYM and its dual type IIB string theory description. The formalism we develop is however quite general and should allow for prompt generalizations, for instance to theories of class $S$ \cite{Gaiotto:2009we}.

Let us summarize the main points of the construction. A key role is played by a topological sector of type IIB string theory compactified on $S^5$, which dominates at long distances: it is a 5d Chern-Simons-like (CS) theory \cite{Aharony:1998qu, Witten:1998wy, Belov:2004ht}%
\footnote{In our notation, we multiply differential forms leaving all wedge products implicit.}
\be
\label{eq:action}
S_\text{CS} =\frac{N}{2\pi} \int  b\; dc \,\equiv\, \frac{N}{4\pi} \int \cB^\sT \epsilon \, d \cB \;.
\ee
Here $b$ and $c$, that we package into $\cB=(b,c)$, are 2-form gauge fields coming from the NS-NS and R-R 2-form potentials, respectively, while $\epsilon = \smat{0 & 1 \\ -1 & 0}$. This theory can also be interpreted as a $\bZ_N$ 2-form gauge theory \cite{Banks:2010zn}, and its boundary conditions encode the global structure of the boundary theory \cite{Witten:1998wy}. The gauge fields $b,c$ are dual to the $\bZ_N \times \bZ_N$ 1-form symmetry of the set of boundary theories (taking into account the possible global structures). The 5d TQFT (\ref{eq:action}) has a global symmetry%
\footnote{The symmetry is $SL(2,\bZ_N)$ on spin manifolds, and a subgroup thereof on non-spin manifolds \cite{Witten:1998wy}. Since we are dealing with supersymmetric theories, we restrict to spin manifolds in this paper.}
$\Gamma = SL(2,\bZ_N)$ that acts linearly on $\cB$ and is generically spontaneously broken by the axiodilaton (on which it acts by fractional linear transformations), as well as a $\bZ_N \times \bZ_N$ global 2-form symmetry that shifts $b,c$. The symmetry defects $V_{M\in \Gamma}$ for $\Gamma$ turn out to be higher gaugings \cite{Roumpedakis:2022aik} of the 2-form symmetry (or a subgroup thereof) on 4d submanifolds, with suitable choices of discrete torsion $\cT(M)$ that we determine.

Quite interesting are the $\Gamma$-twisted sectors $D_M$ that live at the boundary of the symmetry defects $V_M$. For most $M \in \Gamma$, $V_M$ turns out to be an invertible TQFT that produces anomaly inflow and constrains its twisted sector. A minimal representative for $D_M$ with the correct anomaly is a certain 3d TQFT $\cA^{N, -\cT}(\cB)$ coupled to $\cB$, introduced in \cite{Hsin:2018vcg}. The fusion of twisted sectors takes the schematic form
\bea
D_{M_2} \times D_{M_1} &= \cA^{N, - (\cT_2 + \cT_1)} \;  D_{M_2M_1} \;, \\
D_M \times \overline{D}_M &= \cC^{\bZ_N \times \bZ_N} \;,
\eea
where $\cC^{\bZ_N \times \bZ_N}$ is a 3d condensate of the 2-form symmetry. Interestingly, to compute this fusion rules one uses a modified version $\cA^{N,-\cT_1}(\cB) \times_\cB \cA^{N,-\cT_2}(\cB)$ of the stacking of TQFTs, in which the lines in the first factor acquire nontrivial braiding
\be
B_{n_1n_2} = \exp \biggl( \frac{2 \pi i}{N} \, 2^{-1} \, n_1^\sT \, \epsilon \, n_2 \biggr)
\ee
(here $n_1, n_2 \in \bZ_N \times \bZ_N$ parametrize the lines of the two factors, respectively, and $N$ is odd) with the lines in the second factor, due to the interactions with the bulk 5d theory (\ref{eq:action}).

In order to complete the holographic setup and make contact with the self-duality defects of the boundary $\cN=4$ SYM theory, two more steps are necessary. First, one should choose topological boundary conditions $\rho(\cL)$ for (\ref{eq:action}) on AdS$_5$, which are labelled by Lagrangian subgroups $\cL$ of the $\bZ_N \times \bZ_N$ global 2-form symmetry and correspond to a choice of global structure $\rho$ on the boundary. Imposing the boundary condition is equivalent to gauging $\cL$ in the bulk \cite{Kapustin:2010hk, Kapustin:2010if, Kaidi:2021gbs}, which is necessary in order to remove global symmetries from the bulk gravitational theory \cite{Benini:2022hzx}. The topological self-duality defects of the boundary theory eventually are bulk operators placed on top of the boundary. We thus determine the pull-backs $D_{M,\cL}$ of $D_M$ on the boundary and their fusion rules. The answer turns out to depend on whether $\cL$ is invariant under $M$ or not.

Second, recall that for generic values of the axiodilaton $\tau$, the 0-form symmetry $\Gamma$ is spontaneously broken in the full supergravity theory. At the special values $\tau = i, \, e^{2\pi i /3}$ an Abelian subgroup $G = \bZ_4,\, \bZ_6$ generated by $S,\, ST$, respectively, is preserved, but crucially this symmetry is gauged. We study in detail the 5d theory obtained by gauging $G$ in (\ref{eq:action}). The defects $V_{g \in G}$ become transparent, because they implement the gauging, and hence their boundaries $D_{g \in G}$ become genuine 3d topological operators. They are ill-defined in isolation because of the anomaly, but one can form well-defined 3d topological operators $\fD_g$ by stacking $D_g$ with a 3d Gukov-Witten \cite{Gukov:2006jk}, or twist, operator $\GW_g$ for the pure $G$ gauge theory. This mechanism is similar to KOZ \cite{Kaidi:2021xfk}, but the role of the anomaly is played here by the torsion. The effect of gauging has also an effect on the boundary conditions, and we indicate by $\rho^*$ the gauged boundary conditions.

The final fusion rules we find in the boundary theory take the schematic form:
\bea
\fD_{g_2,\, \cL} \times \fD_{g_1,\, \cL} &= \cN_{g_2, g_1} \; \fD_{g_2g_1,\, \cL} \qquad& g_2g_1 &\notin \text{Stab}(\rho^*) \;, \\
\fD_{g_2,\, \cL} \times \fD_{g_1,\, \cL} &= \cC^{\bZ_N} \; \GW_{g_2g_1} \qquad& g_2g_1 &\in \text{Stab}(\rho^*) \;, \\
\GW_{g_2} \times \GW_{g_1} &= \GW_{g_2g_2} \qquad& g_1, g_2 &\in \text{Stab}(\rho^*) \;.
\eea
The explicit form of the TQFT coefficients $\cN$ is given in Section~\ref{sec: fusion gapped boundary}. The third line describes a subcategory of invertible symmetries (that always includes charge conjugation).
Our results reproduce the known duality and triality defects of 4d $\cN=4$ SYM \cite{Choi:2021kmx, Choi:2022zal, Kaidi:2022uux}.

The paper is organized as follows.
In Section~\ref{sec: holography} we recall a few facts about symmetries in holography.
In Section~\ref{sec: CS theory} we study the 5d Chern-Simons theory, its symmetries, and its gapped boundaries.
In Section~\ref{sec:TwistedSectors} we describe the twist operators for the $SL(2,\bZ_N)$ 0-form symmetry, using both a Lagrangian as well as a more formal approach based on minimal TQFTs, we compute their fusion, and the pull-back to gapped boundaries.
Finally in Section~\ref{sec:gaugedTheory} we address how to gauge a discrete Abelian symmetry $G$. That is used to present the final composition laws.
We conclude in Section~\ref{sec:conclusions}. Detailed computations are collected in several appendices.

\bigskip

While this work was nearing completion, the paper \cite{Kaidi:2022cpf}, which deals with a construction similar to ours, appeared on the arXiv.

%%%%%%%%%%%%%%%%%%%%%%%%%%%%%%%%%%%%%%%%%%
%%%%%%%%%%%%%%%%%%%%%%%%%%%%%%%%%%%%%%%%%%

\section{Symmetries and global structures in holography}
\label{sec: holography}

The four-dimensional $\cN=4$ SYM theory with gauge algebra $\su(N)$ is holographically dual to type IIB string theory on asymptotically AdS$_5 \times S^5$ spaces \cite{Maldacena:1997re}. The boundary theory, however, is characterized by a specific gauge \emph{group} with the given algebra, and thus this piece information must be encoded in the bulk theory. As explained by Witten \cite{Witten:1998wy}, kinematical properties of the boundary theory, such as the global form of the gauge group, are captured by the long-distance behavior of the gravitational theory (or equivalently, by the behavior close to the boundary), which is encoded in the terms in the Lagrangian with the lowest number of derivatives, namely in the topological terms. One can more conveniently work with the effective 5d theory in AdS$_5$ obtained by reducing on the internal manifold. The 10d type IIB supergravity action contains the topological term
\be
S_\text{IIB} \,\supset\, \int_{X_{10}} B_2 \; dC_2 \; F_5 \;,
\ee
where $B_2$ is the NS 2-form potential, $C_2$ the RR 2-form potential, and $F_5$ is the field strength of the RR 4-form potential. In compactification on $\cM_5 \times S^5$ with $N$ units of 5-form flux on $S^5$, one obtains at low energies the 5d Chern-Simons action (\ref{eq:action}) \cite{Witten:1998wy, Belov:2004ht}.

The continuous 2-form gauge fields $b,c$ are dual to a $U(1) \times U(1)$ global 1-form symmetry of the boundary theory, whose two factors act on 't~Hooft and Wilson line operators, respectively. This symmetry does not have to act faithfully on the boundary theory: it only acts faithfully on the full set of boundary theories with all possible global structures. This follows from the necessity of choosing boundary conditions. If we choose topological boundary conditions, the action (\ref{eq:action}) restricts $b,c$ to be $\bZ_N \times \bZ_N$ gauge fields \cite{Banks:2010zn} and accordingly restricts the 1-form symmetry. Boundary conditions $\rho(\cL)$ further set to zero a linear combination of $b,c$ along a Lagrangian subgroup $\cL \subset \bZ_N \times \bZ_N$, only leaving a 1-form symmetry of order $N$. Thus, the choice of boundary conditions specifies the global structure of the SYM theory \cite{Aharony:1998qu, Witten:1998wy} and the spectrum of extended (here line) operators \cite{Aharony:2013hda}. For instance, if we set $b=0$ at the boundary, the boundary theory is $SU(N)$. Fundamental strings (that couple to $b$) can end on the boundary producing Wilson line operators in generic representations \cite{Maldacena:1998im, Rey:1998ik}, their $\bZ_N$ charge being measured by the topological operators $e^{i \!\int\! c}$, while 't~Hooft lines only exist with vanishing $\bZ_N$ charge. On the contrary, if we set $c=0$ we obtain the $PSU(N)_0$ theory.%
\footnote{See \cite{Aharony:2013hda} and Section~\ref{sec: CS theory} for the meaning of this notation.}
D1-branes (that couple to $c$) can end on the boundary producing 't~Hooft line operators with generic $\bZ_N$ charge, the latter being measured by $e^{i \!\int\! b}$, while Wilson lines only exist in representations with trivial $N$-ality. One can also choose conformal boundary conditions $b = *\,c$: they give rise to an extra singleton sector \cite{Belov:2004ht, Kravec:2014aza} and describe the theory $U(N)$, for which the 1-form symmetry is indeed $U(1) \times U(1)$.

Type IIB string theory also enjoys an $SL(2,\bZ)$ symmetry. As in any theory of quantum gravity, this must be a \emph{gauge} symmetry. It acts on the axiodilaton field $\tau =C_0+ie^{-\phi}$ by standard fractional linear transformations
\be
\tau \to \frac{ \underline{a} \, \tau + \underline{b}}{\underline{c} \, \tau + \underline{d}} \;,\qquad\qquad \mat{\underline{a} & \underline{b} \\ \underline{c} & \underline{d} } \in SL(2,\bZ)
\ee
(only $PSL(2,\bZ)=SL(2,\bZ)/\bZ _2$ acts on $\tau$) and on $b,c$ as on a doublet $\cB =(b,c)$ in the fundamental representation,
\be
\mat{b \\ c} \to \mat{\underline{a} & \underline{b} \\ \underline{c} & \underline{d}} \mat{b \\ c}  \;.
\ee 
At generic points $\tau$ in the moduli space, $SL(2,\bZ)$ is spontaneously broken to its $\bZ_2$ center, and thus the corresponding gauge field does not appear in the low-energy supergravity description.%
\footnote{The $\bZ_2$ center of $SL(2,\bZ)$, that maps $(b,c) \mapsto (-b, -c)$ but does not act on $\tau$, is however always preserved and then the corresponding $\bZ_2$ gauge field should be included.}
However, special values of $\tau$ are left invariant by a larger subgroup $G \subset SL(2,\bZ)$ which therefore remains unbroken. The corresponding gauge field should then be included in the supergravity description, where it appears as an emergent gauge field for the low-energy observer. Specifically, $G = \bZ_4$ at $\tau = i$ and $G = \bZ_6$ at $\tau = e^{2\pi i /3}$. After compactification on $S^5$, we obtain a discrete $G$ gauge field $a$ in five dimensions, coupled to a $G$ subgroup of the $SL(2,\bZ_N)$ symmetry of $S_\text{CS}$ (\ref{eq:action}). This is an interesting subsector of the full theory on its own. Our aim is to show that $a$ is the gauge field corresponding to the non-invertible symmetries of the boundary theory.%
\footnote{As noted in footnote~\ref{foo: fusion to group}, the non-invertibility of duality and triality defects is \emph{only} up to condensates. It is perhaps then not surprising that the corresponding bulk gauge field is a standard discrete connection for $G$, though coupled to a nontrivial topological sector $S_\text{CS}$. The gauge field holographically dual to symmetries which remain non-invertible also up to condensates would presumably be a more complicated object.}

We conclude this section recalling that the way in which a bulk TQFT affects the symmetry on its boundary is made very clear in the recently introduced formalism of symmetry TFTs \cite{Gaiotto:2020iye, Freed:2022qnc} (see also \cite{Apruzzi:2021nmk, DelZotto:2022ras, vanBeest:2022fss}). Independently of holography, the symmetry TFT approach separates the local dynamics of a QFT from its global structure, by viewing a physical (or absolute) $d$-dimensional theory  as the compound $(d+1)$-dimensional system of a slab of topological theory with two parallel boundaries. One boundary supports the relative \cite{Freed:2012bs} version of the theory. Roughly speaking, a relative theory is a vector of theories encoding all possible global structures. The opposite is a gapped boundary determined by some topological boundary conditions for the TFT. Their choice corresponds to picking a particular state, \ie, selecting one particular absolute theory. We collect some details about the construction in Appendix~\ref{app: symtft}.

%%%%%%%%%%%%%%%%%%%%%%%%%%%%%%%%%%%%%%%%%%
%%%%%%%%%%%%%%%%%%%%%%%%%%%%%%%%%%%%%%%%%%

\section{The 5d Chern-Simons theory and its symmetries}
\label{sec: CS theory}

\begin{figure}[t]
$$
\hspace{-2.4cm}
\begin{tikzpicture}
	\node at (-4.5,0) {};
	\draw [thick] (0,0) arc [start angle = 90, end angle = 430, x radius = 1.5, y radius = 0.7] node[pos = 0.2, above left] {$U_m$};
	\draw [thick] (4.85,-2) +(180:5) node[right] {$U_{m'}$} arc [start angle = 180, end angle = 175.5, radius = 5];
	\draw [thick] (4.85,-2) +(171:5) arc [start angle = 171, end angle = 150, radius = 5];
	\node at (3,-0.8) {${} = \quad B_{mm'}$};
	\draw [thick] (9.3,-2) +(180:5) node[right] {$U_{m'}$} arc [start angle = 180, end angle = 150, radius = 5];
\end{tikzpicture}
\hspace{-1.8cm}
\begin{tikzpicture}
	\draw (-3,-1) -- (1,-1) -- (3,1) node[above] {$\rho(\cL)$} -- (-1,1) -- cycle;
	\filldraw [red] (0, -0.1) circle [radius = 0.1] node[right, black] {$\partial U_l$};
	\draw [thick] (0, -0.1) arc [start angle = 180, end angle = 150, radius = 5] node[right] {$U_{l\in \cL}$};
	\filldraw [white] (0.21, 1.34) circle [radius = 0.12];
	\draw [thick, rotate around = {-10: (0.3, 2.1)}] (0.3, 2.1) arc [start angle = 110, end angle = 445, x radius = 1, y radius = 0.4] node[pos = 0.15, left] {$U_t$};
	\draw [thick, blue!80!black] (-0.15, 0.5) arc [start angle = 110, end angle = 450, x radius = 1, y radius = 0.6] node[pos = 0.15, left, black] {$\wh U_t$};
	\draw [thick, densely dashed, ->, green!50!black] (-0.4, 1.4) arc [start angle = 150, end angle = 190, radius = 1.2];
\end{tikzpicture}
$$
\caption{Left: Antisymmetric braiding $B_{mm'}$ between 2-dimensional defects $U_m$ in 5d Chern-Simons theory. Right: Induced braiding between 2-dimensional defects $\wh U_t$ and line defects $\partial U_{l \in \cL}$ on gapped boundaries $\rho(\cL)$.
\label{fig: 5d linking}}
\end{figure}
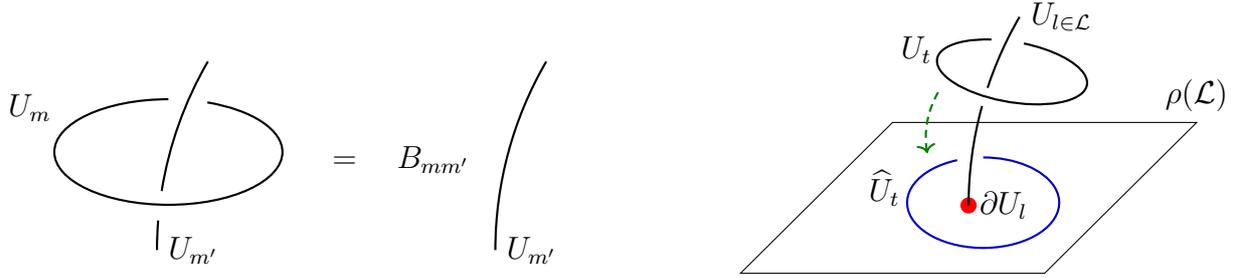

Consider the five-dimensional Chern-Simons action \cite{Witten:1998wy} (see also \cite{Belov:2004ht, MooreCourse:2011, Kravec:2014aza, Gaiotto:2014kfa, Hofman:2017vwr})%
\footnote{This action, as written, is not well defined \cite{Witten:1998wy}. When the spacetime manifold $M_5$ is the boundary of a six-manifold $Z$, one can define $S[Q] = \frac1{4\pi} \int_Z Q(d\cB, d\cB)$. However, the bordism group in five dimensions is non-trivial and thus this cannot be done in general. One could instead use the formalism of Cheeger-Simons differential characters \cite{Cheeger:1985}.
\label{foo: 5d diff characters}}
\be
\label{5d theory general action}
S[Q] = \frac{1}{4 \pi} \int Q(\cB , \, d\cB) \;,
\ee
where $\cB$ is a vector of $2n$ 2-form gauge fields, $Q$ is an integer-valued $2n \times 2n$ non-degenerate antisymmetric matrix (or symplectic form), and we used the notation $Q(x,y) = x^\sT Q y$. We study this theory on spin manifolds. The theory has topological surface operators
\be
\label{def surface operators Um}
U_m = e^{i m^\sT \!\int \cB} \;,
\ee
where $m$ is an integer-valued vector in $\bZ^{2n}$, and the integral is over a 2-dimensional surface. These operators generate an (anomalous) 2-form symmetry. The operators $U_m$ have nontrivial linking (see Figure~\ref{fig: 5d linking} left) given by the antisymmetric braiding matrix
\be
\label{braiding matrix general}
B_{m m'} = e^{2 \pi i \, Q^{-1}\left(m, \/ m' \right)} \;.
\ee
Any operator for which $m = Q \, k$ with $k \in \bZ^{2n}$ is completely transparent and thus trivial. Those operators generate a lattice $\Lambda_Q$, and the 2-form symmetry defect operators are labelled by the elements of the discriminant group
\be
\cD_Q = \bZ^{2n} / \Lambda_Q \;.
\ee
This is the 2-form symmetry of the theory. Notice that $|\cD_Q| = \left| \det Q \right|$.

The case relevant to type IIB string theory compactified on $S^5$ is $n=1$ and $Q = N \epsilon$ with $\epsilon = \smat{0 & 1 \\ -1 & 0}$.
We denote by $b$ and $c$ the two components of $\cB$. The action reads:%
\footnote{We work with an antisymmetric 5d Lagrangian, which is manifestly invariant under $SL(2,\bZ)$ symmetry. One should however keep in mind that, as written, the action is not well defined (see footnote~\ref{foo: 5d diff characters}), and thus conclusions drawn from it should be taken with care. It turns out \cite{Witten:1998wy} that for $N$ odd, the theory is $SL(2,\bZ)$ invariant only on spin manifolds, while on non-spin manifolds it is invariant under the subgroup $\Gamma(2)$ generated by $S$ and $T^2$.}
\be
\label{eq: 5d BF}
S = \frac{N}{4 \pi} \int \cB^\sT \epsilon \, d \cB = \frac{N}{4 \pi} \int \langle \cB , \, d \cB \rangle  = \frac{N}{4\pi} \int \bigl( b\, dc - c\, db \bigr) \;.
\ee
We introduced the antisymmetric Dirac pairing $\langle x,y \rangle = x_\rme y_\rmm - x_\rmm y_\rme$, where $x = (x_\rme, x_\rmm)$ is the expression of a vector in components. When describing the surface operators $U_m$, it might be convenient to package the information about $m$ and the geometric 2-cycle wrapped by $U_m$ into $\gamma \in H_2(M_5, \bZ_N \times \bZ_N)$, or its Poincar\'e-dual cocycle $\text{PD}(\gamma)\in H^3 (M_5, \bZ _N\times \bZ _N )$. In this case $U(\gamma)$ is described by the insertion of
\be
U(\gamma) = \exp\left( i \int \cB^\sT \, \text{PD}(\gamma) \right)
\ee
in the path integral.

In the general case, gapped boundary conditions $\rho(\cL)$ are in bijection with Lagrangian subgroups $\cL$ of $\cD_Q$. A subgroup is called Lagrangian if all its elements are mutually transparent, \ie, if $B_{ll'} = 1$ for all $l,l' \in \cL$, and if any element outside $\cL$ braids non-trivially with at least one $l \in \cL$ (\ie, $\cL$ is maximal). Defining a gapped boundary $\rho(\cL)$ is equivalent to gauging the Lagrangian subgroup $\cL$ of the 2-form symmetry \cite{Kapustin:2010hk, Kapustin:2010if, Kaidi:2021gbs}.%
\footnote{More precisely, gauging the discrete symmetry $\cL$ is equivalent to inserting a network of symmetry defects for $\cL$ in the spacetime manifold. This is also equivalent to removing a tubular neighborhood of the network from the spacetime manifold, and placing the topological boundary condition $\rho(\cL)$ there. Thus, $\rho(\cL)$ is a topological interface between the ungauged theory and the trivial theory obtained by gauging $\cL$ (such a theory is trivial because $\cL$ is Lagrangian).}

Only dyons $U_l$ with $l \in \cL$ may terminate on the gapped boundary, defining in this way topological line operators $\partial U_l$ there. Besides, dyons in $\cL$ are absorbed by the gapped boundary if they are moved to lie within it, in other words the dyons $U_{l \in \cL}$ are completely transparent (they do not contribute to correlation functions) when placed on the gapped boundary. The boundary has non-trivial topological surface operators corresponding to $m \notin \cL$, obtained by moving $U_{m \notin\cL}$ to lie within the boundary, however, because of the property just mentioned, those operators $\wh U_t$ are labeled by conjugacy classes $t \in \cD_Q / \cL \equiv \cS$. The operators $\wh U_t$ are stuck to the gapped boundary (because $t$ would be an ambiguous label in the bulk), and generate a 1-form symmetry $\cS$ there. The charges under that symmetry are carried by the lines $\partial U_{l \in \cL}$, as follows from the 5d braiding (see Figure~\ref{fig: 5d linking} right):
\be
\wh U_t( \partial U_l ) = e^{2 \pi i \, Q^{-1} ( t, l)} \, \partial U_l \;,
\ee
where, with some abuse of notation, we indicated by $t$ any representative of its class in $\cD_Q$.

Some properties become clear in the Lagragian description (\ref{5d theory general action}): a gapped boundary on $X$ is defined by Dirichelet boundary conditions
\be
l^\sT \cB \Big|_X = 0 \quad \text{ (up to gauge transformations)} \qquad\quad\text{for all } l \in \cL \;.
\ee
Introducing a rectangular matrix $L$ whose columns are the generators of $\cL$ in $\bZ^{2n}$, so that $L^\sT Q L = 0$, the boundary condition is $L^\sT \cB \big|_X = 0$ (up to gauge transformations). This can be imposed by a boundary TQFT:
\be
\label{boundary action}
S_{\substack{\text{gapped}\hspace{0.7em} \\ \text{boundary}}}[L] = \frac1{2\pi} \int Q\Bigl( \eta ,\, L^\sT (\cB - d\xi) \Big) \;+\; \text{counterterms} \;,
\ee
where $\eta$ is a 2-form gauge field in $\bR^{2n} / \langle \cL\rangle$, $\xi$ is a 1-form gauge field in $\langle \cL \rangle$, and $\langle \cL \rangle$ is the real span of $\cL$. The counterterms only involve $\cB$, and are fixed by overall gauge invariance.
To give an example, consider the type IIB case $Q = \smat{0 & N \\ -N & 0}$ and take the \emph{electric boundary} $\rho(\cL)$ where $\cL$ is generated by $l = (1,0)$, corresponding to the boundary condition $b \big|_X = 0$ (up to gauge transformations). The boundary action is
\be
S_{\substack{\text{electric}\hspace{0.7em} \\ \text{boundary}}} = \frac{N}{2\pi} \int \biggl[ \eta \, (b - d \xi) - \frac12\, bc \biggr] \;.
\ee
If we introduce a coordinate $r$ transverse to the boundary, place the boundary at $r=0$ and the bulk in the region $r<0$, the full bulk plus boundary system has action
\be
S_{\substack{\text{bulk plus}\hspace{0.1em} \\ \text{boundary}}} = \frac{N}{4\pi} \int_{r<0} \bigl( b \, dc - c \, db \bigr) + \frac{N}{2\pi} \int_{r=0} \biggl[ \eta \, (b - d \xi) - \frac12\, bc \biggr] \;.
\ee
The equations of motion fix the following conditions on the boundary:
\be
b = d\xi \;,\qquad\quad c = \eta \;,\qquad\quad \eta \in H^2(M_5, \bZ_N) \;.
\ee
Thus, $b$ is set to be pure gauge, while $\eta$ is the pull-back of $c$ to the boundary and $c$ remains unconstrained ($c \in H^2(M_5, \bZ_N)$ is already imposed by the bulk EOMs). The system is invariant under the following gauge transformations:
\be
b \,\to\, b + d\alpha_\rme \;,\qquad c \,\to\, c + d\alpha_\rmm \;,\qquad \eta \,\to\, \eta + d\alpha_\rmm \;,\qquad \xi \,\to\, \xi + \alpha_\rme \;.
\ee

Interpreting instead $\cL$ as a subgroup of $\cD_Q$ that is gauged, the dyons $U_{l \in \cL}$ become trivial in the bulk because they are pure gauge and can be absorbed by the network of defects. On the contrary, the operators with $m \notin \cL$ are projected out in the bulk (using the fact that $\cL$ is Lagrangian) and can only exist on the boundary.

In the holographic setup, the 2-form symmetry $\cL$ that we gauge in the bulk dictates what is the spectrum of physical lines in the holographic boundary \cite{Witten:1998wy, Benini:2022hzx}. Thus, the surfaces $U_l$ with $l \in \cL$ become trivial in the bulk, but if they are attached to the holographic boundary, their end-lines $\partial U_l \equiv W_l$ are the physical line operators of the boundary theory (notice that these are no longer topological, due to the holographic boundary conditions).%
\footnote{In the picture in which the bulk with gauged $\cL$ is substituted by a slab of bulk between the holographic boundary and a gapped boundary $\rho(\cL)$, the operators $U_l$ can be stretched between a copy of $W_l$ in the holographic boundary and a copy of $W_l$ in the gapped boundary.}
The 1-form symmetry of the boundary theory under which the lines $W_l$ are charged is generated by the surface operators $\wh U_t$, that can only live on the boundary.

Coming back to type IIB string theory, where $Q = N\epsilon$, the simplest case to discuss is when $N$ is prime. We label the bulk surfaces $U_m$ by $m= (m_\rme, m_\rmm)$, where $m_\rme$ and $m_\rmm$ are the electric and magnetic charges, respectively. The topological sector has $N+1$ gapped boundary conditions:\footnote{See \cite{Bergman:2022otk} for a recent in depth study of gapped boundary conditions in the 5d Chern-Simons theory.}
\begin{itemize}

\item An electric gapped boundary $\rho(e)$, for which $\cL$ is generated by $l=(1,0)$. As a gauging, this is obtained by condensing the electric surfaces $(m_\rme, 0) \in \cL$ (while in terms of a gapped boundary, this is implemented by setting $b=0$ there). It corresponds to the global variant $SU(N)$ of the boundary theory. The Wilson lines $W_{l \in \cL}$ are endpoints of bulk surfaces $U_l$. For instance, the Wilson line in the fundamental representation is the endpoint of a fundamental string \cite{Maldacena:1998im, Rey:1998ik}, which couples as $e^{i \!\int\! b}$ to the NS B-field $b$. The boundary 1-form symmetry is generated by the surfaces $\wh U_t$, and we can take for $\cS \cong \bZ_N$ the representatives $t = (0, t_\rmm)$.

\item $N$ magnetic gapped boundaries $\rho(m)_r$ with $r=0, \dots, N-1$, for which $\cL$ is generated by $l=(r,1)$. They are obtained by condensing the dyonic surfaces $(r m_\rmm, m_\rmm) \in \cL$ (or by setting $rb+c = 0$ on a gapped boundary). They correspond to the global variants $PSU(N)_r$ of the boundary theory \cite{Aharony:2013hda}. The 't~Hooft or dyonic lines are endpoint of bulk surfaces $U_{l \in \cL}$, for instance for $r=0$ the basic 't~Hooft line is the endpoint of a D1-brane, which couples as $e^{i \!\int\! c}$ to the Ramond field $c$. The boundary 1-form symmetry is generated by surfaces $\wh U_t$, represented for instance by $t = (t_\rme, 0)$.
\end{itemize}
If $N$ is not prime, then there is a larger number $\sigma_1(N) = \sum_{k|N}k$ of Lagrangian subgroups of $\bZ_N \times \bZ_N$, corresponding to global variants of the boundary theory of the form $\bigl( SU(N)/\bZ_k \bigr)_r$.

\subsection{Global 0-form symmetries}

The theories (\ref{5d theory general action}) can have 0-form symmetries as well. On spin manifolds, a (unitary) 0-form symmetry $\omega$ is an automorphism of the discriminant group $\cD_Q$ that preserves the quadratic form:
\be
\omega^\sT Q^{-1} \omega = Q^{-1} \mod 1 \;.
\ee
Since $\omega$ is invertible, it maps Lagrangian subgroups to Lagrangian subgroups. We say that a gapped boundary $\rho(\cL)$ is $\omega$-invariant if the corresponding Lagrangian subgroup is:
\be
\omega \, \cL = \cL \;.
\ee

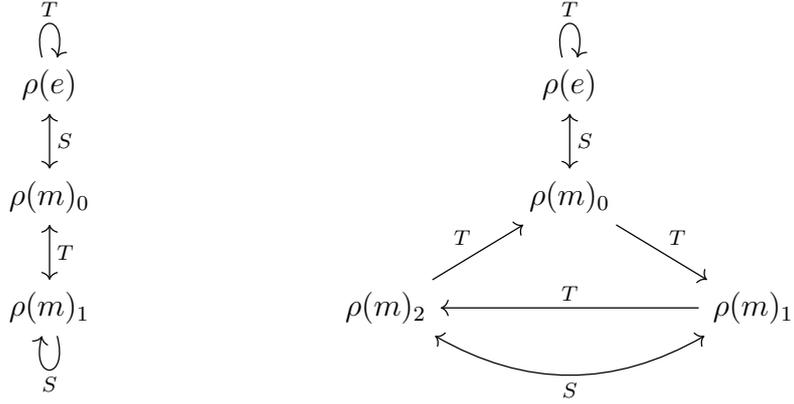
\begin{figure}[t]
$$
\begin{tikzcd}
\rho(e) \arrow[d, leftrightarrow, "S"] \arrow[loop above, "T"] \\
\rho(m)_0 \arrow[d, leftrightarrow, "T"] \\
\rho(m)_1 \arrow[loop below, "S"]
\end{tikzcd}
\hspace{3cm}
\begin{tikzcd}
 & \rho(e) \arrow[d, leftrightarrow, "S"] \arrow[loop above, "T"] \\
 & \rho(m)_0 \arrow[dr, "T"] \\
\rho(m)_2 \arrow[ur, "T"] \arrow[rr, leftrightarrow, bend right=30, "S"'] &  & \rho(m)_1 \arrow[ll, "T"']
\end{tikzcd}
$$
\caption{Action of $PSL(2,\bZ_2) \cong S_3$ (left) and $PSL(2,\bZ_3) \cong A_4$ (right) on Lagrangian subgroups (gapped boundaries).
\label{fig:examplemap}}
\end{figure}

In the type IIB example, the 0-form symmetry group $\Gamma$ is $SL(2,\bZ_N)$, whose generators act on electric and magnetic charges as follows: 
\be
\label{eq:sl25d}
S: (e,m) \mapsto (-m,e) \;,\qquad T: (e,m) \mapsto (e + m, m) \;,\qquad C: (e,m) \mapsto (-e,-m) \;.
\ee
They satisfy $S^2=C$, $T^N = \unit$, and $(ST)^3=C$. If $M$ is the matrix acting on charges, then $M^\sT$ gives the action on the gauge fields $\cB$, as it follows from (\ref{def surface operators Um}). This means that in our conventions
\be
\label{eq:sl25donfields}
S: (b,c) \mapsto (c,-b) \;,\qquad T: (b,c) \mapsto (b,c + b) \;,\qquad C: (b,c) \mapsto (-b,-c) \;.
\ee
All subgroups of $\cD_Q$ are invariant under $C$. For $N$ prime we also have:
\be
\rho(m)_r \,\overset{T}{\longrightarrow}\, \rho(m)_{r+1} \;,\qquad
\rho(e) \,\overset{S}{\longleftrightarrow}\, \rho(m)_0 \;,\qquad
\rho(m)_r \,\overset{S}{\longleftrightarrow}\, \rho(m)_{r_S} \;\;\text{ for } r\neq 0
\ee
where $r_S = - r^{-1}$ in $\bZ_N$, while $\rho(e)$ is invariant under $T$ (see Figure~\ref{fig:examplemap} for two examples).
Lagrangian subgroups form two-terms orbits under $S$, except for $\rho(m)_r$ with $r^2 = -1 \text{ mod } N$ which are invariant. Similarly, they form three-terms orbits under $ST$, except for $\rho(m)_r$ with $r(r+1) = -1 \text{ mod }N$ which are invariant. Gapped boundaries corresponding to $\omega$-invariant subgroups $\cL$ allow for a 0-form symmetry action of the subgroup $G \subset \Gamma$ which stabilizes them.%
\footnote{This is true if $G$ is a normal subgroup of $\Gamma$. This will always be so in the cases of interest to us.}
This is clear from the Lagrangian description of the gapped boundaries, $S[L]$ in (\ref{boundary action}). The action of the 0-form symmetry does not leave the coupling to $\eta$ invariant, but it can be reabsorbed in a redefinition of the generators $L$ of $\cL$.

\subsection{Symmetry defects from higher gauging}

In unitary TQFTs without local operators, all 0-form symmetries are expected to be generated by codimension-1 symmetry defects that are condensations of higher-form symmetries. This statement can be proven in the context of three-dimensional modular tensor categories (MTCs) \cite{Fuchs:2012dt, Roumpedakis:2022aik}, while it seems plausible for higher dimensional TQFTs \cite{Roumpedakis:2022aik}.
In this section we construct the $SL(2,\bZ_N)$ symmetry generators of the 5d CS theory \eqref{eq: 5d BF}, in terms of condensations of the 2-form symmetry on 4d submanifolds. The $\bZ_N \times \bZ_N$ 2-form symmetry generated by the topological surface operators $U_m$ in the 5d bulk becomes a 1-form symmetry on a 4d submanifold $\Sigma$ on which we perform the condensation.

We assume that the fusion algebra of surface operators is strictly associative, and since surfaces cannot braid in 4d, we can condense any subgroup $\mathcal{A} \subset \bZ_N \times \bZ_N$ of the 2-form symmetry. While condensing on a (spin) 4-manifold $\Sigma$, we have the possibility to add discrete torsion in the form of Dijkgraaf-Witten terms \cite{Dijkgraaf:1989pz}.
When we gauge the full group $\bZ_N \times \bZ_N$, the torsion is classified by $\bZ_N^3$ and we label it by $x,y,z \in \bZ_N$. In terms of the background $\Phi \in H^2(\Sigma, \bZ_N \times \bZ_N)$ that we decompose into $\varphi_\rme, \varphi_\rmm \in H^2(\Sigma, \bZ_N)$, the phase of discrete torsion is given by
\be
\label{discrete torsion with cochains}
\Theta_{x,y,z} = \exp \left[ \frac{2\pi i}{2N} \int_\Sigma \Bigl( y \, \fP(\varphi_\rme) + z\, \fP(\varphi_\rmm) + 2x \, \varphi_\rme \cup \varphi_\rmm \Bigr) \right] \;.
\ee
Here $\fP: H^2(\Sigma, \bZ_N) \to H^4(\Sigma, \bZ_{N \gcd(N,2)})$ is the Pontryagin square operation \cite{Whitehead:1949}. For $N$ even, $\fP(\varphi)$ takes values in $\bZ_{2N}$ and on spin manifolds it is an even class, therefore $y,z \in \bZ_N$. For $N$ odd, $\fP(\varphi)$ takes values in $\bZ_N$ and we interpret the exponent as $\frac{2\pi i}N \, 2^{-1} y \int \fP(\varphi_\rme)$ where $2^{-1} = \frac{N+1}2 \text{ mod } N$, and similarly for $z \, \fP(\varphi_\rmm)$, therefore $y,z \in \bZ_N$ once again.
On the other hand, when we gauge a $\bZ_N$ subgroup, the torsion is classified by $\bZ_N$ and then only a combination of $x,y,z$ appears. For simplicity, we will only consider the case that $N$ is a prime number, because then $\bZ_N$ does not contain non-trivial proper subgroups, and all its non-zero elements are invertible.

\begin{figure}[t]
$$
\scalebox{0.85}{\begin{tikzpicture}%
[decoration={markings, mark= at position 0.5 with {\arrow{Stealth}}}] 
\draw[line width =1, color=red, postaction={decorate}] (3,-3) -- (3,-1);
\draw[line width =1, color=red, postaction={decorate}] (3,1) -- (3,3);
\draw[line width =1, color=green!70!black, postaction={decorate}] (3,-1) -- (3,1);
\draw[line width =1, color=blue, postaction={decorate}] (3,-1) -- (6,0);
\draw[line width =1, color=blue, postaction={decorate}] (0,0) -- (3,1);
\draw[fill=black!50!blue] (3,-1) circle (0.1);
\draw[fill=black!50!blue] (3,1) circle (0.1);
\node at (3.3,-2) {$n$};
\node at (3.3,2) {$n$};
\node at (4.5, -0.9) {$m$};
\node at (2.3, -0.4) {$n-m$};
\node at (1.5, 0.9) {$m$};
\draw (0,-3) -- (6,-3) -- (6,3) -- (0,3) -- cycle;
\node at (3,-3.5) {$S^2$};
\node at (-0.5,0) {$T^2$};
\end{tikzpicture}}
\hspace{2cm}
\raisebox{0.5em}{\scalebox{0.9}{\begin{tikzpicture}%
[decoration={markings, mark= at position 0.5 with {\arrow{Stealth}}}]
\draw [dashed] (0,0) +(0:1.81 and 0.7) arc [start angle = 0, end angle = 180, x radius = 1.81, y radius = 0.7];
\draw [line width = 1, color=blue, line cap = round, postaction={decorate}, rotate around = {15:(-0.5, 1.1)}] (-0.5, 1.1) arc [start angle = -90, end angle = -3, x radius = 2.51, y radius = 0.5] node (n1) {} node [pos = 0.45, below right, black] {$m$};
\draw [dashed, line width = 0.5, color=blue, postaction={decorate}, rotate around = {-17:(n1)}] (n1) arc [start angle = 4, end angle = 184, x radius = 1.875, y radius = 0.5] node (n2) {};
\draw [line width = 1, color=blue, line cap = round, postaction={decorate}, rotate around = {22:(n2)}] (n2) arc [start angle = -186, end angle = -110, x radius = 1.8, y radius = 0.5];
\draw [line width = 1, color=red, postaction={decorate}] (-0.5, -0.68) -- (-0.5, 1.1) node [pos = 0.4, left, black] {$n$};
\draw [line width = 1, color=red, postaction={decorate}] (-0.5, 3.2) -- (-0.5, 4.32) node [pos = 0.7, left, black] {$n$};
\draw [line width = 1, color=green!70!black, postaction={decorate}] (-0.5, 1.1) -- (-0.5, 3.2) node [pos = 0.4, left, black] {\footnotesize$n-m$};
\draw [line width = 1, color=brown!90!black] (0.5, -0.2) -- (0.5, 1.25) node [pos = 0.1, right, black] {$l$};
\draw [line width = 1, color=brown!90!black, postaction={decorate}] (0.5, 1.6) -- (0.5, 4.2);
\draw [line width = 1, color=brown!90!black] (0.5, 4.45) -- (0.5, 4.9);
\draw (0,0) +(0:1.81 and 0.7) arc [start angle = 0, end angle = -180, x radius = 1.81, y radius = 0.7];
\draw (-1.81, 0) to (-1.81, 5);
\draw (1.81, 0) to (1.81, 5);
\draw (0, 5) ellipse [x radius = 1.81, y radius = 0.7];
\draw [fill=black!50!blue] (-0.5, 3.2) circle (0.07);
\draw [fill=black!50!blue] (-0.5, 1.1) circle (0.07);
\end{tikzpicture}}}
\hspace{1.5cm}
\raisebox{2em}{\scalebox{0.9}{\begin{tikzpicture}
\draw [line width = 1, color=blue, postaction={decorate, decoration={markings, mark = at position 0.58 with \arrow{Stealth}}}] (0,0) +(78: 1.7 and 0.7) arc [start angle = 78, end angle = 428, x radius = 1.7, y radius = 0.7] node [pos = 0.65, below right, black] {$U_m$};
\draw [line width = 1, color=brown!90!black] (0.5, -2) node [right, black] {$U_l$} -- (0.5, -0.85);
\draw [line width = 1, color=brown!90!black, postaction={decorate, decoration={markings, mark = at position 0.25 with \arrow{Stealth}}}] (0.5, -0.5) -- (0.5, 2.4);
\draw [line width = 8, color=white] (-0.7, -2.4) -- (-0.7, 2);
\draw [line width = 1, color=red, postaction={decorate, decoration={markings, mark = at position 0.3 with \arrow{Stealth}}}] (-0.7, -2.4) -- (-0.7, 2) node [pos = 0.2, left, black] {$U_n$};
\end{tikzpicture}}}
\hspace{1cm}
$$
\caption{The condensation defect on $\Sigma = T^2 \times S^2$ with its network of 2d defects (left) surrounds a topological defect $U_l$ placed on $T^2 \times \{0 \}$ (center), where $0\in B^3$ is the center of the 3-ball whose boundary is $S^2$. Up to a phase (\ref{normal ordering phase}), the network can be resolved into a collection of closed surfaces with no junctions (right).
\label{fig: cylinder}}
\end{figure}
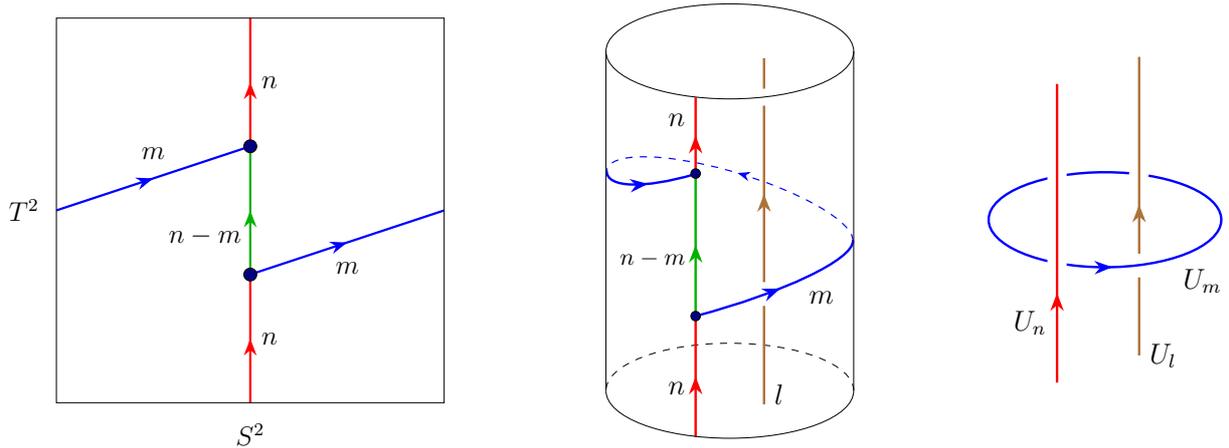

We want to compute the action of the 0-form condensation defects $V$ on the 2-form defects $U_l$. To that purpose, we place $U_l$ along $\bR^2$ and wrap $V$ around them, namely we place $V$ on $\bR^2 \times S^2$ with $S^2$ surrounding $U_l$. It turns out that it is more clear to perform condensation on compact submanifolds, therefore we substitute $\bR^2$ with $T^2$. Eventually, we place $U_l$ on $T^2$ and $V$ on $\Sigma \equiv T^2 \times S^2$ around $U_l$ (as in Figure~\ref{fig: cylinder} center).

To condense on $\Sigma$, we decompose  the 2-form symmetry background $\Phi \in H^2(\Sigma, \bZ_N\times \bZ_N)$ into a pair of backgrounds $\{ \phi^{T^2}, \phi^{S^2} \}$ on the two factors of $\Sigma$, and we denote by $n = (n_\rme, n_\rmm)$ the holonomy of $\phi^{S^2}$ on $S^2$ (representing defects on $T^2$) and by $m = (m_\rme, m_\rmm)$ the holonomy of $\phi^{T^2}$ on $T^2$ (representing defects on $S^2$). Given a class $(x,y,z) \in \bZ_N^3$ representing the choice of discrete torsion (\ref{discrete torsion with cochains}), its contribution to the path integral is
\be
\Theta_{x,y,z}(n,m) = \exp \left[ \frac{2\pi i}{N} \Bigl( x \, \bigl( n_\rme m_\rmm + n_\rmm m_\rme \bigr) + y \, n_\rme m_\rme + z \, n_\rmm m_\rmm \Bigr) \right] = \exp \left[ \frac{2\pi i}{N} \, m^\sT \cT n \right]
\ee
where we introduced the symmetric matrix of discrete torsions 
\be
\cT = \mat{ y & x \\ x & z} \;,
\ee
whose entries are in $\bZ_N$.
We can label the condensation defects of the 5d Chern-Simons theory as $V[\cA, \cT]$, where $\cA$ is the condensed subgroup of $\bZ_N \times \bZ_N$ and $\cT$ is the matrix of discrete torsions. When $\cA=\bZ _N\times \bZ _N$ we omit it, while when $\mathcal{A}$ is one-dimensional we denote it by one of its generators $(p,q)$.

The condensation on $\Sigma$ involves a network of 2-dimensional defects, as in Figure~\ref{fig: cylinder} left. Instead of working with a network (that requires to understand the trivalent junctions), we can resolve it into a pair of 2-dimensional defects: one $U_n$ along $T^2$ on an outer copy of $\Sigma$, and one $U_m$ along $S^2$ on an inner copy of $\Sigma$ (Fig.~\ref{fig: cylinder} right). This operation involves a phase, and is equivalent to a normal ordering prescription. More generally, for $N$ odd we can write
\be
\label{normal ordering phase}
U(\gamma_1 + \gamma_2) = \exp \left[ - \frac{2 \pi i }{N} \, 2^{-1} \, \langle \gamma_1 , \gamma_2 \rangle \right] U(\gamma_1) \, U(\gamma_2) \;.
\ee
(The case of $N$ even is discussed below.) Here $\gamma_i \in H^2(\Sigma, \bZ_N \times \bZ_N)$ represent two defects on $\Sigma$, while $\langle \,,\, \rangle$ is the product of the (symmetric) cup product on $\Sigma$ and the (antisymmetric) Dirac pairing in $\bZ_N \times \bZ_N$. On the right-hand-side, $U(\gamma_1)$ is outer while $U(\gamma_2)$ is inner. This is essentially a square root of the braiding matrix
\be
\label{braiding matrix}
B_{mn} = e^{\frac{2\pi i}N \, \langle m, n \rangle}
\ee
as in Figure~\ref{fig: 5d linking} left. We obtain:
\bea
\label{first definition 4d defects V}
V[\cA,\cT] \, U_l &= \frac1{\bigl| H^2(\Sigma, \cA) \bigr|^{1/2} } \sum_{n,m \in\cA} \Theta_{x,y,z} (n,m) \; e^{- \frac{2\pi i}N 2^{-1} \langle n,m \rangle } \, B_{ml} \; U_{l+n} \\
&= \frac1{|\cA|} \sum_{n,m \in \cA} \exp\biggl[ \frac{2\pi i}N \biggl( m_\rme \Bigl( y\, n_\rme + x\, n_\rmm + \tfrac12 n_\rmm + l_\rmm \Bigr) \\
&\hspace{4.2cm} + m_\rmm \Bigl( x\, n_\rme + z\, n_\rmm - \tfrac12 n_\rme - l_\rme \Bigr) \biggr) \biggr] \, U_{l + n} \;.
\eea
The sum over $m$ produces a delta function for $n$. When this has exactly one solution, the sum over $n$ selects a defect $U_{M l}$ where $M \in SL(2,\mathbb{Z}_N)$ is the group element corresponding to the condensation defect $V[\cA,\cT] \equiv V_M$. The cases in which there are multiple or no solutions, even though they are not relevant to our purposes, will be discussed at the end.

If  $\cA \cong \bZ_N$ is generated by $(p,q)$, we can write 
\be
m_\rme = \mu \, p \;,\qquad m_\rmm = \mu \, q \;,\qquad n_\rme = \nu \, p \;,\qquad n_\rmm = \nu\, q \;.
\ee
Notice that the phase (\ref{normal ordering phase}) trivializes.
The sum over $\mu$ produces a delta function that fixes $p\, l_\rmm - q\, l_\rme + \xi\, \nu = 0$ and selects one value for $\nu$ (as long as $\xi \neq 0$), where $\xi =2pqx+yp^2+zq^2$. This reproduces the action of$\,$%
\footnote{One has $\ds T^k_\cH = \mat{ 1- kpq & kp^2 \\ - kq^2 & 1 + k pq }$, $\nu = k p l_\rmm - k q l_\rme$ and so $\ds T^k_\cH \mat{ l_\rme \\ l_\rmm} = \mat{ l_\rme + \nu p \\ l_\rmm + \nu q}$.
\label{foo: matrix TkH}}
\be
M = T^k_\cH \equiv \cH T^k \cH^{-1}
\ee
for $k = -\xi^{-1}$ and
\be
\cH = \mat{ p & * \\ q & *} \,\in\, SL(2,\bZ_N) \;.
\ee
The three parameters $x,y,z$ enter only in the combination $\xi$, as expected since the discrete torsion is classified by $\bZ_N$. Since $T^k$ leaves invariant the vector $v=(1,0)$, then $T^k_\cH$ leaves invariant the vector $\cH v =(p,q)$, and we obtain the defect implementing $T_\cH^k$ by condensing the algebra generated by $(p,q)$ (with a non-vanishing torsion determined by $k$).
For instance, $T^k$ is obtained by condensing the electric surfaces $(n_\rme, 0)$, while its electromagnetic dual $ST^kS^{-1}$ is realized by condensing the magnetic surfaces $(0,n_\rmm)$. An element of $SL(2,\bZ _N)$ (with $N$ prime) can be written as $\cH T^k \cH^{-1}$ if and only if its trace is $2 \text{ mod }N$. There are $N^2$ such elements, including the identity.%
\footnote{All matrices $M \in SL(2,\bZ_N)$ with $\Tr M = 2$ can be written as $M = \rule{0pt}{1.3em} \smat{1 - \alpha & \beta \\ \gamma & 1+ \alpha}$ with $\alpha^2 = \beta\gamma$ mod $N$. This equation, for $N$ prime, has $N^2-1$ solutions with at least one of $\alpha$, $\beta$, $\gamma$ not zero. One can also easily show that, for $N$ prime, any such matrix $M$ can be written as in footnote~\ref{foo: matrix TkH}. The total number of elements in $SL(2,\bZ_N)$ ($N$ prime) is instead $N^3 -N$.}
Indeed condensation produces $N-1$ defects (as we change the torsion $\xi$) for each of the $N+1$ $\bZ_N$ subgroups of $\bZ_N \times \bZ_N$, besides the identity (which is formally obtained by condensing the trivial subgroup $(0,0)$). We will comment on the case with vanishing torsion below.

The elements of $SL(2,\bZ_N)$ ($N$ prime) with trace different from 2 are obtained by condensing the full $\bZ _N\times \bZ _N$. The sum over $m$ produces a delta function that fixes%
\footnote{When working in $\bZ_N$ with $N$ prime, by fractions we always mean the inverse element mod $N$.}
$\bigl( \cT + \frac\epsilon2 \bigr) n + \epsilon l = 0$ and selects one value of $n$ (as long as $\bigl( \cT + \frac\epsilon2\bigr)$ is invertible). This reproduces the action of
\be
\label{exp of M from T}
M = \Bigl( \cT + \frac\epsilon2 \Bigr)^{-1} \Bigl( \cT - \frac\epsilon2 \Bigr) \;.
\ee
Note that $\det\bigl( \cT \pm \frac\epsilon2 \bigr) = ( 2 - \Tr M \big )^{-1}$, therefore all elements $M \in SL(2,\bZ_N)$ with $\Tr M \neq 2$ can be obtained this way. The relation can be inverted:
\be
\label{exp of T from M}
\cT = \frac\epsilon2 \, \bigl( 1 + M \bigr) \, \bigl( 1 - M \bigr)^{-1} \;.
\ee
Notice that the two factors on the right-hand-side commute. Moreover, in $SL(2,\bZ_N)$ we have $\det(1 \pm M) = 2 \pm \Tr M$. The following relation is also useful:
\be
\label{relation T with M}
\cT + \frac\epsilon2 = \epsilon \, \bigl( 1 - M \bigr)^{-1} \;.
\ee
Explicitly, the discrete torsion that produces the symmetry defect for $M = \smat{ a & b \\ c & d} \in SL(2,\bZ_N)$ with $\Tr M \neq 2$ is $x = \frac{d-a}{2(2-a-d)}$, $y = \frac{c}{2-a-d}$, $z = - \frac{b}{2-a-d}$. Finally, assuming that $\bigl( \cT + \frac\epsilon2 \bigr)$ is invertible, $\det \cT = \Tr(1+M) \bigl[ 4 \Tr (1-M) \big]^{-1}$ therefore $\cT$ is invertible if and only if $\Tr M \neq -2 \text{ mod }N$. The case $\cT =0$ corresponds to $M = -\unit \equiv C$ which is charge conjugation. The case that $\cT$ has rank 1 corresponds (for $N$ prime) to $M = C \cH T^k \cH^{-1}$ where
\be
\cT = \frac{k}4 \mat{ q^2 & -pq \\ -pq & p^2} = \frac{k}4 \, (\epsilon v) \cdot (\epsilon v)^\sT \;,\qquad v = \mat{ p \\ q} \qquad\text{and}\qquad \cH = \mat{ p & * \\ q & *} \;.
\ee
In Table~\ref{tab: examples of condensation} we summarize a few examples. 

\begin{table}[t]
$$
\begin{array}{c|c|c|c}
M \in SL(2,\bZ_N) & M \cdot (l_1, l_2) & \cA & (x,y,z)  \\[.3em]
\hline\hline
\rule{0pt}{1.3em}  \rule{1em}{0pt} C = \smat{-1 & 0 \\ 0 & -1}  & (-l_1,\, -l_2) & \bZ_N \times \bZ_N & (0,0,0) \\[.3em]
\hline
\rule{0pt}{1.3em}  CT^k = \smat{-1 & -k \\ 0 & -1}  & (-l_1 - kl_2,\, -l_2) & \bZ_N \times \bZ_N & (0,0,\frac14 k) \\[.3em]
\hline
\rule{0pt}{1.3em}  \rule{0.6em}{0pt} S = \smat{0 & -1 \\ 1 & 0} & (-l_2 ,\, l_1) & \bZ_N \times \bZ_N & \bigl( 0,  \frac12 ,  \frac12 \bigr) \\[.3em]
\hline
\rule{0pt}{1.3em} ST = \smat{0 & -1 \\ 1 & 1} & (-l_2 ,\, l_1 + l_2) & \bZ_N \times \bZ_N & \bigl( \frac12 ,1,1 \bigr) \\[.3em]
\hline
\rule{0pt}{1.3em} (ST)^2 = \smat{-1 & -1 \\ 1 & 0}  \rule{0.5em}{0pt} & (-l_1-l_2 ,\, l_1) & \bZ_N \times \bZ_N & \bigl(  \frac16 ,  \frac13, \frac13 \bigr) \\[.3em]
\hline
\rule{0pt}{1.3em} T^k  = \smat{ 1 & k \\ 0 & 1} & (l_1 + k l_2,\, l_2 ) & \bigl\langle (1,0) \bigr\rangle & \xi = - k^{-1}
\end{array}
$$
\caption{Examples of $SL(2,\bZ_N)$ condensation defects, obtained by condensing $\cA \subseteq \bZ_N \times \bZ_N$ with torsion $(x,y,z)$, for $N>3$ prime (for $N=2,3$ some of those formulas are different).
\label{tab: examples of condensation}}
\end{table}

\paragraph{Small values of \matht{N}.} Some of the previous formulas are ill-defined for small $N$. For $N=2$, and more generally for $N$ even, we cannot use the normal ordering prescription in (\ref{normal ordering phase}) because $2^{-1}$ is ill-defined. However, notice that the phase that enters in the definition (\ref{first definition 4d defects V}) of the operator $V[\cA,\cT]$ is the product of the torsion and the normal ordering phases:
\be
\exp\biggl[ \frac{2\pi i}N \, m^\sT \mat{ y & \tilde x \\ \tilde x - 1 & z} n \biggr] \,\equiv\, \exp\biggl[ \frac{2\pi i}N \, m^\sT \wt\cT n \biggr] \;,
\ee
where $\tilde x = x +\frac12$ and $\wt\cT = \cT + \frac\epsilon2$. The quantities $\tilde x \in \bZ_N$ and $\wt\cT$ are well defined, even for $N$ even, and we can use them to classify the torsion. The group $SL(2,\bZ_2) \cong PSL(2,\bZ_2) \cong S_3$ (note that $C \cong \unit$) has 6 elements, 4 of which have trace equal to 2~mod~2:
\be
T = \mat{1 & 1 \\ 0 & 1} \;,\qquad STS^{-1} = \mat{1 & 0 \\ 1 & 1} \;,\qquad S = (TS) T (TS)^{-1} = \mat{ 0 & 1 \\ 1 & 0} \;,
\ee
besides the identity. The corresponding defect operators are obtained by condensing the $\bZ_2$ subgroup generated by $(1,0)$, $(0,1)$ and $(1,1)$, respectively, with non-vanishing torsion $\xi = pq + yp^2 + zq^2 = 1$. The remaining elements, 
\be
ST = \mat{0 & 1 \\ 1 & 1} \qquad\text{and}\qquad  (ST)^2 = \mat{1 & 1 \\ 1 & 0} \;,
\ee
have trace equal to 1 and are described by gauging the full $\bZ_2 \times \bZ_2$. The relation between torsion and symmetry action $M$ is as in (\ref{exp of M from T}) and (\ref{relation T with M}), as long as one parametrizes the torsion using $\wt\cT$, therefore $\wt\cT = \epsilon \, (1-M)^{-1}$. One finds that $ST$ is obtained from torsion $(\tilde x, y, z) = (1,1,1)$, while $(ST)^2$ is obtained from $(\tilde x, y, z) = (0,1,1)$. These two values of the torsion are the only possible ones providing a matrix $\wt\cT$ invertible in $\bZ_2$.

For $N=3$, the element $(ST)^2$ in Table~\ref{tab: examples of condensation} has trace equal to 2~mod~3. Indeed we can write $(ST)^2 = \cH T^2 \cH^{-1}$ with $\cH = \smat{1 & 0 \\ 1 & 1}$, and thus the corresponding defect operator is obtained by condensing the $\bZ_3$ subgroup generated by $(1,1)$ with torsion $\xi = 1$.

\paragraph{Fusion.}
The fusion rules of (invertible) condensation defects correctly satisfy the product of $SL(2,\bZ_N)$. The method we describe below is general, however for brevity we only exhibit the product of defects obtained by condensing the full group $\bZ_N \times \bZ_N$. The defect operators $V[\cT]$ on $\Sigma$ defined in (\ref{first definition 4d defects V}) can be rewritten as
\be
\label{eq:V}
V[\cT] = \frac1{N^2} \sum_{n,m \,\in\, \bZ_N \times \bZ_N} \exp\biggl[ \frac{2\pi i}N \, m^\sT \biggl( \cT + \frac\epsilon2 \biggr)n \biggr] \, U_n[T^2] \, U_m[S^2] 
\ee
where we indicated whether the two-dimensional defects $U$ are placed on $T^2$ or $S^2$, and rightmost operators are inner. Using the braiding matrix (\ref{braiding matrix}), we obtain
\begin{align}
& V[\cT_2] \, V[\cT_1] = \\
&\quad = \frac{1}{N^4} \sum_{\substack{n,m \\ n', m'}} \exp\biggl[ \frac{2\pi i}N \biggl( m^\sT \biggl( \cT_2 + \frac\epsilon2 \biggr)n + m^{\prime\sT} \biggl( \cT_1 + \frac\epsilon2 \biggr) n' + m^\sT \epsilon n' \biggr) \biggr] \, U_{n+n'}[T^2] \, U_{m + m'}[S^2] \;. \nn
\end{align}
Setting $n = l - n'$, $m = k - m'$ and performing the sum over $m'$ produces a delta function on $(\cT_1 + \cT_2) n' = \bigl( \cT_2 + \frac\epsilon2 \bigr) l$. When $(\cT_1 + \cT_2)$ is invertible, one eliminates $n'$ obtaining
\be
\label{fusion of defects}
V[\cT_2] \times V[\cT_1] = V[\cT_{21}]
\ee
with
\be
\label{exp for T21}
\cT_{21} = \cT_2 - \biggl( \cT_2 - \frac\epsilon2 \biggr) \bigl( \cT_1 + \cT_2 \bigr)^{-1} \biggl( \cT_2 + \frac\epsilon2 \biggr) \;.
\ee
The relation (\ref{exp for T21}) can be rewritten as
\be
\cT_{21} + \frac\epsilon2 = \biggl( \cT_1 + \frac\epsilon2 \biggr) \bigl( \cT_1 + \cT_2 \bigr)^{-1} \biggl( \cT_2 + \frac\epsilon2 \biggr) \;.
\ee
Together with (\ref{relation T with M}), with a little bit of algebra, it implies $M_{21} = M_2 M_1$ as expected.

When $\cT _1+\cT _2=0$, the sum over $m'$ and $n'$ sets $l=k=0$. We conclude that
\be
V[\cT] \times V[-\cT] = \unit \;,
\ee
in agreement with the fact that $M(-\cT) = M(\cT)^{-1}$. The case in which $\cT_1 + \cT_2$ has rank 1 can be treated in a similar way. In particular, from (\ref{relation T with M}) it follows that
\be
(1-M_2) \, \epsilon^{-1} (\cT_1 + \cT_2) (1-M_1) \, \epsilon^{-1} = (1-M_2 M_1) \, \epsilon^{-1} \;.
\ee
Taking the determinant on both sides and using that $\det (1-M) = \Tr (1-M)$ for $M$ in $SL(2, \bZ)$, we conclude that $(\cT_1 + \cT_2)$ is invertible if and only if $M_2 M_1$ has trace different from 2~mod~$N$, whilst $(\cT_1 + \cT_2)$ has rank 1 if and only if $M_2 M_1$ has trace equal to 2~mod~$N$ but is not the identity, and thus the corresponding symmetry operator is described by the condensation of a subgroup $\cA \cong \bZ_N$.

\paragraph{Degenerate torsion and non-invertible surfaces.}
Besides the $SL(2,\bZ _N)$ symmetry defects, higher gauging can produce projectors when the symmetry we condense on a submanifold could also be condensed in the bulk \cite{Roumpedakis:2022aik}. This is the case when the condensed group is non-anomalous and the discrete torsion would be allowed in 5d. One example is the condensation of $\cA = \bigl\langle (p,q) \bigr\rangle \cong \bZ _N$ with vanishing torsion. From the analysis that follows (\ref{first definition 4d defects V}) we see that the delta function introduced by the sum over $\mu$ either has no solution, or has $|\cA|=N$ solutions, and the operator $V[\cA,0]$ acts on the surfaces $U_l$ as
\be
\label{projectors on lines}
V[\cA,0] \, U_l = \begin{cases}
0 & \text{if } l \notin \cA \;, \\
\sum _{n \in \cA} U_n \phantom{\quad} & \text{if } l \in \cA \;.
\end{cases}
\ee
This is consistent with the following non-invertible composition law \cite{Roumpedakis:2022aik}:
\be 
V[\cA,0]\times V[\cA,0]=\bigl| H^2(\Sigma, \cA) \bigr|^{1/2} \; V[\cA,0] \;,
\ee
where the coefficient on the right-hand-side is the partition function of a TQFT.

Besides, we obtain a non-invertible surface when we condense the full $\bZ_N \times \bZ_N$ 2-form symmetry (which is anomalous in the bulk) with a torsion matrix $\cT$ such that $\bigl( \cT + \frac\epsilon2 \bigr)$ is not invertible. Notice that, since $\cT$ is symmetric and $\epsilon$ antisymmetric, if $\bigl( \cT + \frac\epsilon2 \bigr)$ is non-invertible  then it has rank 1.%
\footnote{This is also true for $N=2$, because the matrix $\wt\cT = \smat{ y & \tilde x \\ \tilde x -1 & z}$ cannot be zero.}
In that case, there exist two integer vectors $v_{1,2} \in \bZ_N \times \bZ_N$ such that
\be
\label{degenerate T}
\cT + \frac\epsilon2 = (\epsilon v_1) \cdot (\epsilon v_2)^\sT \qquad\text{and}\qquad v_1^\sT \epsilon \, v_2 = 1 \;.
\ee
The second condition comes from the antisymmetric part of the matrix. The sum over $m$ in (\ref{first definition 4d defects V}) gives a delta function on the solutions to $\bigl( \cT +\frac\epsilon2 \bigr) n = - \epsilon l$, that takes the form
\be
\label{internal equation for projector}
\bigl( v_2^\sT \epsilon \, n \bigr) \, v_1 = l \;.
\ee
Let $\cA_{1,2} \cong \bZ_N$ be the two subgroups of $\bZ_N \times \bZ_N$ generated by $v_{1,2}$, respectively. If $l \notin \cA_1$ then (\ref{internal equation for projector}) has no solution for $n$. On the contrary, if $l \in \cA_1$ then the solutions are $n = -l + \nu v_2$ with any $\nu \in \bZ_N$. We obtain:
\be
V[\cT] \, U_l = \begin{cases}
0 & \text{if } l \notin \cA_1 \;, \\
\sum _{n \in \cA_2} U_n \phantom{\quad} & \text{if } l \in \cA_1 \;.
\end{cases}
\ee
In fact, given two different $\bZ_N$ subgroups $\cA_1 \neq \cA_2$ (then, for $N$ prime, $\cA_1 \cap \cA_2 = (0,0)$ necessarily), one easily checks the composition law
\be
V[\cA_2, 0] \, V[\cA_1, 0] = V[\cT]
\ee
where, on the right-hand-side, $\cT$ is given by (\ref{degenerate T}).

\subsection{Continuum description of symmetry defects}
\label{sec: continuum descr defects}

In view of describing the twisted sectors of the $SL(2,\bZ_N)$ symmetry, it is useful to reformulate the previous discussion in terms of continuum Lagrangians. We take $N$ odd. When the condensed group is $\cA = \bZ_N \times \bZ_N$, the defect $V[\cT]$ is described by a 4d TQFT with two dynamical 2-forms $\Phi = (\varphi_\rme, \varphi_\rmm)$ and four 1-forms $\Psi = (\psi_\rme, \psi_\rmm)$, $\Gamma = (\gamma_\rme, \gamma_\rmm)$ with action \cite{Kapustin:2014gua}:
\be
\label{eq: defect lagrangian full gauging}
S[\cT] = \frac{N}{2\pi} \int_\Sigma \biggl[ \cB^\sT \bigl( \Phi + d\Gamma \bigr) + \Phi^\sT d \Psi + \frac12 \, \Phi^\sT \cT \Phi\biggr] \;.
\ee
The torsion is parametrized by the symmetric matrix $\cT$ with entries in $\bZ_N$ and such that $\cT + \frac\epsilon2$ is invertible.%
\footnote{After integrating over $\Psi$, the periods of $\Phi$ are multiples of $\frac{2\pi}N$. Thus on spin manifolds $\Sigma$, shifts of the entries of $\cT$ by $N$ leave $e^{iS}$ invariant \cite{Hsin:2018vcg}.}
On the other hand, when $\cA \cong \bZ_N$ is generated by $(p,q)$ we only keep one 2-form $\varphi$ and two 1-forms $\psi$, $\gamma$ with action:
\be
\label{eq: defect lagrangian gauging subgroup}
S\bigl[ \bigl\langle (p,q) \bigr\rangle ,\, \xi \bigr] = \frac{N}{2\pi} \int_\Sigma \biggl[ \bigl( p b + q c \bigr) \, \bigl( \varphi + d \gamma \bigr)+ \varphi \, d\psi + \frac\xi2 \, \varphi \varphi \biggr] \;.
\ee
The torsion is parametrized by a non-vanishing $\xi \in \bZ_N$.

Integrating over $\Psi$ and $\Gamma$ in (\ref{eq: defect lagrangian full gauging}) forces $\Phi$ and the pull-back of $\cB$ to be in 
$H^2(\Sigma, \bZ_N \times \bZ_N)$. Then $\Phi$ can be identified with the Poincar\'e dual to a 2-cycle $\sigma \in H_2(\Sigma, \bZ_N \times \bZ_N)$. Since $\Phi$ couples to $\cB$, $\Phi = \text{PD}(\sigma)$ represents a two-dimensional defect $U[\sigma]$ wrapped on $\sigma$, and the theory (\ref{eq: defect lagrangian full gauging}) reproduces higher gauging of the $\bZ_N \times \bZ_N$ 2-form symmetry on $\Sigma$ with torsion $\cT$. A similar discussion applies to (\ref{eq: defect lagrangian gauging subgroup}).
The action \eqref{eq: defect lagrangian full gauging} is invariant under the following gauge transformations:
\be
\label{eq:shiftsymmetry}
\cB \to \cB +d\alpha \;,\qquad \Phi \to \Phi +d\lambda \;,\qquad \Psi \to \Psi - \cT  \lambda - \alpha + d\mu \;,\qquad \Gamma \to \Gamma - \lambda + d\nu \;.
\ee
Considering $\cB$ as a background field, the theory \eqref{eq: defect lagrangian full gauging} is of a different type depending on whether $\cT$ is an invertible matrix over $\bZ_N$ or not.

If $\cT$ is invertible in $\bZ_N$, then \eqref{eq: defect lagrangian full gauging} is an invertible TQFT.
Indeed, adapting the discussion in \cite{Kapustin:2014gua} to our case, all closed surfaces $\exp\bigl( i m^\sT \!\!\oint\! \Phi \bigr)$ are gauge invariant and implement a $\bZ_N \times \bZ_N$ 1-form symmetry, however, because of the equation of motion $\cT\Phi = - d\Psi$, when $m$ is in the image of the map $\cT: \bZ _N^2 \to \bZ _N^2$, the surface acts trivially. Therefore only $(\bZ_N \times \bZ_N) / \im \cT$ acts faithfully, and if $\cT$ is invertible in $\bZ_N$ then there is no faithful action at all. On the other hand, the line integrals of $\Psi$ might not be gauge invariant by themselves and need to be the boundary of an open surface $\cD$:  $\exp\bigl( ik^\sT \!\!\oint_\ell\! \Psi +  ik^\sT \cT \!\int_\cD\! \Phi \bigr)$ with $\ell = \partial \cD$. They become pure line operators when the surface is transparent, \ie, when $\cT k = 0 \text{ mod }N$. Hence the 2-form symmetry of the theory is $\ker \cT \subset \bZ_N \times \bZ_N$, which is trivial if $\cT$ is invertible in $\bZ_N$.
Summarizing, if $\cT$ is invertible in $\bZ_N$ then the theory \eqref{eq: defect lagrangian full gauging} has no topological operators, and is thus an invertible TQFT. This implies that we could integrate out the fields $\Phi$ and $\Psi$. Their equations of motion say that $\Phi \in H^2(\Sigma, \bZ_N \times \bZ_N)$ and $\cT^{-1}(\cB + d\Psi) + \Phi = \check\Phi$, where $\check\Phi \in H^2(\Sigma, \bZ_1 \times \bZ_1)$ is a gauge field with integer periods, while $\cT^{-1}$ is the inverse of $\cT$ in $\bZ_N$. Substituting into the action, one obtains
\be
\label{anomaly first version}
S_\text{invertible}[\cT] = \frac{N}{2\pi} \int_\Sigma \biggl[ \cB^\sT d \wt\Gamma - \frac12 \, \cB^\sT \cT^{-1} \cB \biggr] \;,
\ee
up to total derivatives and multiples of $2\pi$, where $\wt\Gamma = \Gamma - \cT^{-1} \Psi$ transforms as $\wt\Gamma \to \wt\Gamma + \cT^{-1} \alpha$ under gauge transformations.

If, on the contrary, $\cT$ is a non-invertible matrix, then the 4d theory is a non-trivial TQFT with surface and line operators labeled by $(\bZ_N \times \bZ_N) / \im \cT$ and $\ker \cT$, respectively. Recall that this case corresponds to $SL(2,\bZ_N)$ matrices $M$ with $\Tr M = -2 \text{ mod }N$, which are of the form $M = C \cH T^k \cH^{-1}$. In the special case $\cT = 0$ (that corresponds to $M=C$) the 4d theory (\ref{eq: defect lagrangian full gauging}) is a pure $\bZ_N\times \bZ_N$ gauge theory, whose 1-form symmetry is coupled to the background field $\cB$.

\begin{figure}[t]
\centering
\begin{tikzpicture}
	\draw [->] (-3.5, 0) -- (3.5, 0) node[right] {$r$};
	\draw [->] (0, -0.2) node[below] {$0$} -- (0, 3) node[left] {$\cB$} node[pos = 0.2, right] {$V[\cT]$};
	\draw [blue, thick] (-3.4, 0.3) node[above right, black] {$\cB_\text{L} = M^\sT \cB_\text{R}$} to (-0.5, 0.3) to[out = 0, in = -95] (0, 1.4) to[out = 85, in = 180] (0.5, 2.5) to (3.4, 2.5) node[below left, black] {$\cB_\text{R}$};
	\node at (-2, 2) {Interior};
	\node at (2.5, 1) {Exterior};
\end{tikzpicture}
\caption{The 4d symmetry defect $V[\cT]$ induces a discontinuity in the gauge field $\cB$ across its surface. Compared with the setup of Figure~\ref{fig: cylinder} center, the region $r<0$ is the interior of the cylinder while $r>0$ is the exterior.
\label{fig: B-field jump}}
\end{figure}
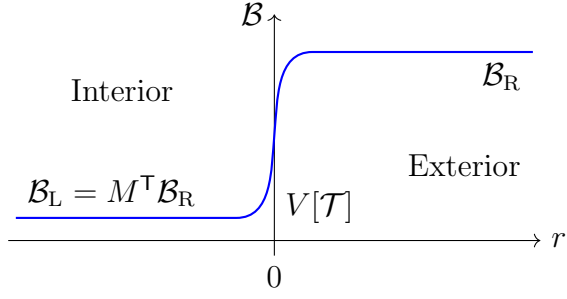

We can verify that $S[\cT]$ in (\ref{eq: defect lagrangian full gauging}) implements the correct transformation of 2d defects $U_l$. We introduce a coordinate $r$ transverse to the 4d defect, such that $\Sigma = \{r=0\}$, and consider the bulk-plus-defect action 
\be
\label{eq: full 5d lagrangian}
S_{\substack{\text{bulk plus} \\ \text{defect} \hspace{1em}}} =\frac{N}{4\pi} \int  \cB^\sT \epsilon \,  d\cB + \frac{N}{2\pi} \int _{r=0}  \biggl[ \cB^\sT \bigl( \Phi+ d\Gamma \bigr) + \Phi^\sT d\Psi + \frac12 \, \Phi^\sT \cT \Phi \biggr] \;.  
\ee
Integrating out the gauge field $\Phi$ we obtain an effective description of the interface, which induces a discontinuity
\be
\label{eq: LR matching}
\cB_L = M^\sT \cB _R
\ee
in the gauge field $\cB$ ($L,R$ stand for left/right at $r<0$ and $r>0$, respectively). Here $M$ is transposed because the $SL(2,\bZ)$ action on fields is dual to the one on charges, that we previously denoted by $M$.
Indeed, imagine placing a 2d defect operator $U_l$ in the region $r<0$ (see Figure~\ref{fig: B-field jump}) which, compared with our previous setup in Figure~\ref{fig: cylinder} center, would be the interior region. The expectation value of the operator is $\exp\bigl( i l^\sT \!\!\int\! \cB_L \bigr) = \exp\bigl( i l^\sT M^\sT \!\!\int\! \cB_R \bigr)$. Thus, for an external observer, the compound system of the 4d defect on $\Sigma$ wrapping the 2d operator $U_l$ appears as a 2d operator $U_{Ml}$.
Let us determine $M$ from (\ref{eq: full 5d lagrangian}). After choosing a gauge $(\lambda, \alpha)$ in which $\Gamma$ and $\Psi$ are zero, the equations of motion for $\cB$ and $\Phi$ read
\begin{align}
\label{relation between B and Phi}
0 &= \bigl( \cB + \cT \Phi \bigr) \, \delta(r) dr \\
\epsilon \, d\cB &= - \Phi \, \delta(r) dr \;.
\end{align}
The gauge field $\Phi$ acts as a source for $\cB$.
Working in a gauge in which $\cB_{r i}=0$, we have $\partial _r \cB (r) = \epsilon \, \Phi \, \delta (r)$. This differential equation can be solved: $\cB (r) = \cB_L + \epsilon \, \Phi \, \theta (r)$, where $\cB_L$ is the value of $\cB$ for $r<0$. Multiplying by $\delta(r)$, integrating in a neighbourhood of $r=0$ and using $\delta(r) = \partial_r \theta(r)$, we obtain $\cB(0) = \cB_L + \frac\epsilon2 \, \Phi = - \cT \Phi$. The second equality follows from (\ref{relation between B and Phi}). Finally, evaluating at $r>0$ we find $\cB_R = \cB_L + \epsilon \, \Phi$ which implies
\be
\cB_R = \biggr[ 1 - \epsilon \Bigl( \cT + \frac\epsilon2 \Bigr)^{-1} \biggr] \, \cB_L \;.
\ee
This discontinuity, when written in terms of $M$ using (\ref{relation T with M}), is exactly (\ref{eq: LR matching}). If $\cT$ is invertible, one can repeat the computation using (\ref{anomaly first version}) obtaining the same result.

When $\cA \cong \bZ _N$ one should use the defect Lagrangian \eqref{eq: defect lagrangian gauging subgroup} with only one gauge field $\varphi$. For instance, when the defect action is coupled to $b$ (\ie, $(p,q) = (1,0)$) and with torsion $\xi\neq 0$, the equation of motion from $c$ simply sets $db=0$ implying $b_L = b_R$, while the equation of motion from $b$, after substituting for the solution $\varphi = - \xi^{-1}b(0)$, gives
\be
c_L = c_R - \xi^{-1} b_R \;.
\ee
This corresponds to the action of $T^k$ with $k = -\xi^{-1}$.

\paragraph{Fusion of defects.}
We can derive the fusion of defects --- that we already analyzed around (\ref{fusion of defects}) in terms of the discrete formalism --- using continuum Lagrangians. We place two defects, with action as in (\ref{eq: defect lagrangian full gauging}), along two codimension-1 surfaces $\Sigma_{1,2}$ at positions $r_{1,2}$ with $r_1 < r_2$. They act as sources for the bulk gauge fields $\cB$:
\be
\epsilon \, d \cB = - \bigl( \Phi_1 + d\Gamma_1 \bigr) \, \delta(r-r_1)dr - \bigl( \Phi_2 + d\Gamma_2 \bigr) \, \delta(r - r_2)dr \;.
\ee
Since $d\Phi_{1,2} = 0$ from the equations of motion, we can solve the equation as
\be
\label{backreaction defects in fusion}
\cB = \cB_0 + \epsilon \, \bigl( \Phi_1 + d\Gamma_1 \bigr) \, \theta(r-r_1) + \epsilon \, \bigl( \Phi_2 + d\Gamma_2 \bigr) \, \theta(r - r_2) \;.
\ee
Here $\cB_0$ is a background value for $\cB$, before adding the effect of the defects. It turns out that a crucial role in computing the fusion is played by the slab of bulk theory in between the two defects, which produces a phase factor. There are two contributions. One comes from substituting (\ref{backreaction defects in fusion}) in the bulk action:
\begin{align}
& \frac{N}{4\pi} \int_{r_2} (\Phi_1 + d\Gamma_1)^\sT \epsilon \, (\Phi_2 + d\Gamma_2) \, \theta(r_2 - r_1)
+ \frac{N}{4\pi} \int_{r_1} (\Phi_2 + d\Gamma_2 )^\sT \epsilon \, (\Phi_1 + d\Gamma_1 ) \, \theta(r_1 - r_2) \nn \\
&  =  \frac{N}{4\pi} \int_{r_2} \bigl( \Phi_1 + d\Gamma_1 \bigr)^\sT \epsilon \, \bigl( \Phi_2 + d\Gamma_2 \bigr) \;.
\end{align}
Another one comes from substituting (\ref{backreaction defects in fusion}) in the two defect actions. The defect at \mbox{$r=r_2$} produces $-\frac{N}{2\pi} \int_{r_2} (\Phi_1 + d\Gamma_1 )^\sT \epsilon \, (\Phi_2 + d\Gamma_2 )$, while the one at $r=r_1$ does not give any contribution. In those substitutions we did not include the background $\cB_0$, that we will couple to the final effective action. Collecting the contributions, we obtain the following action for the product of defects:
\be
\label{eq:4dfusion}
S[\cT_{21}]= S[\cT_1] + S[\cT_2] - \frac{N}{4 \pi} \int \bigl( \Phi_1 + d\Gamma_1 \bigr)^\sT \epsilon \, \bigl( \Phi_2 + d\Gamma_2 \bigr) \;.
\ee
We can interpret the effect of the last term in the path integral as a phase due to the braiding between 2-dimensional defects $U_m$. To write out the effective action, we identify $r_1 = r_2 = 0$ and simply write $\cB_0 \to \cB$ for the background field. We also change variables to $\Phi = \Phi_1 + \Phi_2$, $\wt\Psi = \Psi_1 - \Psi_2$ and $\Gamma = \Gamma_1+\Gamma_2$. We obtain:
\bea
\label{action with interaction}
S[\cT_{21}] &= \frac{N}{2\pi} \int_\Sigma \biggl[ \cB^\sT ( \Phi + d\Gamma ) + \Phi^\sT \Bigl( d\Psi _2 + \frac\epsilon2 \, d\Gamma_1 \Bigr) +\frac12 \, \Phi^\sT \cT _2 \Phi \biggr] + S_\text{int}(\Phi, \Phi_1) \\
S_\text{int} &= \frac{N}{2\pi} \int_\Sigma \biggr[ \Phi _1^\sT \biggl( d\wt\Psi - \Bigl(\cT _2 + \frac\epsilon2 \Bigr) \Phi - \frac\epsilon2 \, d\Gamma \biggr) + \frac12 \, \Phi_1^\sT \bigl( \cT _1 + \cT _2 \bigr) \Phi _1 - \frac12 \, d\Gamma_1^\sT \epsilon\, d\Gamma \biggr] \;.
\eea
The field $\Phi_1$, which is forced to be a cochain in $H^2(\Sigma, \bZ_N \times \bZ_N)$ by the equations of motion, does not directly couple to the bulk. The last term is a total derivative that vanishes on closed manifolds.

When $(\cT_1 + \cT_2)$ is invertible in $\bZ_N$, then $\Phi_1$ appears quadratically and can be integrated out, obtaining:
\begin{align}
\label{eq:composition action}
S[\cT_{21}] &= \frac{N}{2\pi} \int_\Sigma \biggl[ \cB^\sT ( \Phi + d\Gamma ) + \Phi^\sT d\Psi + \frac12 \, \Phi^\sT \cT_{21} \Phi \nn \\
&\hspace{2cm} - \frac12 \, d \biggl( \Bigl( \wt\Psi - \frac\epsilon2\, \Gamma \Bigr)^\sT \bigl( \cT _1 + \cT _2 \bigr)^{-1} d\Bigl( \wt\Psi - \frac\epsilon2 \, \Gamma \Bigr) \biggr) \biggr] \;,
\end{align}
where $\cT_{21}$ is the matrix (\ref{exp for T21}), we defined $\Psi = \Psi_2 + \frac\epsilon2 \Gamma_1 + \bigl( \cT_2 - \frac\epsilon2 \bigr) \bigl( \cT_1 + \cT_2 \bigr)^{-1} \bigl( \wt\Psi - \frac\epsilon2\Gamma \bigr)$, and $(\cT_1 + \cT_2)^{-1}$ is the inverse in $\bZ_N$. The last term is a total derivative and can be ignored on closed manifolds. We reproduce the action of a single defect with discrete torsion $\cT_{21}$, which corresponds to $M_{21} = M_2 M_1$.

When $\cT_2 = - \cT_1 \equiv \cT$, then $\Phi_1$ is a Lagrange multiplier imposing $\Phi = \bigl( \cT + \frac\epsilon2 \bigr)^{-1} d \bigl( \wt\Psi -\frac\epsilon2 \Gamma \bigr)$ and the defect Lagrangian, up to total derivatives,
simply becomes
\be
\label{eq:Identity_4ddefect}
S = \frac{N}{2\pi} \int_\Sigma \cB^{\sT} d \wh\Gamma \;,
\ee
where $\wh\Gamma = \Gamma + \bigl( \cT + \frac\epsilon2 \bigr)^{-1}\bigl( \wt\Psi - \frac\epsilon2 \Gamma\bigr)$.
On closed manifolds, this reproduces the result \mbox{$V[\cT] \times V[-\cT] = \unit$}. Indeed the action \eqref{eq:Identity_4ddefect} simply imposes that the pullback of $\cB$ be in $H^2(\Sigma,\mathbb{Z}_N \times \mathbb{Z}_N)$  without any discontinuity between the L and R regions.

The other cases can be dealt with in a similar way. When $\cT_1 + \cT_2$ has rank one, the component of $\Phi_1$ living in the kernel of $\cT_1 + \cT_2$ acts as a Lagrange multiplier, setting to zero one component of $\Phi$, while the component in the cokernel produces the torsion term for the remaining component of $\Phi$. Fusions involving defects from the condensation of $\cA \cong \bZ_N$ can be studied similarly.

%%%%%%%%%%%%%%%%%%%%%%%%%%%%%%%%%%%%%%%%%%
%%%%%%%%%%%%%%%%%%%%%%%%%%%%%%%%%%%%%%%%%%

\section{Twisted sectors and non-invertible defects}
\label{sec:TwistedSectors}

Whenever a theory has a discrete 0-form symmetry $\Gamma$, one can consider its twisted sectors. In particular, there exist codimension-2 operators that live at the boundary of the codimension-1 defect operators implementing $\Gamma$. We call them the codimension-2 operators in the twisted sector. Gauging a (non-anomalous) subgroup $G \subset \Gamma$, the corresponding defects become transparent and the codimension-2 operators at their boundary get promoted to genuine operators of the gauged theory.%
\footnote{When $G$ is Abelian, these are the codimension-2 operators charged under the $(d-2)$-form symmetry $\wh G$ dual to $G$ ($\wh G$ is the Pontryagin dual) and implemented by the Wilson lines of $G$.}
For instance, in 2d CFTs the twisted sectors are described by local operators at the end of defect (or twist) lines, and their inclusion in the gauged theory is required by modular invariance. In 3d TQFT the twisted sectors are described by line operators at the end of defect surfaces, and the modular tensor category (MTC) of lines gets promoted to a $G$-crossed MTC \cite{Barkeshli:2014cna}, also in order to assure modularity.

The situation in higher dimensions is less well understood. In this section we study the twisted sectors of the 5d Chern-Simons theory, exploiting the Lagrangian description of codimension-1 symmetry defects that implement $SL(2,\bZ_N)$. In particular, we describe the 3d twist defects $D[\cT]$ and $D[\cA,\xi]$ (or more compactly $D_M$) at the boundary of 4d symmetry defects $V[\cT]$ and $V[\cA,\xi]$ (or $V_M$), respectively.

\subsection[{Lagrangian description of $D[\mathcal{T}]$}]{Lagrangian description of \matht{D[\cT]}}
\label{sec: Lag descr D[T]}

We can obtain a Lagrangian description of the 3d twisted-sector operators --- that we dub $D[\cT]$ --- at the boundary of 4d $SL(2,\bZ_N)$ symmetry defect operators $V[\cT]$ from the Lagrangian description (\ref{eq: defect lagrangian full gauging}) of the latter.%
\footnote{A similar discussion would apply to the defects $D[\cA,\xi]$ at the boundary of $V[\cA,\xi]$, derived from (\ref{eq: defect lagrangian gauging subgroup}).}
As we will see in a moment, it is convenient to perform an integration by parts of the couplings $\cB^\sT d\Gamma$ and $\Phi^\sT d\Psi$ and use the following equivalent Lagrangian for the 4d defect operators $V[\cT]$:
\be
\label{new defect Lagrangian}
S[\cT] = \frac{N}{2\pi} \int_\Sigma \biggl[ \cB^\sT \Phi + \Gamma^\sT d\cB + \Psi^\sT d\Phi + \frac12 \, \Phi^\sT \cT \Phi\biggr] \;.
\ee
In the presence of a boundary $Y = \partial\Sigma$, this action is not invariant under the gauge transformations (\ref{eq:shiftsymmetry}), rather, it shifts by a boundary term (up to integer multiples of $2\pi$):
\be
\label{eq:non gauge invariance with boundary}
S \,\to\, S + \frac{N}{2\pi} \int_Y \biggl[ \cB^\sT ( \lambda - d\nu)  + \Phi^\sT(\alpha + \cT\lambda - d\mu) + \alpha^\sT d\lambda + \frac{1}{2} \, \lambda^\sT \cT d \lambda \biggr] \;.
\ee
This can by canceled by the following boundary action:
\be
\label{eq:lagrangian twisted sector}
S_\text{twist}[\cT] = \frac{N}{2\pi} \int_Y \biggl[ \cB^\sT \Gamma + \Phi^\sT \Psi + \Gamma^\sT d\Psi - \frac12 \, \Gamma^\sT \cT d\Gamma \biggr] \;.
\ee
The reason why we wrote the 4d action as in (\ref{new defect Lagrangian}) is that the 4d fields $\Gamma$ and $\Psi$ only appear as Lagrange multipliers with no derivatives, and thus their path-integrals at different spacetime points are independent. On the contrary, they appear dynamically (with derivatives) in the 3d action (\ref{eq:lagrangian twisted sector}) and therefore their restrictions to $Y$ can be treated as independent 3d fields, or edge modes. From the 3d point of view, the fields $\cB$ and $\Phi$ appear as background fields (that can be integrated afterwards in 5 and 4 dimensions, respectively).%
\footnote{Using the equivalent action (\ref{eq: defect lagrangian full gauging}) one obtains the boundary action $S'_\text{twist} = \frac{N}{2\pi} \int_Y \bigl[ \Gamma^\sT d\Psi - \frac12 \Gamma^\sT \cT d\Gamma \bigr]$ in which the couplings to $\cB$ and $\Phi$ are not manifest.}
The coupled 4d-3d system is gauge invariant. We call the 3d defect defined by $S_\text{twist}[\cT]$ a twist defect $D[\cT]$ associated to the $SL(2,\bZ_N)$ element $M(\cT)$ (\ref{exp of M from T}).

The actions (\ref{new defect Lagrangian}) and (\ref{eq:lagrangian twisted sector}) are invariant under all elements $M' \in SL(2,\bZ_N)$ that commute with $M$, if we supplement the transformation $\cB \to M^{\prime\sT} \cB$ with%
\footnote{\label{foo: invariance of T}%
Invariance of the last term follows from the fact that $M'$ commutes with $M$ if and only if $M^{\prime\sT} \cT M' = \cT$.} 
\be
\Phi \,\to\, M^{\prime -1} \Phi \;,\qquad\qquad \Gamma \,\to\, M^{\prime -1}\Gamma \;,\qquad\qquad \Psi \,\to\, M^{\prime\sT} \Psi \;.
\ee
Such an invariance is expected since, in general, acting with a 0-form symmetry $h$ on a twisted sector $D_g$ gives an element of $D_{hgh^{-1}}$. This will be important when gauging a subgroup of $SL(2,\bZ_N)$.

Let us analyze the content of the three-dimensional theory $D[\cT]$. For simplicity, we only consider the cases in which $\cT$ is invertible in $\bZ_N$, or $\cT=0$. We start with the former. Setting $\cB = \Phi =0$, \eqref{eq:lagrangian twisted sector} is the action of an Abelian Chern-Simons theory with four gauge fields, whose level matrix $K$ and its inverse are
\be
K = N \mat{ - \cT & \unit \\ \unit & 0} \;,\qquad\qquad K^{-1} = N^{-1} \mat{ 0 & \unit \\ \unit & \cT} \;.
\ee
There are $\lvert \det K \rvert = N^4$ line operators, given by $e^{i \!\int ( n^\sT \Gamma + m^\sT \Psi)}$ with $n,m \in \bZ_N \times \bZ_N$. Not all of them, however, are genuine 3d line operators in the coupled 4d-3d system (keeping the 5d bulk as a background), rather some of them live at the end of a bulk surface $e^{i n^\sT \!\!\int\! \Phi}$. This follows from the gauge transformations \eqref{eq:shiftsymmetry}.
A basis of genuine line operators is given by
\be
\label{lines of ANT}
W_n = \exp \biggl[ i n^\sT \int \Bigl( \cT \Gamma - \Psi \Bigr) \biggr] \;.
\ee
We have chosen the parametrization such that $W_n$ has charge $n = (n_\rme, n_\rmm)$ under the $\bZ_N \times \bZ_N$ 1-form symmetry that couples to $\cB$.%
\footnote{Indeed, under the transformation $\cB \to \cB + d\alpha$, $\Psi \to \Psi - \alpha$, the operator gets a phase $W_n \to e^{in^\sT \!\!\int\! \alpha} W_n$.}
These lines have spin
\be
\label{spins of ANT}
\theta[W_n] = \exp \biggl(-\frac{\pi i}{N} \, n^\sT \cT n \biggr) \;,
\ee
and give a $\bZ_N \times \bZ_N$ generalization of the $\cA^{N,p}$ minimal TQFTs introduced in \cite{Hsin:2018vcg} (see Appendix F there and Appendix~\ref{app:ANp} here). Indeed, these lines have braiding $B_{ab} = \frac{\theta_{a + b}}{\theta_a\theta_b} = \exp\bigl[ -\frac{2\pi i}N a^\sT \cT b \bigr]$ and, taken in isolation, give rise to a consistent MTC with unitary S-matrix $S_{ab} = \frac1N B_{ab}$. We will use the notation $\cA^{N,-\cT}(\cB)$ to denote the theory of these lines:
\be
\cA^{N, -\cT}(\cB) \;\subset\; D[\cT] \;.
\ee
There is some redundancy in the nomenclature of the theories $\cA^{N,-\cT}$: for all matrices $\cQ$ invertible in $\bZ_N$, the theory $\cA^{N, -\cQ^\sT \cT \cQ}$ (where the product of matrices is in $\bZ_N$) is equivalent to $\cA^{N,-\cT}$ up to a relabelling of the lines $n \to \cQ n$. They are distinguished, however, by how they couple to $\cB$. We will refer to the theory (\ref{spins of ANT}) in which $W_n$ has charge $n$ as $\cA^{N,-\cT}(\cB)$. Notice that this theory is not coupled to the 4d field $\Phi$.

The remaining lines are not genuine in the coupled 4d-3d system, and are generated by
\be
L_m = \exp\biggl[ - im^\sT \biggl( \int_{\partial X} \Psi + \cT \int_X \Phi \biggr) \biggr] 
\ee
in addition to $W_n$, where $X$ is a two-dimensional open surface ending on $D[\cT]$. The twisted sector, as an isolated 3d theory, is formed by \emph{both} genuine and non-genuine line operators. We chose the generators $L_m$ such that in 3d (\ie, switching the background $\Phi$ off) they have trivial braiding with $W_n$. Indeed, the twisted sector can be decomposed as
\be
D[\cT] = \cA^{N,-\cT}(\cB) \times \cA^{N,\cT}(\cB + \cT\Phi) \;,
\ee
where the two factors are the MTCs of $W_n$ and $L_m$, respectively.%
\footnote{One could also consider the non-genuine operators $\ell_m = \exp\bigl[ i m^\sT \bigl(  \int_{\partial X} \Gamma + \int_X \Phi \bigr) \bigr]$ which do not couple to $\cB$, however they have vanishing spin and do not form a MTC by themselves.}
However, as we will see in Section~\ref{sec:gaugedTheory}, once a subgroup of the $SL(2,\bZ_N)$ 0-form symmetry is gauged in the bulk, some of the 4d operators become transparent and only the subcategory $\cA^{N,-\cT}(\cB)$ of genuine operators survives.

The $\bZ_N \times \bZ_N$ 1-form symmetry of $\cA^{N,-\cT}$ is anomalous, since the lines $W_n$ that generate it have non-trivial braiding. Turning on the background field $\cB$ coupled to the 1-form symmetry, the anomaly is canceled \cite{Hsin:2018vcg} by the following four-dimensional inflow action:%
\footnote{In the conventions of \cite{Hsin:2018vcg}, the 1-form symmetry is generated by the lines $\wt W_n \equiv W_{-\cT^{-1}n}$ which have charge $-\cT^{-1}n$ and spin $\exp\bigl( - \frac{\pi i}{N} \, n^\sT \cT^{-1} n \bigr)$. This theory, that \cite{Hsin:2018vcg} would call $\cA^{N, -\cT^{-1}}$, has an anomaly that is canceled by (\ref{anomaly}).}
\be
\label{anomaly}
I_\cT(\cB) = \frac{N}{2\pi} \int_\Sigma \biggl[ \cB^\sT d \wt\Gamma - \frac12 \, \cB^\sT \cT^{-1} \cB \biggr] \;,
\ee
where the dynamical field $\wt\Gamma$ imposes $\cB \in H^2(\Sigma,\bZ_N \times \bZ_N)$ on shell, and $\cT^{-1}$ is the inverse of $\cT$ in $\bZ_N$. This implies that $\cA^{N,-\cT}(\cB)$ is not invariant under the gauge transformation $\cB \to \cB + d\alpha$, $\wt\Gamma \to \wt\Gamma + \cT^{-1} \alpha$ but rather its path integral picks up a phase:
\be
\label{eq:ausiliario}
\exp\biggl[ - \frac{i N}{2 \pi} \int_Y \biggl( \alpha^\sT d\wt\Gamma + \frac12 \alpha^\sT \cT^{-1} d\alpha \biggr) \biggr] \;.
\ee
Indeed one can check that the anomaly inflow action (\ref{anomaly}) for $\cA^{N,-\cT}(\cB) \times \cA^{N, \cT}(\cB + \cT\Phi)$, if supplemented by the condition that $\Phi \in H^2(\Sigma, \bZ_N \times \bZ_N)$, coincides with the 4d action (\ref{eq: defect lagrangian full gauging}) for the defect $V[\cT]$. Alternatively, one can start with the action (\ref{new defect Lagrangian}) for $V[\cT]$ and integrate out $\Phi$. This is possible because, as stressed after (\ref{eq:shiftsymmetry}), the theory is trivial as long as $\cT$ is invertible in $\bZ_N$. We already did this computation in (\ref{anomaly first version}): one is left with the invertible TQFT (\ref{anomaly}) in the 4d bulk and $\cA^{N,-\cT}$ on the 3d boundary. Either way, the coupled 4d-3d system is anomaly free.

The case of $\cT=0$, which describes the charge conjugation operator $V_C$, needs a separate discussion. Contrary to the previous case, there is no consistent MTC that describes the lines $W_n$ decoupled from $\Phi$. Those lines have trivial spin and braiding among themselves. This phenomenon was already observed in \cite{Hsin:2018vcg} and is a consequence of the non-invertibility of the 4d 2-form gauge theory for $\Phi$.
The action for the twisted sector $D[\cT=0] \equiv D_C$ is
\be
\label{eq:twistC}
S_\text{twist}[\cT=0] = \frac{N}{2 \pi} \int_{Y} \Bigl[ \cB^\sT \Gamma + \Phi^\sT \Psi + \Gamma^\sT d\Psi \Bigr] \;.
\ee
This is a 3d $\bZ_N \times \bZ_N$ gauge theory (described by the 3d fields $\Gamma, \Psi$) coupled to the backgrounds fields $\cB$ and $\Phi$ for the two copies of the $\bZ_N \times \bZ_N$ 1-form symmetry, and we denote it by $(\bZ_N \times \bZ_N)_0(\cB, \Phi)$.

\paragraph{Degeneracies.}
We ask what is the degeneracy of the twisted sectors, \ie, how many boundaries an $SL(2,\bZ_N)$ symmetry defect $V$ can have. In three-dimensional TQFTs with a 0-form symmetry $\Gamma$, the number of simple lines in a twisted sector labeled by $g \in \Gamma$ is equal to the number of $g$-invariant simple lines in the untwisted sector \cite{Barkeshli:2014cna}.
In our case, the 5d CS theory has no genuine codimension-2 operators (besides the trivial one), therefore we expect every twisted sector to be unique.
One could argue that we should also consider the operators obtained by fusing $D[\cT]$ with codimension-2 condensation defects obtained from the bulk 2-form symmetry.

We can show that for defects $V[\cT]$ obtained by condensing the full $\bZ_N \times \bZ_N$ 2-form symmetry, the boundary $D[\cT]$ is left invariant by every such fusion, up to stacking with a decoupled TQFT. Indeed, fusing $D[\cT]$ with a 2d symmetry defect $U(\gamma)$ with \mbox{$\gamma \in H_2 (Y, \bZ_N \times \bZ _N)$} is equivalent to adding the following coupling to the action (\ref{eq:lagrangian twisted sector}) of $D[\cT]$:
\be
\label{eq:defectinsertion}
\delta S_\text{twist}[\cT] = \int_Y \cB^\sT \Gamma_\gamma \;,\qquad\qquad \Gamma_\gamma = \text{PD}(\gamma)
\ee
where $\text{PD}(\gamma) \in H^1(Y, \bZ_N \times \bZ_N)$ is the Poincar\'e dual to $\gamma$ on $Y$. Given a continuum description of the class $\Gamma_\gamma$, for instance through a delta 1-form, the extra coupling can be reabsorbed by the field redefinition $\Gamma \to \Gamma- \frac{2\pi}N \Gamma_\gamma$, $\Phi \to \Phi + \frac{2\pi}N d\Gamma_\gamma$, $\cB \to \cB - \frac{2\pi}N \cT d\Gamma_\gamma$, which however produces a phase
\be
\label{phase from Bockstein}
\exp\biggl( - \frac{2\pi i}N \int \frac12 \, \Gamma_\gamma^\sT \cT d\Gamma_\gamma \biggr) = \exp\biggl( - \frac{2\pi i}N \int \frac12 \, \Gamma_\gamma^\sT \, N\cT \, \beta(\Gamma_\gamma) \biggr) \equiv Q_{N\cT}(\Gamma_\gamma) \;.
\ee
Notice that, in the continuum description on the left-hand-side, $d\Gamma_\gamma$ is a class with values in $N$ times $\bZ \times \bZ$ rather than identically zero. On the right-hand-side we wrote the phase in a more precise way in terms of $\Gamma_\gamma \in H^1(Y,\bZ_N \times \bZ_N)$ and the Bockstein map associated to the short exact sequence $0 \to \bZ_N \xrightarrow{N} \bZ_{N^2} \xrightarrow{\text{mod }N} \bZ_N \to 0$ so that $\beta(\Gamma_\gamma) \in H^2(Y, \bZ_N \times \bZ_N)$. The integrals in (\ref{phase from Bockstein}) are well defined on generic manifolds if $N\cT$ is an even matrix, and on spin manifolds if $N\cT$ is a more general integer matrix. Hence
\be
\label{eq:fusionUgamma}
U(\gamma) \times D[\cT] = e^{i Q_{N\cT}(\Gamma_\gamma)} \, D[\cT] \;.
\ee
A similar effect has already been appreciated in dealing with $N$-ality defects in \cite{Kaidi:2021xfk,Choi:2022zal}.

Now, a 3d condensation defect for the $\bZ_N \times \bZ_N$ 2-form symmetry can be thought of as a 3d $\bZ_N \times \bZ_N$ Dijkgraaf-Witten (DW) theory, possibly with torsion $\cP$, coupled to the dynamical field $\cB$. The coupling is precisely (\ref{eq:defectinsertion}) with $\Gamma_\gamma$ substituted by the dynamical gauge field of the DW theory. We denote the 3d condensation defect as $\cC^{\bZ_N \times \bZ_N}_\cP$, and omit the subscript when there is no torsion. Stacking the condensation defect on $D[\cT]$ replaces the coupling to $\cB$ with the torsion term $Q_{N\cT}(\Gamma_\gamma)$: this produces a shift  $\delta\cP = - N\cT$ of the torsion of the DW theory. It turns out (see below) that if $N$ is odd and the theory is spin, then shifts of the torsion components by multiples of $N$ give equivalent theories, and so in our case the shift is immaterial. We conclude that
\be
\label{condensation_twist}
\cC^{\bZ_N \times \bZ_N}_\cP \times D[\cT] = (\bZ_N \times \bZ_N)_\cP \; D[\cT] \;.
\ee
The factor on the right-hand-side is a decoupled Dijkgraaf-Witten TQFT. A similar argument applies to any other 3d condensate in which only a subgroup of $\bZ_N \times \bZ_N$ is condensed (possibly with torsion): they can all be absorbed by $D[\cT]$. We conclude that there is no degeneracy in these twisted sectors.

When, on the other hand, the defect $V[\cA, \xi]$ is obtained by condensing a subgroup $\cA$ of $\bZ_N \times \bZ_N$, then only condensates of surfaces in $\cA$ can similarly be absorbed by $D$, while more general surfaces in $\bZ_N \times \bZ_N$ cannot and give rise to a genuine degeneracy of the twisted sector. Since surfaces in $\cA$ are absorbed, the degeneracy is given by all condensates (with torsion) of the quotient group $(\bZ_N \times \bZ_N)/\cA$ (or its subgroups).

The last case is the 4d indentity interface $V_\unit$, on which we do not gauge any symmetry. Its sector, which is the untwisted sector, consists of all possible 3d condensates in $\bZ_N \times \bZ_N$.

\subsubsection{Dijkgraaf-Witten theories}
\label{sec: DW theories}

The 3d $\bZ_N^k$ Dijkgraaf-Witten theories can be described by the following Abelian Chern-Simons action:
\be
\label{DW action}
S_\text{DW}[\cT] = \int_Y \biggl[ \frac{N}{2\pi} \, x^\sT dy + \frac{1}{4\pi} \, y^\sT \cP dy \biggr]
\ee
where $x,y$ are $k$-dimensional vectors of Abelian vector fields and $\cP$ is a $k\times k$ symmetric integer matrix. The theory is bosonic if $\cP$ is even (\ie, if its diagonal entries are even), otherwise it is spin. The level matrix is $K = \smat{0 & N\,\unit \\ N\,\unit & \cP}$. The theory has $N^{2k}$ lines labelled by $n \in \bZ_N^{2k}$ with spin
\be
\label{eq:braidingDW}
\theta[n] = \exp\bigl( \pi i \, n^\sT K^{-1} n \bigr) \qquad\text{where}\qquad K^{-1} = \frac1{N^2} \mat{ -\cP & N\,\unit \\ N\,\unit & 0} \;.
\ee
In all cases, a shift of $\cP$ by $N$ times an even integer matrix gives an equivalent theory, \ie, the diagonal entries of $\cP$ are defined modulo $2N$ while the off-diagonal entries modulo $N$. This follows from the field redefinition $\smat{x \\ y} \to \smat{\unit & \cQ \\ 0 & \unit} \smat{x \\ y}$ where $\cQ$ is an integer matrix, or equivalently, from the relabelling $n \to \smat{ \unit & 0 \\ \cQ & \unit} n$ of the lines. If $N$ is odd, in addition, theories in which the entries of $\cP$ differ by multiples of $N$ are equivalent as spin theories.%
\footnote{This is not true, in general, if $N$ is even. A counterexample for $k=1$ is the family of four $\bZ_2$ theories.}
This follows from the fact that the relabelling $n \to  \smat{ \unit & 0 \\ 2^{-1} \cQ & \unit} n$ (where $2^{-1}$ is the inverse in $\bZ_N$) preserves the spin modulo a sign, which can be cancelled by fusing with the transparent fermion.

The coupling of the electric 1-form symmetry to a $\bZ_N^k$ background field $\cB$ is described by
\be
\label{DW action with B}
S_\text{DW}[\cT](\cB) = \int_Y \biggl[ \frac{N}{2\pi} \, \bigl( \cB^\sT y + x^\sT dy \bigr) + \frac{1}{4\pi} \, y^\sT \cP dy \biggr] \;,
\ee
invariant under $\cB \to \cB + d\alpha$, $x \to x-\alpha$. The lines labelled by $n = (n_\rme, n_\rmm)$ have charge $-n_\rme$ under the electric 1-form symmetry. This statement persists under shifts of the components of $\cP$ by multiples of $N$.

\subsection{Fusion of twist defects}

We now study the fusion of two twist defects $D[\cT_1]$ and $D[\cT_2]$. As expected, the fusion is compatible with the group product rule $M_{21} = M_2 M_1$ of 4d $SL(2,\bZ_N)$ defect operators $V[\cT]$, \ie, of twisted sectors, however we would like to understand which condensates and decoupled TQFTs can be generated.

As already discussed in Section~\ref{sec: continuum descr defects} for the fusion of 4d defects, the 5d bulk provides a crucial contribution to the fusion of 3d twist defects as well. The bulk contribution was computed in (\ref{eq:4dfusion}), thus the total action for the system of two 4d defects with boundary located on the same 3d (spin) manifold $Y$ is
\be
\label{fusion 3d extra terms}
S = S[\cT_1] + S[\cT_2] - \frac{N}{4\pi}\int ( \Phi_1 + d\Gamma_1 )^\sT \epsilon \, (\Phi_2 + d\Gamma_2) + S_\text{twist}[\cT_1] + S_\text{twist}[\cT_2] \;,
\ee
where, this times, we use the 4d action (\ref{new defect Lagrangian}) for the symmetry defects.

The computation in the 4d bulk is similar to the one we did in Section~\ref{sec: continuum descr defects}. One introduces $\Phi = \Phi_1 + \Phi_2$, $\Gamma = \Gamma_1 + \Gamma_2$, $\wt\Psi = \Psi_1 - \Psi_2$, and eliminates $\Phi_2$, $\Gamma_2$, $\Psi_1$. If $(\cT_1 + \cT_2)$ is an invertible matrix in $\bZ_N$, the field $\Phi_1$ can be integrated out leaving the bulk theory
\be
S_\text{bulk} = \frac{N}{2\pi} \int_\Sigma \biggl[ \cB^\sT \Phi + \Gamma^\sT d\cB + \Psi^\sT d\Phi + \frac12 \Phi^\sT \cT_{21} \Phi \biggr] \;,
\ee
where $\cT_{21}$ is given in (\ref{exp for T21}) and $\Psi = \Psi_2 + \frac\epsilon2 \Gamma_1 + \bigl( \cT_2 - \frac\epsilon2 \bigr) (\cT_1 + \cT_2)^{-1} \bigl( \wt\Psi - \frac\epsilon2 \Gamma \bigr)$. This is the theory $S[\cT_{21}]$. There are leftover boundary terms, that together with $S_\text{twist}[\cT_1] + S_\text{twist}[\cT_2]$ give
\bea
\label{boundary theory from fusion}
S_\text{boundary} &= \frac{N}{2\pi} \int_Y \biggl[ \cB^\sT \Gamma + \Phi^\sT \Psi + \Gamma^\sT d\Psi_2 - \frac12 \Gamma^\sT \cT_2 d\Gamma + \Gamma_1^\sT d\Bigl( \wt\Psi + \bigl( \cT_2 - \tfrac\epsilon2 \bigr) \Gamma \Bigr) \\
&\qquad\qquad\quad - \frac12 \Gamma_1^\sT (\cT_1 + \cT_2) d\Gamma_1 - \frac12 \bigl( \wt\Psi - \tfrac\epsilon2 \Gamma\bigr)^\sT (\cT_1 + \cT_2)^{-1} d\bigl( \wt\Psi - \tfrac\epsilon2 \Gamma \bigr) \biggr] \;.
\eea
The gauge transformations of the new fields are
\be
\cB \to \cB + d\alpha \;,\quad \Phi \to \Phi + d\lambda \;,\quad \Psi \to \Psi - \cT_{21} \lambda - \alpha + d\mu \;,\quad \Gamma \to \Gamma - \lambda + d\nu
\ee
where $\lambda = \lambda_1 + \lambda_2$. The theory (\ref{boundary theory from fusion}) is not trivial and we cannot integrate other fields out. We perform a more rigorous analysis of it below, but for now, in order to understand the physics, let us perform an approximate computation. We introduce a new 1-form field 
\be
H =\Psi _1- \cT _1 \Gamma _1-\Psi _2+\cT _2\Gamma _2= \widetilde{\Psi} + \cT_2 \Gamma- \left(\cT_1+\cT_2\right)\Gamma_1 \;.
\ee
This combination is special because it is invariant under the gauge transformations (\ref{eq:shiftsymmetry}) parametrized by $\lambda_1$, $\lambda_2$, $\alpha$. We eliminate $\Gamma_1$ in favor of $H$: this is not a legit operation since $(\cT_1 + \cT_2)$ is not a unimodular integer matrix, but let us proceed anyway and treat $(\cT_1 + \cT_2)^{-1}$ as the inverse in $\bQ$. Up to total derivatives, we obtain
\be
\label{fusion twist defects+ decoupled TQFT}
S_\text{boundary} \,\sim\,
\frac{N}{2\pi} \int_Y \biggl[ \cB^\sT \Gamma + \Phi^\sT \Psi + \Gamma^\sT d\Psi - \frac12 \Gamma^\sT \cT_{21} d\Gamma \biggr] - \frac{N}{4\pi} \int_Y H^\sT (\cT_1 + \cT_2)^{-1} dH \;.
\ee
The first term is the expected action $S_\text{twist}[\cT_{21}]$ of the twisted sector $D[\cT_{21}]$. The second term is a decoupled TQFT, described by a Chern-Simons action with fractional level-matrix. Perturbatively, it behaves as the theory $\cA^{N, -\cT_1 - \cT_2}$ (while it is not well defined at the non-perturbative level).

If $(\cT_1+\cT_2)$ is not invertible in $\bZ_N$ then the procedure has to be slightly changed. Let us discuss the case $\cT_2 = - \cT_1 \equiv \cT$, corresponding to the fusion of a defect with its ``inverse''. This case is interesting because the fusion of two defects in inverse twisted sectors must produce an operator in the untwisted sector, which however contains all three-dimensional condensation defects. Starting with (\ref{fusion 3d extra terms}) and performing the field redefinitions to $\Phi$, $\Gamma$, $\wt\Psi$, in the 4d bulk one finds $\Phi_1$ to be a Lagrange multiplier imposing $\Phi = d  \bigl( \cT + \frac\epsilon2\bigr)^{-1} \bigl( \wt\Psi - \frac\epsilon2 \Gamma \bigr)$. It is convenient to define
\bea
\wh\Gamma &= \Gamma + \bigl( \cT + \tfrac\epsilon2 \bigr)^{-1} \bigl( \wt\Psi - \tfrac\epsilon2 \Gamma \bigr) \\
\wh\Psi &= \Psi_2 + \cT \Gamma_1 + \cT \bigl( \cT + \tfrac\epsilon2 \bigr)^{-1} \bigl( \wt\Psi - \tfrac\epsilon2 \Gamma \bigr) \;.
\eea
Then the bulk action simply reduces to the completely trivial theory
\be
S_\text{bulk} = \frac{N}{2\pi} \int_\Sigma \wh\Gamma^\sT d\cB
\ee
that describes the identity operator $V_\unit$. The boundary terms instead give
\be
\label{3d theory from D Dbar}
S_\text{boundary} = \frac{N}{2\pi} \int_Y \biggl[ \cB^\sT \wh\Gamma + \wh\Gamma^\sT d\wh\Psi - \frac12 \, \wh\Gamma^\sT \cT d\wh\Gamma \biggr] \;.
\ee
The fields $\wh\Gamma$, $\wh\Psi$ are invariant under the gauge transformations $\lambda_1, \lambda_2$, indeed this 3d theory does not need to be attached to any 4d theory. On the other hand, $\wh\Psi \to \wh\Psi - \alpha$ under gauge transformations of $\cB$ (while $\wh\Gamma$ is invariant). The action (\ref{3d theory from D Dbar}) describes a 3d $\bZ_N \times \bZ_N$ Dijkgraaf-Witten theory with torsion equal to $-N\cT$, in which a $\bZ_N \times \bZ_N$ 1-form symmetry is coupled to $\cB$ --- as in (\ref{DW action with B}). Alternatively, this can be though of as a 3d condensation defect for the $\bZ_N \times \bZ_N$ global 2-form symmetry of the 5d bulk theory: $\wh\Psi$ forces $\wh\Gamma \in H^1(Y, \bZ_N \times \bZ_N)$, then $e^{i \frac{N}{2\pi} \!\int\! \cB^\sT \wh\Gamma}$ is a two-dimensional operator of the 5d theory placed on the Poincar\'e dual to $\wh\Gamma$ within $Y$, and the last term in (\ref{3d theory from D Dbar}) produces a phase weighing the sum over surfaces. We dubbed such a 3d condensation defect $\cC^{\bZ_N \times \bZ_N}_{-N\cT} \equiv \cC^{\bZ_N \times \bZ_N}$, since we are considering $N$ odd.
Therefore, the fusion of a twist defect with its ``inverse'' is given by
\be
\label{product of inverse twisted sectors}
D[\cT] \times D[-\cT] = D[\cT] \times \overline{D}[\cT] = \cC^{\bZ_N \times \bZ_N} \;.
\ee

\paragraph{A more rigorous analysis of twisted sectors.} The analysis of the fusion of twist defects we performed in (\ref{fusion twist defects+ decoupled TQFT}) using the Lagrangian formulation, while suggesting the correct result, was imprecise. We can obtain a more rigorous and precise derivation by studying the algebra of topological operators.

\begin{figure}[t]
\centering
\begin{tikzpicture}
	\fill[fill=white!70!green, opacity=0.8] (-1, -1.5) -- (0,0) -- (2,0) -- (1, -1.5) -- cycle;   % V1
	\draw[color=red, line width=2] (-1,-1.5) -- (1,-1.5);
	\draw (-1, -1.5) -- (0,0) -- (2,0) -- (1, -1.5);
	\draw[color=blue, line width=2] (-0.5,-1.5) to [bend left= 90] (2.5,-2);   % blue lines
	\draw[color=blue, line width=2] (1.8, -1.08) to [bend right= 30] (3, 2); 
	\fill[fill=white!70!green, opacity=0.8] (0,0) -- (1,-2) -- (3,-2) -- (2,0) -- cycle;   % V2
	\draw[color=red, line width=2] (1,-2) -- (3,-2);
	\draw (1, -2) -- (0,0) -- (2,0) -- (3, -2);
	\filldraw[color=white, bottom color=white!60!green, top color=white!100!green] (0,0) -- (0,2) -- (2,2) -- (2,0) -- cycle; % V21
	\draw (0,0) --(2,0);
	\draw (0,0) -- (0,2); \draw (2,0) -- (2,2);
	\draw[fill=black] (-0.5,-1.5) node[below] {$W^{(1)}$} circle (0.05);   % text
         \draw[fill=black] (2.5,-2) node[below] {$W^{(2)}$} circle (0.05);
         \node[above] at (1,2) {$V_{21}$};
         \node[left] at (-0.6,-0.75) {$V_1$};
         \node[right] at (2.5,-1) {$V_2$};
	\node at (4.5, 0) {\LARGE$\leadsto$};
 \begin{scope}[shift={(6.5,0)}]
 \filldraw[color=white, bottom color=white!60!green, top color= white!100!green] (0,-2) -- (3,-2) -- (3,2) -- (0,2) -- cycle;
 \draw[color=red, line width=2] (0,-2) -- (3,-2);
 \draw[color=blue, line width=2] (0.5,-2) arc(180:0:1 and 2);
 \draw[color=blue, line width=2] (1.5, 0) -- (1.5,2);
  \draw[fill=black] (0.5,-2) node[below] {$W^{(1)}$} circle (0.05);
   \draw[fill=black] (2.5,-2) node[below] {$W^{(2)}$} circle (0.05);
   \node[above] at (1.5,2) {$V_{21}$};
  \node at (3.5,0) {$=$};
  \begin{scope}[shift={(4,0)}]
 \filldraw[color=white, bottom color=white!60!green, top color= white!100!green] (0,-2) -- (3,-2) -- (3,2) -- (0,2) -- cycle;
  \draw[color=red, line width=2] (0,-2) -- (3,-2);
   \draw[color=blue, line width=2] (1.5, 0) -- (1.5,2);      % blue line: straight piece
   \draw[color=blue, line width=2] (2, -0.5) arc (0:180:0.5 and 0.5);    % blue line: upper arc
   \draw[color=blue, line width=2] (2.5, -2) to [out = 160, in = -90] (1, -0.5);   % blue line: piece of W1
   \fill[fill=white!70!green] (1.5, -1.5) circle [radius = 0.2];
   \draw[color=blue, line width=2] (0.5, -2) to[out = 20, in = -90] (2, -0.5);   % blue line: piece of W2
  \draw[fill=black] (0.5,-2) node[below] {$W^{(2)}$} circle (0.05);    % text
  \draw[fill=black] (2.5,-2) node[below] {$W^{(1)}$} circle (0.05);
     \node[above] at (1.5,2) {$V_{21}$};
 \end{scope}
 \end{scope}
\end{tikzpicture}
\caption{Braiding between lines $W^{(1)}$ and $W^{(2)}$ from bulk ordering. We represented the lines $W^{(i)}$ by black points, the 3d twist sectors $D[\cT_i]$ by red lines, the surfaces $U(\gamma)$ by blue lines, and the 4d condensation defects $V_i$ by green surfaces. Left: bulk definition of fusion. Right: two different ordering procedures, related by the half-braiding phase of the bulk 5d theory. In canonical quantization, time runs horizontally.
\label{Fig:braiding12}}
\end{figure}
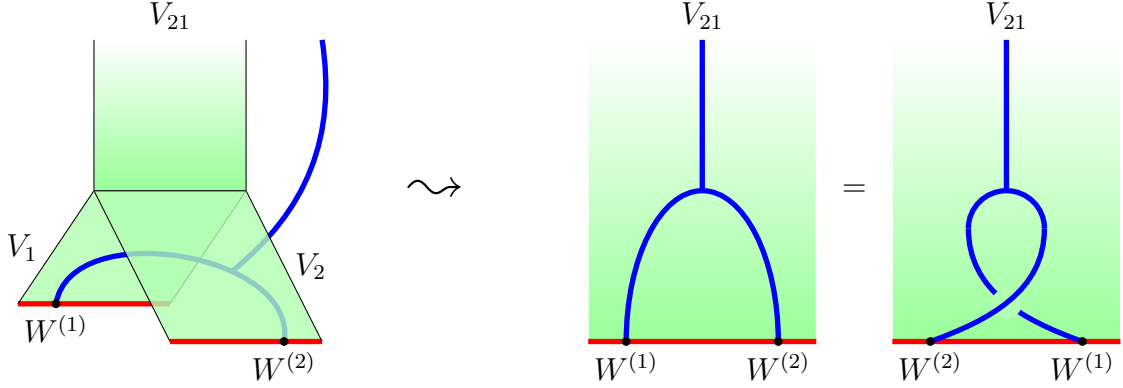

As discussed in Section~\ref{sec: Lag descr D[T]}, if $\cT$ is invertible in $\bZ_N$ then the twist operator $D[\cT]$ hosts a MTC of local line operators (which are not coupled to the 4d defect) forming the minimal TQFT $\cA^{N, -\cT}(\cB)$. When we fuse two twist operators $D[\cT_1]$ and $D[\cT_2]$, the set of local line operators is not simply the stacking of the two TQFTs because of the bulk contribution. Taken separately, the two minimal TQFTs have lines $W^{(1)}_{n_1}$ and $W^{(2)}_{n_2}$, respectively. The 5d dynamical bulk field $\cB$, however, generates a non-trivial braiding between the two sets of lines:
\be
\label{extra braiding}
B_{W^{(1)}\!, W^{(2)}} = \exp\biggl( \frac{2\pi i}N \, n_1^\sT \frac\epsilon2 n_2 \biggr)
\ee
where we are taking $N$ odd. This follows from the boundary term $-\frac{N}{2\pi} \int_Y d\Gamma_1^\sT \frac\epsilon2 d\Gamma_2$ in (\ref{fusion 3d extra terms}) and the expression (\ref{lines of ANT}) for the local lines. It can also be understood as follows. In canonical quantization, the braiding matrix appears as a non-trivial commutator 
\be 
W^{(1)} \, W^{(2)} = B_{W^{(1)} \!, W^{(2)}} \, W^{(2)} \, W^{(1)} \;,
\ee
where the operators are time ordered. If $W^{(i)}$ were local lines in the full theory, this would be trivial because the lines would live on separate defects. However, in the full theory $\cB$ is dynamical and thus both $W^{(1)}$ and $W^{(2)}$, which are coupled to $\cB$, must be the end-lines of suitable bulk surfaces $U(\gamma)=e^{i \!\int _{\gamma} \cB}$. Likewise, also the product $W^{(1)} W^{(2)}$ must be attached to a bulk surface with the correct charge (see Figure~\ref{Fig:braiding12}). Commuting the order in which the end-lines are fused has the effect of half-braiding the attached bulk surfaces, which is captured by the normal ordering phase $\exp\bigl(\frac{2 \pi i}{N} \, 2^{-1} \langle n_1, n_2 \rangle \bigr)$ we already introduced in (\ref{normal ordering phase}). This is precisely the braiding \eqref{extra braiding}.

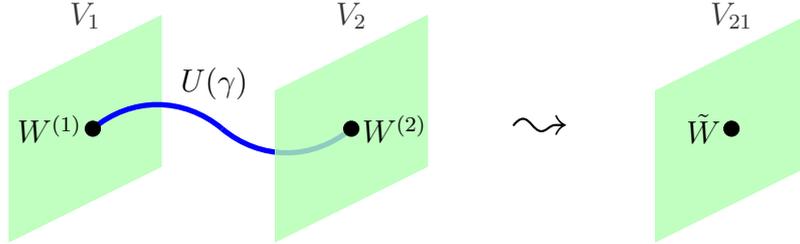
\begin{figure}[t]
\centering
\begin{tikzpicture}
	\filldraw[white!70!green, opacity=0.8] (-0.5, 0) -- (1.5, 1) -- (1.5, 3) node[black, shift={(-1,0)}] {$V_1$} -- (-0.5, 2) -- cycle;   % left green square
	\draw[blue, line width=2] (0.6, 1.5) to [out=40, in=140] (2.3, 1.5) node[black, shift={(-0.1, 0.6)}] {$U(\gamma)$};  % blue line
	\draw[blue, line width=2] (2.3, 1.5) to [out=-40, in=220] (4, 1.5);
	\filldraw[white!70!green, opacity=0.8] (3,0) -- (5,1) -- (5,3) node[black, shift={(-1,0)}] {$V_2$} -- (3,2) -- cycle;  % right green square
	\filldraw (0.6, 1.5) circle (0.1) node[left] {$W^{(1)}$};  % black dots
	\draw[fill = black] (4,1.5) circle (0.1) node[right] {$W^{(2)}$};
	\filldraw[white!70!green, opacity=0.8] (8, 0) -- (10, 1) -- (10, 3) node[black, shift={(-1,0)}] {$V_{21}$} -- (8, 2) -- cycle;   % green square after fusion
	\filldraw (9, 1.5) circle (0.1) node[left] {$\tilde W$};
	\node at (6.5, 1.5) {\LARGE$\leadsto$};
\end{tikzpicture}
\caption{The lines $\tilde W$ of $V_{21}$ that are decoupled from $\cB$ can be seen as products of end-lines that are attached to a surface $U(\gamma)$ stretched between the two defects $V_1$ and $V_2$. Once the defects are fused, the lines $\tilde W$ become local in $V_{21}$.
\label{fig:w12}}
\end{figure}

We indicate the product of the two sectors $\cA^{N, -\cT_i}$ deformed by the extra braiding (\ref{extra braiding}) as $\cA^{N, -\cT_2} \times_\cB \cA^{N, - \cT_1}$, in order to distinguish it from the standard decoupled tensor product. We label the lines of this theory by $\cN = (n_1, n_2)$. The spin of the lines of $W^{(1)}$ and $W^{(2)}$ is undeformed, while the spin of product of lines can be computed using $\theta_{a+b} = \theta_a \theta_b B_{ab}$. We obtain
\be
\label{spin in product theory}
\theta[ W_\cN] = \exp\Bigl( \pi i \,\, \cN^{\sT} \cK_{21} \cN \Bigr) \,\qquad\qquad \cK_{21} = \frac1N \mat{ -\cT_1 & \frac\epsilon2 \\ - \frac\epsilon2 & -\cT_2 } \;.
\ee
The line $W_\cN$ has charge $n_1 + n_2$ under the $\bZ_N \times \bZ_N$ 1-form symmetry coupled to $\cB$. We can identify a subset of lines that are decoupled from $\cB$ and, under certain conditions, form a consistent, independent, and local 3d MTC. These are the lines with $\cN = (l, -l)$: they exist without an attached bulk surface, and can be thought of as sitting at opposite ends of a $\cB$ surface before fusion, see Figure~\ref{fig:w12}. The spin of these lines is $\exp\bigl( - \frac{\pi i}N \, l^\sT(\cT_1 + \cT_2) l \bigr)$ and thus, as long as $(\cT_1 + \cT_2)$ is invertible in $\bZ_N$, they form the consistent MTC $\cA^{N, -\cT_1 - \cT_2}$. The remaining lines are coupled to $\cB$. We can identify a subset that has trivial braiding with the lines of $\cA^{N, -\cT_1 - \cT_2}$. They are given by $\cN_\eta = (\xi, -\xi+\eta)$ with $\xi = (\cT_1 + \cT_2)^{-1} \bigl( \cT_2 + \frac\epsilon2 \bigr) \eta$ and their spin is $\exp\bigl( - \frac{\pi i}N \, \eta^\sT \cT_{21} \eta \bigr)$ where the matrix $\cT_{21}$ is the one in (\ref{exp for T21}). Since the line $\cN_\eta$ has charge $\eta$ under the $\bZ_N \times \bZ_N$ 1-form symmetry, they form the MTC $\cA^{N, -\cT_{21}}(\cB)$. Hence we arrive to the result
\be
\label{eq:bulkstacking}
\cA^{N,-\cT_2}(\cB) \times_\cB \cA^{N, -\cT_1}(\cB) = \cA^{N, -\cT_1 - \cT_2} \times \cA^{N, -\cT_{21}}(\cB) \;.
\ee
The product on the right-hand-side is the standard tensor product. The result is in accord with the factorization theorem of \cite{Hsin:2018vcg}. We have thus shown that:
\be
\label{eq:fusiontwistsectors}
D[\cT_2] \times D[\cT_1] = \cA^{N, - \cT_1 - \cT_2} \; D[\cT_{21}] \;,
\ee
as long as as both $(\cT_1 + \cT_2)$ and $\cT_{21}$ are invertible in $\bZ_N$, as suggested by (\ref{fusion twist defects+ decoupled TQFT}).

The result could be confronted with the known composition of minimal TQFTs $\cA^{N,p}$  \cite{Choi:2022zal}, namely
$ \cA^{N,p} \times \cA^{N,q} = \cA^{N,\, p + q} \times \cA^{N,\, (p^{-1} + q^{-1})^{-1}}$
valid when $\gcd(p+q,N)=1$. While we found an equivalent expression for the decoupled lines on the right-hand-side, the lines coupled to $\cB$ fuse differently because of the bulk dynamics.

Let us mention two cases in which the decomposition (\ref{eq:bulkstacking}) fails. One case is when $\cT_1 + \cT_2 = 0$, namely when we consider the fusion $D[\cT] \times D[-\cT]$ in the untwisted sector. Set $\cT_2 = - \cT_1 = \cT$. The lines decoupled from $\cB$ have vanishing spin and form a Lagrangian subgroup of $\bZ_N \times \bZ_N$, signaling that $\cA^{N,-\cT_2} \times_\cB \cA^{N, -\cT_1}$ must be a Dijkgraaf-Witten theory. Indeed, exploiting (\ref{spin in product theory}), we can exhibit the set of lines $E_n = \smat{ \unit \\ -\unit} \bigl( \cT - \frac\epsilon2 \bigr)^{-1} n$ decoupled from $\cB$ and with vanishing spin, a set of lines $M_m = (m, 0)$ with charge $m$ under $\cB$ and with spin $\exp\bigl( \frac{\pi i}N m^\sT \cT m\bigr)$, and show that the two sets have canonical braiding $\exp\bigl( \frac{2\pi i}N n^\sT m \bigr)$. This is precisely the content of the theory (\ref{3d theory from D Dbar}). We thus reproduce the result (\ref{product of inverse twisted sectors}).

Another special case is when $\cT_{21} = 0$, namely when we consider two defects $V[\cT_2]$, $V[\cT_1]$ that fuse into the charge-conjugation defect $V_C \equiv V[\cT=0]$. The two torsion matrices must be related by $\cT_1 = - \frac\epsilon2 \cT_2^{-1} \frac\epsilon2$. When this happens, the product $\cA^{N, -\cT_2} \times_\cB \cA^{N, -\cT_1}$ is not a MTC because it contains a subcategory of transparent lines --- the lines in the $C$ twisted sector which couple only to $\cB$ and not to $\Phi$. In this case, we can study the fusion of the full twisted sectors, including the lines coupled to $\Phi$ (see Appendix~\ref{app:changeconj}). The final result is:
\be
\label{fusion to C}
D[\cT_1] \times D[\cT_2] = \bigl( \bZ_N  \times \bZ_N \bigr)_0(\Phi, \Phi_1) \; D[0] \;.
\ee
We can compare this result with our standard computation by using the factorization (also proven in the same appendix):%
\footnote{Note that $\cT_1 + \cT_2 = \cT_1 - \frac\epsilon 2 \cT_1^{-1} \frac\epsilon2 = \bigl( \cT_1 + \frac\epsilon2 \bigr) \cT_1^{-1} \bigl( \cT_1 - \frac\epsilon2 \bigr)$ which is invertible under our assumptions.}
\be
\bigl( \bZ_N \times \bZ_N \bigr)_0(\Phi, \Phi_1) = \cA^{N, -\cT_1 - \cT_2} \times \cA^{N, \cT_1 + \cT_2}(\Phi, \Phi_1) \;.
\ee
Thus the result coincides with the remaining cases if we discard the term coupling to $\Phi_1$ (which is integrated out in the 4d bulk computation).

Summarizing, we have obtained the following bulk fusion rules:
\bea
U(\gamma) \times D[\cT] &= e^{i Q_{N\cT}(\Gamma_\gamma)} \; D[\cT] \;, \\
D[\cT_2] \times D[\cT_1] &= \cA^{N, \,-\cT_1 - \cT_2} \; D[\cT_{21}] \;, \\
D[\cT] \times \overline{D}[\cT] &= \cC^{\bZ_N \times \bZ_N} \;.
\eea
We now extend our analysis to the physically relevant case of fusion on a gapped boundary.

\subsection{Fusion on gapped boundaries}
\label{sec: fusion gapped boundary}

In Section~\ref{sec: CS theory} we discussed gapped boundaries $\rho(\cL)$ of the bulk 5d theory.
These are defined by choosing a Lagrangian subgroup $\cL \subset \cD_Q \equiv \bZ_N \times \bZ_N$. The boundary condition sets to 1 the surface operators $U_n$ with $n \in \cL$, which are then screened on the boundary:
\be
U_n \big|_X =1 \qquad \text{if } n \in \cL \;.
\ee
In terms of fields, one imposes Dirichelet boundary conditions $\l^\sT \cB \big|_X = 0$ (up to gauge transformations) for all $l \in\cL$. For $N$ odd prime, the $N+1$ Lagrangian subgroups of $\cD_Q$ are all isomorphic to $\bZ_N$ and are generated by a single vector $l$. Thus the gapped boundaries are implemented by
\be
l^\sT \cB \big|_X \,\equiv\, b_l  = 0 \qquad\text{(up to gauge transformations)} \;.
\ee
In Section~\ref{sec: CS theory} we introduced the 1-form symmetry group $\cS = \cD_Q/\cL$ of the gapped boundary. Here we also introduce the lattice $\cL_\perp$ dual to $\cL$ with respect to the Dirac pairing, and the vector $l_\perp = \epsilon\, l$ that generates $\cL_\perp$. It satisfies $l_\perp^\sT l = 0 \text{ mod } N$.%
\footnote{The lattice $\cL$ could be self-dual, in which case $l_\perp = s \, l$ for some $s \in \mathbb{Z}$. In particular, for $N$ prime, the self-dual lattices are in one-to-one correspondence with the roots $s^2 = -1$ and are generated by $l = (1,s)$.}
Using this vector we can solve the boundary conditions by setting:
\be
\cB \big|_X = \tilde{b}_\perp \, l_\perp \;.
\ee
Notice that this is a condition on the field and not on the charges.

In this section we want to understand the fate of various types of defects once they are placed on the gapped boundary, or when they terminate on it.
We already discussed the case of the 2d surfaces $U_p$ with $p \in \cL$: they can terminate on the gapped boundary, and become trivial if they are placed on top of it. On the other hand, if we fuse a 4d defect $V_M$ (implementing the action of $M \in SL(2,\bZ_N)$ on the gauge field $\cB$) with a gapped boundary $\rho(\cL)$ we obtain a new gapped boundary $\rho(M\cL)$.

Let us now discuss the properties of the twist defects $D[\cT]$ on a gapped boundary. Focusing on the case that $V[\cT]$ comes from the condensation of $\bZ_N \times \bZ_N$ and that $\cT$ is invertible in $\bZ_N$, in Section~\ref{sec: Lag descr D[T]} we discussed the 3d sector $\cA^{N,-\cT}(\cB)$ of lines $W_n$ on $D[\cT]$ that are decoupled from $V[\cT]$ but that cancel its anomaly (\ref{anomaly}). Those lines are charged under $\cB$, and thus are the end-lines of surfaces $U_n$ in the fully dynamical theory. The subsector of lines $W_{n = sl}$ ($s \in \bZ_N$) with charge proportional to $l$ are attached to 2d surfaces of $b_l$, and form a consistent MTC $\cA^{N, -t_l}$ for a 1-form symmetry $\cL \cong\bZ_N$, where $t_l \in \bZ_N$ is
\be
t_l = l^\sT \cT l \;,
\ee
provided that $t_l \neq 0$, namely, that the boundary $\rho(\cL)$ is \emph{not} invariant under $V[\cT]$.%
\footnote{If $\cL$ is invariant under $M$, then $Ml = sl$ for some $s \in \bZ_N$. Note that $s\neq 0$, and since we are considering here defects $V_M$ such that $\Tr M \neq 2 \text{ mod }N$, then $s\neq 1$. From (\ref{exp of T from M}) one finds $l^\sT \cT l = \frac{1+s}{1-s} \, l^\sT \frac\epsilon2 l = 0$. On the contrary, if $l^\sT \cT l = 0$ then $\cT l = r\, \frac\epsilon2 l$ for some $r \in \bZ_N$ and here $r\neq-1$. From (\ref{exp of M from T}) one finds $Ml = \frac{r-1}{r+1}\, l$. This shows that $\cL$ is invariant under $M$ if and only if $t_l = 0$. Besides, when $\cT$ is invertible and thus $\Tr M \neq -2 \text{ mod }N$, then a similar argument also shows an if and only if $t_\perp = 0$. \label{foot:tperp}}
(The case that $\cL$ is invariant under $V[\cT]$ will be dealt with in Section~\ref{subsec:invbdy}.)
On the gapped boundary we set $b_l = 0$ (up to gauge transformations), therefore this sector becomes a decoupled TQFT. This allows us to define a minimal boundary twist defect $D_\cL[\cT]$, obtained by discarding the decoupled TQFT $\cA^{N, -t_l}$.%
\footnote{The operation of discarding $\cA^{N, - t_l}$ can be implemented as $D_\cL = \bigl[ D\big|_\text{boundary} \times \cA^{N, t_l} \bigr]/ \bZ_N$ \cite{Hsin:2018vcg}.}
The lines that braid trivially with $\cA^{N, -t_l}$ can be generated by $n = \cT^{-1}l_\perp$ and form a MTC $\cA^{N, - t_\perp}$ with $t_\perp \in \bZ_N$ defined as
\be
t_\perp = l_\perp^{\sT} \cT^{-1} l_\perp \;.
\ee
These lines are coupled to the gauge field $b_\perp \equiv l_\perp^\sT \cT^{-1} \cB$.%
\footnote{The splitting of $\cB$ into $b_l$ and $b_\perp$ is well defined as long as the boundary is not invariant under $V[\cT]$. Otherwise, $\cT l \propto l_\perp$ and so $b_\perp \propto b_l$ which vanishes on the gapped boundary.}
(We omit the dependence of $t_\perp$ and $b_\perp$ on $\cT$ in order not to clutter.)
We have then proved the factorization:
\be
\label{factorization on gapped boundaries}
\cA^{N, -\cT}(\cB) = \cA^{N, -t_l}(b_l) \times \cA^{N, -t_\perp}(b_\perp) \;.
\ee
When we move the twist defect $D[\cT]$ on top of a gap boundary, the first factor on the \rhs{} decouples yielding $D[\cT] \big|_\text{boundary} = \cA^{N, -t_l} \times D_\cL[\cT]$. We obtain:
\be
\label{twist defect on gapped boundary}
D_\cL[\cT] = \cA^{N, - t_\perp}(b_\perp) \qquad\qquad \text{for $M\cL \neq \cL$} \;.
\ee
Notice that $M\cL = \cL$ if and only if $\cT\cL = \cL_\perp$ (see footnote~\ref{foot:tperp}). As we will see, this definition of $D_\cL[\cT]$ is consistent under fusion. Notice also that the twist defect $D_\cL[\cT]$, as opposed to $D[\cT]$, is stuck on the gapped boundary.

As a check of (\ref{factorization on gapped boundaries}), one can take the anomaly inflow action (\ref{anomaly}) and impose the boundary condition $b_l = 0$. This can be done by parametrizing a gauge field in the quotient group as $\cB = t_\perp^{-1} b_\perp l_\perp $, which yields:
\be
\label{anomaly on gapped boundary}
I(b_\perp) = \frac{N}{2\pi} \int_\text{4d} \biggl[ b_\perp d \wt\gamma - \frac12 \, b_\perp t_\perp^{-1} b_\perp \biggr]
\ee
as expected (here $\wt\gamma = t_\perp^{-1} l_\perp^\sT \wt\Gamma$). Thus, the theory $D_\cL[\cT]$ is the minimal one required to cancel the anomaly on the gapped boundary.

In order to compute the fusion $D_\cL[\cT_2] \times D_\cL[\cT_1] $ on a gapped boundary, we need to understand how to impose the boundary condition on the product theory $\cA^{N, -\cT_2} \times_\cB \cA^{N,- \cT_1}$. Following our previous reasoning, the lines $W^{(1)}_{s_1l}$ and $W^{(2)}_{s_2l}$ with charges in $\cL$ are the end-lines of surfaces of $b_l$ but decouple from $\cB$ on the boundary. Since they are all in the same Lagrangian subgroup of $\bZ_N \times \bZ_N$, the two groups maintain trivial mutual braiding even after the deformation by $\cB$. We have thus identified a subset of lines that couple to $b_l$ and form the MTC $\cA^{N,- t_{2,l}}(b_l)  \times \cA^{N,- t_{1,l}}(b_l)$, where $t_{j,l} = l^\sT \cT_j l$. The lines $W_\cN$ that braid trivially with that subset, as we will see, form a MTC $\cA^{N, -\cR_{21}}$ coupled to $\cB$ for some matrix $\cR_{21}$:
\be
\label{decomposition on gapped boundary 1}
\cA^{N, -\cT_2}(\cB) \times_\cB \cA^{N,- \cT_1}(\cB) = \cA^{N,- t_{2,l}}(b_l)  \times \cA^{N,- t_{1,l}}(b_l) \times \cA^{N, -\cR_{21}}(\cB) \;.
\ee
On the gapped boundary, the first two factors on the right-hand side decouple and moreover are precisely the two factors that are discarded in the definition of $D_\cL[\cT_1]$ and $D_\cL[\cT_2]$. After imposing $b_l \big|_X = 0$, the third factor only couples to a projection of $\cB$. As we will see, such a projection is the very one predicted by fusion, namely to $b_\perp = l_\perp^\sT \cT_{21}^{-1} \cB$. Besides, we expect the MTC  $\cA^{N, -\cR_{21}}(b_\perp)$  to be the product of a MTC $\cN_{21}$ that does not couple to $b_\perp$, and the MTC $\cA^{N, - t_{21}^\perp}(b_\perp)$ (where $t_{21}^\perp = l_\perp^\sT \cT_{21}^{-1} l_\perp$) that lives on the twisted sector $D_\cL[\cT_{21}]$. We will verify this expectation, and show that
\be
\cA^{N, -\cR_{21}}(b_\perp) = \cN_{21} \times \cA^{N, - t^\perp_{21}}(b_\perp) \;.
\ee
These relations imply the fusion rules
\be
D_\cL[\cT_2] \times D_\cL[\cT_1] = \cN_{21} \; D_\cL[\cT_{21}] \;,
\ee
where the decoupled TQFT $\cN_{21}$ plays the role of a fusion coefficient.

Let us compute $\cN_{21}$. The $N^2$ lines $W_\cN$ of $\cA^{N, - \cR_{21}}$, that braid trivially with the first two factors on the \rhs{} of (\ref{decomposition on gapped boundary 1}), have charges $\cN = (\xi_1, \xi_2)$ determined by solving the equations
\bea
\label{system for xi1 xi2}
\cT_1 \, \xi_1 - \frac{\epsilon}{2} \, \xi_2 &= a_1 \, l_\perp \\
\cT_2 \, \xi_2 + \frac{\epsilon}{2} \, \xi_1 &= a_2 \, l_\perp
\eea
for some coefficients $a_{1,2} \in \bZ_N$ that depend on the line. In fact, one can use $a_1,a_2$ to parametrize the solutions. We first consider the simple case $\cT_1 = \cT_2$, then the generic case, and finally the exceptional case $\cT_1 = - \cT_2$.

\paragraph{Case \matht{\cT_1 = \cT_2 \equiv \cT}.} This case computes the square of a defect $D_\cL[\cT]$.
Noticing from (\ref{exp for T21}) that $\cT_{21} = \frac12 \bigl( \cT + \frac{\epsilon}{2} \cT^{-1} \frac{\epsilon}{2} \bigr)$, we find:
\be
\xi_1 = \frac12 \, \cT_{21}^{-1} \Bigl( a_1 + a_2 \, \frac\epsilon2 \, \cT^{-1} \Bigr) l_\perp \;,\qquad
\xi_2 = \frac12 \, \cT_{21}^{-1} \Bigl( a_2 - a_1 \, \frac\epsilon2 \, \cT^{-1} \Bigr) l_\perp \;.
\ee
The charge of a line under $\cB$ is $\xi_1 + \xi_2$. One can check that the lines with $a_1 = a_2$ have charge proportional to $\cT_{21}^{-1} l_\perp$, and so they couple to $b_\perp$.
With some algebra%
\footnote{One should use that $\cT_{21} \epsilon \cT = \cT \epsilon \cT_{21}$. It also implies that such a matrix is antisymmetric.}
and (\ref{spin in product theory}), one can check that those lines braid trivially with the lines with $a_1 = -a_2$. This suggests to label the lines in terms of $a,c \in \bZ_N$ and set $a_1 = a-c$, $a_2 = a+c$. The spin of a line labelled by $(a,c)$ is found to be
\be
\theta\bigl[ W_{(a,c)} \bigr] = \exp\biggl( - \frac{\pi i}N \, t_{21}^\perp \, \bigl( a^2 + c^2 \bigr) \biggr) \;,
\ee
where $t_{21}^\perp = l_\perp^\sT \cT_{21}^{-1} l_\perp$. As long as $t_{21}^\perp \neq 0$, such lines form the theory $\cA^{N, - \cR_{21}}$ with
\be
\cR_{21} = \mat{ t_{21}^\perp & 0 \\ 0 & t_{21}^\perp } \;.
\ee
The subset of lines $(a,0)$ form the MTC $\cA^{N, - t_{21}^\perp}(b_\perp)$, as expected. The lines $(0,c)$ have charges under $\cB$ proportional to $\cT_{21}^{-1} \epsilon \cT^{-1} l_\perp$, which has vanishing contraction with $l_\perp^\sT$ and thus is proportional to $l$. On the gapped boundary $b_l = 0$ and hence these lines form a decoupled MTC
\be
\cN_{21} = \cA^{N, -t_{21}^\perp} \;.
\ee
We have obtained the fusion rule
\be
D_\cL[\cT] \times D_\cL[\cT] = \cA^{N, - t_{21}^\perp} \; D_\cL[\cT_{21}] \;.
\ee
Notice that this fusion rule is the same (with the same $\cN_{21}$) on all gapped boundaries $\rho(\cL)$ belonging to the same orbit under $V[\cT]$. This follows from footnote~\ref{foo: invariance of T}.

\paragraph{Generic case.} In order to treat the general case it is convenient to parametrize the lines $(\xi_1, \xi_2) = (v, \eta -v)$ in terms of two vectors $v, \eta$, so that the charge of a line under $\cB$ is $\eta$, and redefine the numbers $a_1 = p+q$, $a_2=q$. The equations (\ref{system for xi1 xi2}) become
\bea
\bigl( \cT_1 + \cT_2 \bigr) v - \bigl( \cT_2 + \tfrac\epsilon2 \bigr) \eta &= p \, l_\perp \\
 \cT_2 \, \eta - \bigl( \cT_2 - \tfrac\epsilon2 \bigr) v &= q\, l_\perp \;.
\eea
Defining $\Gamma = \bigl( \cT_2 - \frac\epsilon2 \bigr) \bigl( \cT_1 + \cT_2 \bigr)^{-1}$, the solutions are
\be
\label{solution for v and eta}
v = q \; \Gamma^\sT \cT_{21}^{-1} l_\perp + p \, \Bigl[ \bigl( \cT_1 + \cT_2 \bigr)^{-1} + \Gamma^\sT \cT_{21}^{-1} \Gamma \Bigr] l_\perp \;,\qquad
\eta = q \; \cT_{21}^{-1} l_\perp + p \; \cT_{21}^{-1} \Gamma l_\perp
\ee
and can be labelled by $q,p \in \bZ_N$. Substituting in (\ref{spin in product theory}), the spins of the lines are
\be \label{eq: generalW}
\theta\bigl[ W_{(q,p)} \bigr] = \exp\biggl( - \frac{\pi i}N \; (q,p) \cR_{21} \Bigl( \!\! \begin{array}{c} q \\[-.5em] p \end{array} \!\! \Bigr) \biggr)
\qquad\text{with}\qquad \cR_{21} = \mat{ t_{21}^\perp & c_\mathrm{o} \\ c_\mathrm{o} & c_\mathrm{d} } \;,
\ee
where
\be
c_\mathrm{o} = l_\perp^\sT \, \cT_{21}^{-1} \Gamma \, l_\perp \;,\qquad
c_\mathrm{d} = l_\perp^\sT \, \Bigl[   \bigl( \cT_1 + \cT_2 \bigr)^{-1} + \Gamma^\sT \cT_{21}^{-1} \Gamma  \Bigr] \, l_\perp = l_\perp^\sT \, \bigl( \cT_1 + \tfrac\epsilon2 \cT_2^{-1} \tfrac\epsilon2 \bigr)^{-1} \, l_\perp \;.
\ee
The subset of lines $(q, 0)$ have charges $\eta$ proportional to $\cT_{21}^{-1} l_\perp$ and thus couple to $b_\perp$. Their spins show that they form the MTC $\cA^{N, - t_{21}^\perp}(b_\perp)$. On the other hand, the subset of lines $(q, p)$ with $q = - (t_{21}^\perp)^{-1} c_\mathrm{o} \, p$ braid trivially with the former subset and constitute the theory $\cN_{21}$. Their charges $\eta$ are such that $l_\perp^\sT \eta = 0$, therefore they are decoupled from $\cB$ on the gapped boundary. Their spins show that
\be
\cN_{21} = \cA^{N, - n_{21}} \qquad\text{with}\qquad n_{21} = c_\mathrm{d} - (t_{21}^\perp)^{-1} c_\mathrm{o}^2 = (t_{21}^\perp)^{-1} \det \cR_{21} \;. \label{eq: n21formula}
\ee
One should recall that, in the absence of a coupling to $\cB$, the theories $\cA^{N, -p} $ and $\cA^{N, - pr^2}$ are equivalent for any invertible $r \in \bZ_N$, and thus for $N$ odd prime the only physical information in $n_{21} \neq 0$ is whether it is a quadratic residue or not. This is detected by the Legendre symbol $n_{21}^{(N-1)/2} \text{ mod }N \in \{1, -1\}$.%
\footnote{In the higher-rank case the situation is similar. For $N$ odd prime, one can always bring a symmetric matrix $\cT$ with values in $\bZ_N$ to a diagonal form $U^\sT \cT U = \diag(t_1, \dots, t_r)$ using an invertible matrix $U$ (see, \eg, \cite{Hahn:2008}). The TQFT is then characterized by the number of $+1$ and $-1$ Legendre symbols of the $t_i$'s.}

\paragraph{Case \matht{\cT_2 = -\cT_1 \equiv \cT}.} This is the case leading to condensation.
The equations for lines in $\cA^{N, -\cR_{21}}$ are just $\cT \xi_1 + \frac{\epsilon}{2} \xi_2 = - a_1 l_\perp$ and $\cT \xi_2 + \frac{\epsilon}{2} \xi_1 = a_2 l_\perp$. The general solution is 
\be
\xi_1 = \biggl[ a \, \Bigl( \cT + \frac{\epsilon}{2} \Bigr)^{-1} - c \, \Bigl( \cT - \frac{\epsilon}{2} \Bigr)^{-1} \biggr] \, l_{\perp} \;,\qquad
\xi_2 = \biggl[ a \, \Bigl( \cT + \frac{\epsilon}{2} \Bigr)^{-1} + c \, \Bigl( \cT - \frac{\epsilon}{2} \Bigr)^{-1} \biggr] \, l_{\perp}
\ee
where we redefined $a_1 = c-a$ and $a_2 = c+a$.
For these lines:
\be
\theta\bigl[ W_{(a,c)} \bigr] = \exp\biggl( - \frac{2\pi i}{N} \, a c \; 2 l_\perp^\sT \Bigl(\cT + \frac{\epsilon}{2} \Bigr)^{-1} l_\perp \biggr) \;.
\ee
Lines with either $a$ or $c=0$ have vanishing spin, which indicates that we are dealing with a DW type theory. The lines with $a=0$ (electric) do not couple to $\cB$ since they have $\xi_1+\xi_2 = 0$. Redefining $a \to \bigl[2 l_\perp^\sT \bigl(\cT + \frac{\epsilon}{2} \bigr)^{-1} l_\perp \bigr]^{-1}$ gives the canonical braiding $B_{ac}= e^{\frac{2 \pi i}{N} ac}$.
Thus
\be
\cA^{N, - \cR_{21}}(b_\perp) = (\bZ_N)_0(b_\perp) = \cC^{\bZ_N} \;.
\ee
We conclude that:
\be
D_{\cL}[\cT] \times \overline{D}_{\cL}[\cT] =  \cC^{\bZ_N} \;.
\ee
The condensate $\cC^{\bZ_N}$ is for the 1-form symmetry $\cS = (\bZ_N \times \bZ_N)/\cL \cong \bZ_N$ that exists on the gapped boundary.

\paragraph{Examples.} We can now apply our formalism to the known cases of duality and triality defects. We consider a generic boundary $\rho(\cL)$, but assume that it is not invariant under any symmetry defect appearing below (apart from $C$, which leaves every boundary invariant).
For the application to self-duality defects, we must compute the fusion $D_{\cL}[S] \times D_{\cL}[S] $. This is a special case, since the \rhs{} involves charge conjugation. The explicit computation is done in Appendix~\ref{app:changeconj} (see also the comments in Section~\ref{subsec:invbdy}).
 The complete fusion gives a coefficient which is a product of DW theories, these all admit a universal boundary condition which allows us to set them to one. This corresponds to the Dirichlet boundary of the DW theory. After this we find:
\bea
D_{\cL}[S] \times D_{\cL}[S] &= (\bZ_N)(\tilde{b}_\perp,\phi_{\perp}) \; D_\cL^{\text{triv}}[0] \\
D_{\cL}[S] \times \overline{D}_{\cL}[S] &= \cC^{\bZ_N} \;.
\eea
For triality defects we compute:%
\footnote{To get to the result we use the property $\cA^{N, pr^2} = \cA^{N,p}$ for $\gcd(r,N)=1$.}
\bea
D_{\cL}[ST] \times D_{\cL}[ST] &= \cA^{N, -p_{ST}} \; D_{\cL} \bigl[ (ST)^2 \bigr] \\
D_{\cL}[CST] \times D_{\cL}[CST] &= \cA^{N, -p_{ST}} \; D_{\cL} \bigl[ (ST)^2 \bigr] \\
D_{\cL}\bigl[ (ST)^2 \bigr] \times D_{\cL} \bigl[ (ST)^2 \bigr] &= \cA^{N, \; p_{ST}}\; D_{\cL} [CST]
\eea
where
\be
p_{ST} = \begin{cases} 1 &\text{if } \cL = \cL(e) \;, \\ r^2 + r + 1 &\text{if } \cL = \cL(m)_r \;. \end{cases}
\ee
These fusions agree with those computed in \cite{Choi:2022zal} on the electric boundary.%
\footnote{One uses that $\cA^{N,1} = U(1)_N$ \cite{Hsin:2018vcg}. Notice that the conventions of \cite{Choi:2022zal} defined in their eqns. (6.7)--(6.9) differ from ours, and their defects are the orientation reversal of ours, leading to a sign change in the level.}
Notice that $p_{ST} \neq 0 \text{ mod }N$ as long as the boundary Lagrangian subgroup $\cL$ is not invariant under $ST$.

\paragraph{Other defects.} Another interesting case is when the 4d defects $V_M = V[\cA, \xi]$ are obtained by condensing a $\bZ_N$ subgroup of $\bZ_N \times \bZ_N$, corresponding to the elements $M \in SL(2,\bZ_N)$ that are conjugate to $T^k$ for some $k$. For simplicity let us consider the case $M = T^k$ with $k = 1, \dots, N-1$. The twisted sectors $D_{T^k \!,\, \cL}$ are described by the minimal theories $\cA^{N, k^{-1}}(b)$ for $\bZ_N$ coupled to the bulk field $b$. For the lines in these theories there is no extra contribution to the braiding when we stack the theories, and thus they fuse in the standard way:
\be
D_{T^k \!,\, \cL} \times D_{T^{k'} \!,\, \cL} = \cA^{N,\, k^{-1} + k^{\prime -1}} \; D_{T^{k+k'} \!,\, \cL}
\ee
as long as $k+k' \neq 0 \text{ mod }N$. This formula is in agreement with the fusion law of $N$-ality defects found in \cite{Kaidi:2021xfk, Choi:2022zal}.
Notice that these twist sectors are not unique since they can be fused to 3d condensates for the magnetic symmetry. However, on the magnetic boundaries $\cL(m)$ (on which the twisted sector $D_{T^k}$ hosts a minimal theory) we can take the condensates to be generated by the magnetic symmetry $l(m) \in \cL(m)$.\footnote{We can always arrive at this choice since any two magnetic lattices differ by electric ones, which can be absorbed by $D_{T^k}$}. On the magnetic boundary these condensates however become all decoupled DW theories since $l(m)^\sT \cB \big|_X = 0$.

\subsubsection{Twist defects and boundary-changing operators}

\begin{figure}
\begin{center}
\begin{tikzpicture}
	\filldraw[color=white!100!teal, left color=white!75!teal, right color=white!100!teal] (0,0) -- (0,3) -- (2,3) -- (2,0) -- cycle;
	\draw[color=green, line width=2] (0,3) -- (0,0);
	\draw (0,1.5) -- (2,1.5);
	\draw[fill=black] (0,1.5) circle (0.1); 
	\node at (-0.5,0.75) {$\rho(\cL)$};
	\node at (-0.5,2.25) {$\rho(\cL)$}; 
	\node at (-0.7,1.5) {$D_{M, \cL}$};
	\node at (2.4,1.5) {$V_M$};
	\begin{scope}[shift={(5,0)}]
	\filldraw[color=white!100!teal, left color=white!75!teal, right color=white!100!teal] (0,0) -- (0,3) -- (2,3) -- (2,0) -- cycle;
	\draw[color=green, line width=2] (0,0) -- (0,1.5);
	\draw[color=black!50!green, line width=2] (0,3) -- (0,1.5);
	\draw (0,1.5) -- (2,1.5);
	\draw[fill=black] (0,1.5) circle (0.1); 
	\node at (-0.5,0.75) {$\rho(\cL)$};
	\node at (-0.7,2.25) {$\rho(M\cL)$}; 
	\node at (-0.5,1.5) {$\cU_M$};
	\node at (2.4,1.5) {$V_M$};
	\end{scope}
	\begin{scope}[shift={(10,0)}]
	\filldraw[color=white!100!teal, left color=white!75!teal, right color=white!100!teal] (0,0) -- (0,3) -- (2,3) -- (2,0) -- cycle;
	\draw[color=black!50!green, line width=2] (0,3) -- (0,2.25);
	\draw[color=green, line width=2] (0,2.25) -- (0,0);
	\draw (0,2.25) arc (90:-90: 0.75 and 0.75);
	\draw[fill=black] (0,2.25) circle (0.1); 
	\draw[fill=black] (0,0.75) circle (0.1); 
	\node at (-0.7,0.75) {$D_{M,\cL}^\dag$};
	\node at (-0.5,2.25) {$\cU_M$};
	\node at (1.1,1.5) {$V_M$};
	\node at (2.5,1.5) {$=$};
	\begin{scope}[shift={(3,0)}]
	\filldraw[color=white!100!teal, left color=white!75!teal, right color=white!100!teal] (0,0) -- (0,3) -- (2,3) -- (2,0) -- cycle;
	\draw[color=green, line width=2] (0,0) -- (0,1.5);
	\draw[color=black!50!green, line width=2] (0,3) -- (0,1.5);
	\draw[fill=black] (0,1.5) circle (0.1); 
	\node at (0.5,1.5) {$\varphi_M$};
	\end{scope}
	\end{scope}
\end{tikzpicture}
\end{center}
\vspace{0.7cm}
\begin{center}
\begin{tikzpicture}
	\filldraw[color=white!80!teal, fill=white!80!teal] (0,0) -- (0,3) -- (2,3) -- (2,0) -- cycle;
	\draw[color=green, line width =2] (0,0) -- (0,1);
	\draw[color=black!50!green, line width =2] (0,2) -- (0,1);
	\draw[color=green, line width =2] (0,2) -- (0,3);
	\draw (0,1) -- (2,1);
	\draw (2,3) -- (2,0);
	\draw[fill=black] (0,1) circle (0.1);
	\draw[fill=black] (0,2) circle (0.1);
	\draw[fill=black] (2,1) circle (0.1);
	\node at (-0.5,0.5) {$\rho(\cL)$};
	\node at (-0.7,1.5) {$\rho(M \cL)$};
	\node at (-0.5,2.5) {$\rho(\cL)$};
	\node at (2.7,0.5) {$R[M\tau]$};
	\node at (2.5,2) {$R[\tau]$};
	\node at (0.4,0.6) {$\cU_M^\dagger$};
	\node at (0.4,1.7) {$\varphi_M$};
	\node at (1.2,1.3) {$V_M$};
	\node at (3.5,1.5) {$\Rightarrow$};
	\node at (-1.8,1.5) {$\Leftarrow$};
	\begin{scope}[shift={(4.5,0)}]
	\draw[color=green, line width =2] (0,0) -- (0,1);
	\draw[color=black!50!green, line width =2] (0,2) -- (0,1);
	\draw[color=green, line width =2] (0,2) -- (0,3);
	\draw[fill=black] (0,1) circle (0.1);
	\draw[fill=black] (0,2) circle (0.1);
	\node at (0.8,0.5) {$A_\rho[M\tau]$};
	\node at (0.7,1.5) {$A_{\rho_M}[\tau]$};
	\node at (0.6,2.5) {$A_\rho[\tau]$};
	\node at (2,1.5) {$\Rightarrow$};
	\draw[color=green, line width =2] (3,0) -- (3,3);
	\draw[fill=black] (3,1.5) circle (0.1);
	\node at (3.6,2.25) {$A_\rho[\tau]$};
	\node at (3.8,0.75) {$A_{\rho}[M\tau]$};			
	\end{scope}
	\begin{scope}[shift={(-5.5,0)}]
	\filldraw[color=white!80!teal, fill=white!80!teal] (0,0) -- (0,3) -- (2,3) -- (2,0) -- cycle;
	\draw[color=green, line width =2] (0,0) -- (0,3);
	\draw (2,0) --(2,3);
	\draw (0,1.5) -- (2,1.5);
	\draw[fill=black] (0,1.5) circle (0.1);				\draw[fill=black] (2,1.5) circle (0.1);	
	\node at (-0.5,2.25) {$\rho(\cL)$};
	\node at (-0.5,0.75) {$\rho(\cL)$};	
	\node at (1.2,1.8) {$V_M$};	
	\node at (2.5,2.25) {$R[\tau]$};
	\node at (2.7,0.75) {$R[M\tau]$};	
	\node at (0.6,1.1) {$D_{M, \, \cL}$};						
	\end{scope}
\end{tikzpicture}
\end{center}
\caption{The way in which the symmetry TFT implements the construction of \cite{Choi:2021kmx,Kaidi:2022uux}. Above: definition of various morphisms. Below: construction of the duality interface $D_{M, \, \cL}$.
\label{fig:morphismsboundary}}
\end{figure}
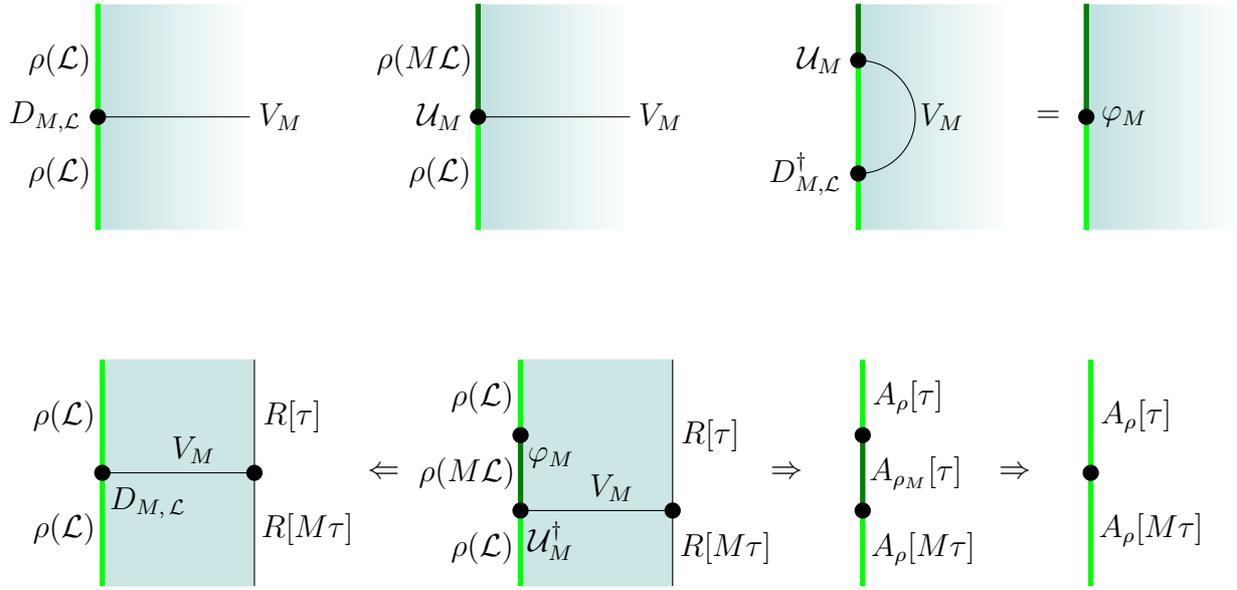

Consider starting with a twist defect $D_M$ (attached to a 4d symmetry defect $V_M$) in the bulk and moving it on top of a gapped boundary $\rho(\cL)$. We are here interested in the case that $\rho(\cL)$ is not invariant under $M$. As discussed before (\ref{twist defect on gapped boundary}), the defect $D_M$ on the boundary decomposes into $D_{M, \,\cL}$ and a decoupled TQFT. We conclude that $D_{M,\,\cL}$ is an interface between two copies of $\rho(\cL)$, or using categorical terms, it defines a morphism $D_{M, \,\cL}: M \times \rho(\cL) \to \rho(\cL)$. This is depicted in Figure~\ref{fig:morphismsboundary} left. 
On the other hand, if we bring the symmetry defect $V_M$ on top of the boundary we obtain an action $V_M \times \rho(\cL) = \rho(M\cL)$. Thus, we can construct an interface between $\rho(\cL)$ and $\rho(M\cL)$ by fusing the boundary with $V_M$ only on a half-space and then letting $V_M$ escape in the bulk, as in Figure~\ref{fig:morphismsboundary} center. This defines a morphism $\cU_M: M \times \rho(\cL) \to \rho(M\cL)$.
Since both interfaces sit at the end of a symmetry defect $V_M$, it is possible to define a local boundary-changing operator as the morphism $\varphi_M = \cU_M \circ D_{M,\,\cL}^\dag: \rho(\cL) \to \rho(M\cL)$, as in Figure~\ref{fig:morphismsboundary} right.

Recall that, in the ungauged theory, one can expect to define only duality \emph{interfaces}. The interface is a composite object given by a discrete gauging operation composed with an invertible duality transformation. 
On the TQFT side this is described by acting with $\cU_M^\dagger$ to map the boundary to $\rho(M\cL)$ and then using $\varphi_M$ to go back to $\rho(\cL)$. After compactifying the slab of symmetry TFT this gives an interface : $ A_\rho[\tau] \to A_{\rho_M}[\tau] \to A_{\rho}[M \tau]$ between absolute theories. Shrinking the middle part of the drawing gives the duality interface. On the other hand the fusion $ \varphi_M \times \cU_M^\dagger = D_{M, \, \cL}$ holds, since $D_{M, \, \cL}$ is unique as a twist defect on the gapped boundary. 
We can thus identify the defect $D_{M, \, \cL}$ on the bottom-left of Figure \ref{fig:morphismsboundary} with the duality interface in the absolute theory $A_\rho[\tau]$.

\subsubsection{Boundaries with a stabilizer}
\label{subsec:invbdy}

Let us also discuss the properties of a twist defect $D_\cL[\cT]$ on a gapped boundary $\rho(\cL)$ that is invariant under the corresponding symmetry defect $V[\cT]$. This means that $M$ is in the stabilizer $H$ of $\cL$ in $SL(2,\bZ_N)$. We can gather information on the degrees of freedom living on $D_\cL[\cT]$ by computing the anomaly inflow. The invertible TQFT living on $V[\cT]$ is (\ref{anomaly}). On the gapped boundary we parametrize%
\footnote{Here the normalization is different than before (\ref{anomaly on gapped boundary}), because $t_\perp = 0$ when $\cL$ is invariant under $M$.}
$\cB = \tilde b_\perp l_\perp$  and obtain
\be
I_{\cT} \big|_{\rho(\cL)} = \frac{N}{2 \pi} \int \biggl[ \tilde\gamma_\perp \, d\tilde b_\perp - \frac12 \, t_\perp \, \tilde{b}_\perp \tilde{b}_\perp \biggr]
\ee
(where $\tilde\gamma_\perp = l_\perp^\sT \wt\Gamma$). Since now $t_\perp = 0$ (see footnote~\ref{foot:tperp}), the anomaly is trivialized.

What happens to the lines in the twisted sector can be understood using the minimal theory description. We start with the twist defect $D[\cT]$ in the bulk and push it onto an invariant boundary $\cL$. Normally we would now separate the degrees of freedom which decouple on the boundary, which form a $\cA^{N, - t_l}(b_l)$ factor. This is generated by lines $L_s \equiv W_{n = s \; l}$.
If the boundary is invariant then $t_l=0$ and this procedure is ill defined as the $L_s$ all have vanishing spin. They thus form a Lagrangian subalgebra. This means that $\cA^{N, -\cT}$ should rather be thought of as a DW theory coupled to $\tilde{b}_\perp$. 
Since the lines with trivial spin are also uncharged under $\tilde{b}_\perp$ this can be thought of as a condensate:
\be
\cA^{N, \cT}(\cB) \big|_X = \cC^{\bZ_N} \;.
\ee
To be more precise we can choose a generator $u$ of $\cS$. Since by definition $ u^\sT l_\perp \neq 0$ lines $\tilde{L}_r \equiv W_{n = r u}$ are charged under $\tilde{b}_\perp$. 
These lines have spin:
\be
\theta \left[ \tilde{L}_r\right] = \exp\left( - \frac{\pi i}{N} \; r^2 \; u^\sT \cT u \right)
\ee
and braid with the electric lines $L_s$:
\be
B_{r, s} = \exp\left(- \frac{2 \pi i}{N} \; r s \; u^\sT \cT l \right) \neq 1 \, ,
\ee
since on the invariant boundary $\cT \cL = \cL_\perp$. Properly redefining $L_s$ we can make this braiding into canonical one. As we have already commented there is no canonical choice for $u$, since we are free to shift it by vectors in $\cL$. The shift $u \to u + l$ does not affect the braiding with $L_r$ but it does affect the spin of $\tilde{L}_s$:
\be
\theta \left[ \tilde{L}_r\right] \to \theta \left[ \tilde{L}_r\right] \exp\left( - \frac{2 \pi i}{N} \; r^2 \; u^\sT \cT l\right)
\ee
For $N$ odd and on spin manifolds we can use this to set $\theta[L_r]$ to one.

Since the defect $V[\cT]$ has trivial anomaly on $\rho(\cL)$, it can end there without adding new degrees of freedom. Therefore the twist defect $D_\cL[\cT]$ is trivial (invertible) on an invariant boundary:
\be
D[\cT] \big|_X = \cC^{\bZ_N} \; D_\cL^{\text{triv}}[\cT] \, ,
\ee
where the superscript is useful to remember this fact.

The same phenomenon appears if we consider a fusion $D_\cL[\cT_1] \times D_\cL[\cT_2]$ in which $V[\cT_{21}]$ leaves the boundary $\rho(\cL)$ invariant, but neither $V[\cT_1]$ nor $V[\cT_2]$ do. We proceed as in the usual case by separating out the lines coupling to $b_l$ from both terms in the fusion. This is a well defined procedure since $t_1^l, \, t_2^l \neq 0$ (due to $\cL$ not being invariant under neither $\cT_1$ nor $\cT_2$).
Based on the previous remarks we expect $\cA^{N, - \cR_{2,1}}$ to also be a condensate. It is clear that the theory contains a Lagrangian algebra generated by $W_{(q,0)}$ in \eqref{eq: generalW}. In the generic discussion these lines were coupled to $b_\perp$, however if the boundary is invariant they are not.%
\footnote{The charge under the gauge symmetry for $\tilde{b}_\perp$ is $q l_\perp^\sT \cT_{2,1}^{-1} l_\perp$, which vanishes when the boundary is invariant.}
These form the set of ``electric'' lines. The magnetic lines $W_{(0,p)}$ instead couple to $\tilde{b}_\perp$, but have nontrivial spin:
\be
\theta\left[W_{(0,p)}\right] = \exp \left(- \frac{\pi i}{N} c_\mathrm{d} \; p^2 \right) \, ,
\ee
As before, we can redefine the magnetic lines by summing a multiple of the electric ones to set this to zero.
Notice that the discussion here is also consistent with the example of $\cT_2 = - \cT_1$ discussed before, when the final result is a condensate and the identity defect leaves all boundaries invariant.

We are now in a position to write down the full result of the boundary fusion for $D_\cL[\cT]$:
\bea
D_\cL[\cT_2] \times D_\cL[\cT_1] &= \cN_{21} \, D_\cL[\cT_{21}]  \;,\qquad& &\text{if } V[\cT_{2,1}]|\rho(\cL)\rangle \neq | \rho(\cL) \rangle \;,  \\
D_\cL[\cT_2] \times D_\cL[\cT_1] &= \cC^{\bZ_N} \; D^{\text{triv}}_\cL [\cT] \;,\qquad& &\text{if } V[\cT_{2,1}]|\rho(\cL)\rangle = | \rho(\cL) \rangle \;, \\
D_\cL[\cT] \times \overline{D_\cL[\cT]} &= \cC^{\bZ_N} \;.
\eea
This will have a more natural interpretation in the gauged theory. In that case we will see that anomaly cancellation forces the Gukov Witten operator $\GW[\cT]$ to exist only as a bound state with the twist defect $D[\cT]$ for $V[\cT]$. When the boundary $\cL$ is $\cT$-invariant there is no anomaly to cancel and $\GW[\cT]$ can exist as a genuine defect on the gapped boundary. The fusion rule above tell us that, when two bound operator fuse onto an invariant one, such fusion is always accompanied by the appearence of a condensation defect.
This is consistent with the fact that defects $D_\cL[\cT]$ absorb surface defects $e^{i \!\int\! b_\perp}$, which survive on the gapped boundary. In the absence of the condensation defect the \rhs{} cannot absorb such lines and fusion would be inconsistent.

%%%%%%%%%%%%%%%%%%%%%%%%%%%%%%%%%%%%%%%%%%
%%%%%%%%%%%%%%%%%%%%%%%%%%%%%%%%%%%%%%%%%%

\section{The gauged theory}
\label{sec:gaugedTheory}

Finally, we discuss the effect of gauging a discrete subgroup $G \subset SL(2,\bZ_N)$ in the bulk TQFT. In the application to $\cN=4$ SYM, the only relevant groups (including the action of charge conjugation) are $\bZ_4$ and $\bZ_6$ generated by $S$ and $ST$, respectively. Notice that they are both Abelian. The construction we present below applies to a generic Abelian $G$, while the non-Abelian case requires modifications that might be important in discussing theories of class $S$ (we comment on that in the conclusions).

We will first describe abstractly the spectrum of operators in the gauged theory. We follow the rules for discrete gauging described for 3d MTCs in \cite{Barkeshli:2014cna} and recently extended to higher dimensions in \cite{Bhardwaj:2022yxj}. Particular care will be needed in describing the Gukov-Witten operators of $G$ gauge theory, as they get dressed by the corresponding twist defects $D[\cT]$. We will present the construction of these operators, that we dub $\fD[\cT]$. Finally, we will study gapped boundaries $|\rho^*\rangle$ in the gauged theory in terms of orbits of boundaries $|\rho\rangle$ in the ungauged theory. This allows for a simple derivation of the fusion rules. We will also comment on the differences arising when the boundary has a nontrivial stabilizer.

In the following we will restrict to the study of twisted sectors $D[\cT]$ for which $M(\cT)$ is an element of $G$. Together with the assumption that $G$ is Abelian, this ensures that different twisted sectors do not mix among each other and that the genuine codimension-2 operators $\fD[\cT]$ of the gauged theory are still labelled by group elements.%
\footnote{In the general case they are labelled by conjugacy classes under the adjoint action of $G$.}

\subsection{Spectrum of bulk operators}

The spectrum of topological operators in the gauged theory can be obtained, at least at a formal level, by applying standard rules for gauging a discrete 0-form symmetry to the ungauged theory. These are nicely summarized in \cite{Bhardwaj:2022yxj}.
Let us start with the surface defects $U_n$ that implement the 2-form symmetry. These operators are in general not gauge invariant, as $G$ acts on them nontrivially. We can build gauge-invariant combinations by considering orbits under $G$:
\be
U^*_{[n]} = \frac{1}{\bigl\lvert \text{Stab}(n) \bigr\rvert} \, \sum_{g \in G} U(g \, n) \;,
\ee
where $\text{Stab}(n)$ is the stabilizer group for $n$ as an element of $\bZ_N \times \bZ_N$. When $n$ admits a nontrivial stabilizer, the surface $U^*_{[n]}$ supports nontrivial line defects labelled by representations of $\text{Stab}(n)$. In the cases considered here, that is $N$ prime and $G= \bZ_4$ or $\bZ_6$, the only surface with a nontrivial stabilizer is the identity, while all others ones do not host any line.

As an example, in the case of the $\bZ_4$ subgroup of $SL(2,\bZ_N)$ generated by $S$, a dyon $(e,m)$ is mapped to an orbit
\be
[e,m] = (e,m) + (m,-e) + (-e,-m) + (-m,e) \;.
\ee
These objects are non-invertible and their fusion is
\be
[e,m] \times [e',m'] = [e + e', m + m'] + [e+m',m-e'] + [e-e',m-m'] + [e-m',m+e'] \;.
\ee

\begin{figure}[t]
\centering
\begin{tikzpicture}
	\draw (0,0) -- (0,3) node[pos = 0.16, left] {$g$} node[pos = 0.9, right] {$h^{-1} g h$};
	\draw (0,1) arc (90:-90: 1 and 1);
	\draw (0,-1) arc (-90:-270: 1.5 and 1.5);
	\draw[fill=white] (0,0) circle (0.1);
	\node[below] at (0,-0.15) {$D[\cT]$};
	\node[right] at (1,0) {$h$};
	\node[left] at (-1.5,0) {$h^{-1}$};
	\draw[fill=red] (0,1) circle (0.05);
	\draw[fill=red] (0,2) circle (0.05);
	\node at (3, 0) {\Huge$\leadsto$};
	\node at (3,1) {$\large\text{gauging}$};
	\draw[fill=white] (5,0) circle (0.1);
	\node[below] at (5,-0.15) {$\fD[\cT]$};
\end{tikzpicture}
\caption{In the gauged theory, a twist defect $D[\cT]$ (with $M(\cT) \equiv g \in G$) is dressed by codimension-1 surfaces of $G$ labelled by $h$.
\label{fig: dressingop}}
\end{figure}
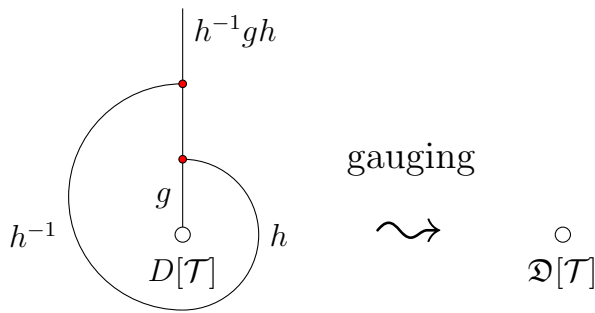

More interesting is the situation for codimension-2 operators. We have already discussed that in the ungauged theory, genuine 3d operators are necessarily condensation defects.
After the discrete gauging the situation is different. The twist defects $D[\cT]$ for surfaces $V[\cT]$ generating $G$ become ``liberated'' --- in the sense that they become genuine 3d operators --- since the surfaces $V[\cT]$ are transparent in the gauged theory.
One could think of the liberated defects as arising from the``lassoed'' configuration shown in Figure~\ref{fig: dressingop} after summing over $G$. Since $G$ is Abelian, each twist sector is left fixed by the action of the lassos and it gives rise to a single genuine operator $\fD[\cT]$.%
\footnote{When instead $G$ is non-Abelian, twist defects also combine into orbits and the situation is more subtle.}
The action of a lasso $V_{M'}$ reduces to a 0-form symmetry action on $D[\cT]$, which maps $W_n \mapsto W_{M'n}$. This is indeed a symmetry of the theory, since
\be
M^{\prime\sT} \, \cT \, M'  = \cT
\ee
and thus it preserves the braiding.
Summing over such action means that the 0-form symmetry on the defect is gauged, so we would like to conclude that $\fD[\cT]$ is $D[\cT]/G$. 

This description is slightly imprecise, because $D[\cT]$ lives at the boundary of $V[\cT]$. Indeed, the gauging process can be thought of as coupling the original system to a discrete $G$ gauge theory. Its gauge field $a \in H^1(M_5,G)$ couples minimally to the 0-form symmetry defects $V_{M \in G}$ of the original theory (more details in Section~\ref{sec:hybrid}).
In this setup, inserting a twist defect $D[\cT]$ is only consistent at locations where $a$ is not closed: it must instead satisfy $\delta a = g$ schematically. Another way of saying this is that $a$ exhibits a nontrivial holonomy $g$ around the 3-cycle $Y$ on which $D[\cT]$ lies. This is the description of Gukov-Witten defect operators in $G$ gauge theory, that we indicate as $\GW_g$.
We infer that a more precise definition of the new operators is:
\be
\fD[\cT] = \GW_{M(\cT)} \times D[\cT]/G \;.
\ee
The appearance of this ``bound state'' has a simple explanation: In the original theory, the defect $D[\cT]$ was not gauge invariant due to anomaly inflow from $V[\cT]$. The $\GW$ operator is not gauge invariant either, as it carries the anomaly of $V[\cT]$. Their combination is a well defined operator in the gauged theory. This is a close cousin of the mechanism described in \cite{Kaidi:2021xfk}. 
We also learn that $\fD[\cT]$ is charged under the dual $\wh{G}$ 3-form symmetry.

The exception is the twist defect $D[\cT=0] \equiv D_C$ for charge conjugation. In this case there is no anomaly inflow and therefore the GW operator for $C$ defines a genuine, group-like object in the gauged theory. This  suggests that we should interpret the contributions from $D_C$ arising upon fusion as decoupled condensates after gauging.

The following table summarizes the properties of some objects in the gauged theory:
\begin{center}
	\begin{tabular}{c|c|c|c}
		Original object & Gauged object & Emergent lines &  Grouplike?  \\[.3em] \hline\hline
		$(0,0)$ & $[0,0]$ & $\text{Rep}(G)$ & YES \\ \hline
		$(e,m)$ & $[e,m] = \oplus_{g \in G} \, g(e,m)$ & none & NO  \\ \hline
		$D[\cT]$ & $\fD[\cT] = \GW_{M(\cT)} \times D[\cT]/G$ & $\text{Rep}(G)$ & NO  \\ \hline
		$D_C$ & $\GW_C$ & none & YES 
	\end{tabular}
\end{center}

\subsection{Hybrid formulation of the gauged theory}
\label{sec:hybrid}

In order to give a Lagrangian description of the gauging of the subgroup $G \subset SL(2,\bZ_N)$ in the 5d Chern-Simons theory, we employ a sort of hybrid formulation in which the Chern-Simons theory is described by continuum gauge fields, while the gauge field for $G$ is described using singular cochains (see, \eg, \cite{HatcherBook, Kapustin:2013qsa} or the appendix in \cite{Benini:2018reh}).

First of all, on the spacetime manifold $M_5$ one chooses a simplicial triangulation. This is made of vertices or 0-simplices $p_i$ with an arbitrary ordering for the index $i$, edges or \mbox{1-simplices} $p_{ij}$ (with $i<j$) connecting the vertices $p_i$ and $p_j$, 2-simplices $p_{ijk}$ (with $i<j<k$) bounded by $p_{ij}$, $p_{jk}$ and $p_{ik}$, and so on. All simplices are contractible, and $M_5$ is the union of all 5-simplices. A gauge field $a$ for the discrete gauge group $G$ is a 1-cochain $a \in C^1(M_5, G)$ that assigns an element $a_{ij} \in G$ to each 1-simplex $p_{ij}$ (with $i<j$), with the constraint that $da = \unit$. We use multiplicative notation and define the differential as $(da)_{ijk} = a_{jk} a_{ik}^{-1} a_{ij}$ (with $i<j<k$). We will only consider the case that $G$ is Abelian. Gauge transformations then map $a_{ij} \mapsto (d\lambda)_{ij} a_{ij}$ where $d\lambda_{ij} = \lambda_j \lambda_i^{-1}$ and $\lambda \in C^0(M_5, G)$ in a 0-cochain. The gauging of $G$ is described by a sum over $a \in H^1(M_5, G)$ in cohomology.

\begin{figure}[t]
\centering
\begin{tikzpicture}
	\draw [fill = green!20!white, opacity = 0.7] (90: 2)  arc [start angle =   40, end angle = 260, radius = 1.843];
	\fill [green!20!white, opacity = 0.7] (0,0) -- (90:2) -- (210:2) -- cycle;
	\draw [fill = blue!10!white, opacity = 0.7] (210:2) arc [start angle = 160, end angle = 380, radius = 1.843];
	\fill [blue!10!white, opacity = 0.7] (0,0) -- (210:2) -- (330:2) -- cycle;
	\draw [fill = red!10!white, opacity = 0.7] (330:2) arc [start angle = 280, end angle = 500, radius = 1.843];
	\fill [red!10!white, opacity = 0.7] (0,0) -- (330:2) -- (90:2) -- cycle;
	\draw [blue] (0,0) -- node[pos = 0.7, right, black] {\small$U_{jk}$} (90:2);
	\draw [blue] (0,0) -- node[pos = 0.85, above = 0.1, black] {\small$U_{ik}$} (210:2);
	\draw [blue] (0,0) -- node[pos = 0.65, below, black] {\small$U_{ij}$} (330:2) ;
	\filldraw (-90: 1.8) circle [radius = 0.05] node[shift={(-90:0.35)}] {\small$p_i$} node[shift={(-60:1.2)}] {\small$U_i$};
	\filldraw (30: 1.8) circle [radius = 0.05] node[above right] {\small$p_j$} node[shift={(100:1.2)}] {\small$U_j$};
	\filldraw (150: 1.8) circle [radius = 0.05] node[above left] {\small$p_k$} node[shift={(80:1.2)}] {\small$U_k$};
	\draw [dashed] (-90:1.8) to node[pos = 0.75, right] {\small$p_{ij}$} (30:1.8) to node[pos = 0.75, above] {\small$p_{jk}$} (150:1.8) to node[pos = 0.75, left] {\small$p_{ik}$} cycle;
	\filldraw [red] (0,0) circle [radius = 0.08] node [shift={(30:0.5)}, black] {\small$U_{ijk}$};
\end{tikzpicture}
\hspace{2.5cm}
\raisebox{0.8cm}{\begin{tikzpicture}
	\draw[color=blue] (0,0) -- (90:2);
	\draw[color=blue] (0,0) -- (-30:2);
	\draw[color=blue] (0,0) -- (210:2);
	\filldraw [red] (0,0) circle [radius = 0.08];
	\draw[dashed] (-90:1) -- (30:1) -- (150:1) --cycle;
	\node at (-90:1.3) {$i$};
	\node at (30:1.3) {$j$};
	\node at (150:1.3) {$k$};
	\node at (-90:2.3) {$\cB_i$};
	\node at (30:2.3) {$\cB_j$};
	\node at (150:2.3) {$\cB_k$};
	\node at (90:2.4) {$a_{jk}$};
	\node at (210:2.4) {$a_{ik}$};
	\node at (-30:2.4) {$a_{ij}$};
\end{tikzpicture}}
\caption{Left: representation of the simplicial triangulation and the covering by closed sets $U_i$, $U_j$, $U_k$ near a triple intersection. Right: assignment of fields $\cB_i$ and $a_{ij}$.
\label{fig:opencover}}
\end{figure}
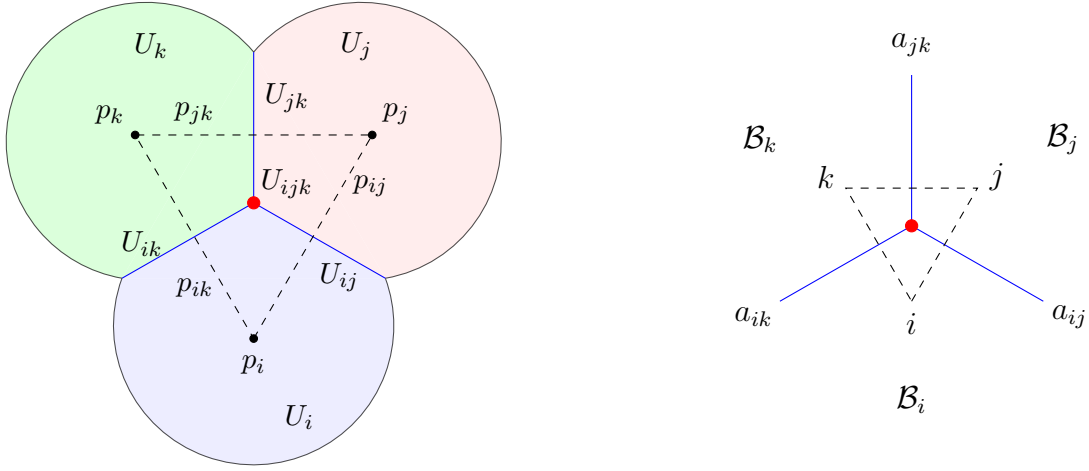

Then we construct a covering of $M_5$ by closed patches that is dual to the triangulation, as follows. Each patch $U_i$ is a 5d contractible manifold with boundary that contains the \mbox{0-simplex} $p_i$. Then each non-empty intersection $U_{i_1 \dots i_k} = U_{i_1} \cap \dots \cap U_{i_k}$ (with $i_1 < \ldots < i_k$ and $k=2,\ldots,6$) is a $(6-k)$-dimensional contractible manifold with boundary that intersects the $(k-1)$-simplex $p_{i_1 \dots i_k}$ at one point. We give a graphical representation of this covering in Figure~\ref{fig:opencover}.

On every patch $U_i$ we define gauge fields $\cB_i$ with values in an Abelian group $\cA$ (either continuous or discrete), and along the intersections $U_{ij}$ we glue them using a group homomorphism $\theta: G \to \Aut(\cA)$ and the gauge field $a$:%
\footnote{Besides, one could also have gauge transformations of $\cB_i$, but we keep them implicit here.}
\be
\label{eq:twistedbc}
\cB_i = \theta(a_{ij}) \, \cB_j \qquad\qquad \text{across } U_{ij} \;.
\ee
The gauge field $\cB$ is thus a piecewise-smooth field with $\cB \big|_{U_i} = \cB_i$. Closeness of $a$ guarantees that each $\cB_i$ can be smooth and have a well-defined limit at triple intersections $U_{ijk}$. In particular, we can always find a gauge in which $a_{ij} = a_{jk} = a_{ik} =1$ around a given triple intersection $U_{ijk}$, and in that gauge $\cB$ can be smooth at the intersection.

The construction is quite general. In our case $\cB _i$ are continuous 2-form gauge fields valued in $\cA = U(1)^2$, while $SL(2,\bZ)$ has the natural action on $\cA$ and $G \subset SL(2,\bZ)$ is an Abelian subgroup. We should now understand how to construct the action. Integrating \mbox{$S= \frac{N}{4\pi} \int \langle \cB, \, d\cB \rangle$} with the discontinuous gluing conditions (\ref{eq:twistedbc}) leads to singularities, in particular the derivative $d\cB$ has delta-function singularities along the surfaces $U_{ij}$. To remedy, we introduce a covariant derivative $d_a$ that removes those singularities:
\be
d_a \cB = d\cB - \sum_{U_{ij}} \delta^{(1)}(U_{ij}) \bigl( \cB_j - \cB_i \bigr) = d\cB - \sum_{U_{ij}} \delta^{(1)}(U_{ij}) \, \sigma(a_{ij}) \, \cB_j \,\equiv\, \Bigl( d - \delta^{(1)}_a \sigma(a) \Bigr) \cB \;,
\ee
where $\delta^{(1)}(U_{ij})$ is a delta-1-form, $\sigma(a) \equiv \unit - \theta(a)$, and in the last expression we used a more compact notation. In this way, $d\cB$ is a piecewise-smooth field such that $d\cB \big|_{\mathring{U}_i} = d\cB_i$ with discontinuities across $U_{ij}$ but no delta-function singularities. We can then construct the action
\be
\label{covariantized action}
S = \frac{N}{4\pi} \int \langle \cB, d_a \cB \rangle = \sum_{U_i} \frac{N}{4\pi} \int_{U_i} \langle \cB_i, d\cB_i \rangle \;.
\ee
The covariant derivative $d_a$ can be integrated by parts, and the action is invariant under gauge transformations of $a$.

\begin{figure}[t]
\centering
\raisebox{-0.8em}{\begin{tikzpicture}
	\draw[color=blue] (0,0) -- (90:2) node[above, black] {\small$a_{ik} = g^{-1} g_{ik}$};
	\draw[color=blue] (0,0) -- (-30:2) node[below, black] {\small$a_{jk} = g_{jk}$};
	\draw[color=blue] (0,0) -- (210:2) node[below, black] {\small$a_{ij} = g_{ij}$};
	\node[circle, draw=black, fill=white, inner sep = 2] at (0,0) {$g$};
	\draw[dashed] (-90:1) -- (30:1) -- (150:1) --cycle;
	\node at (150:1.3) {\small$i$};
	\node at (-90:1.3) {\small$j$};
	\node at (30:1.3) {\small$k$};
	\node at (140:2) {$\cB_i$};
	\node at (40:2) {$\cB_k$};
	\node at (-100:2) {$\cB_j$};
\end{tikzpicture}}
\hspace{1.2cm}
\begin{tikzpicture}
	\draw [blue] (0,0) -- (210:1.5) node[below, black] {\small$g_{ij}$};
	\draw [blue] (0,0) -- (-30:1.5) node[below, black] {\small$g_{jk}$};
	\draw [blue] (0,0) -- (90:2.5) node[pos = 0.3, left, black] {\small$g_{ik}$} node[above, black] {\small$g^{-1} g_{ik}$};
	\draw [blue] (0,1.5) arc [start angle = 90, end angle = -10, x radius = 1.5, y radius = 0.9] node [pos = 0.5, shift={(60:0.5)}, black] {\small$g^{-1}$} node [color = black, circle, draw = black, fill = white, inner sep = 2] {\small$g$};
	\filldraw [red] (0,0) circle [radius = 0.05];
	\filldraw [red] (0,1.5) circle [radius = 0.05];
\end{tikzpicture}
\hspace{1.2cm}
\begin{tikzpicture}
	\draw [blue, thick] (0,0) -- (90:2);
	\draw [blue!80!white, dotted, thick] (0,0) -- (-30: 1.5) node [below, black] {\small$1$};
	\draw [blue!80!white, dotted, thick] (0,0) -- (210: 1.5) node [below, black] {\small$1$};
	\node[circle, draw=black, fill=white, inner sep = 2] at (0,0) {$g$};
	\draw[dashed] (-90:1.3) -- (30:1.3) -- (150:1.3) --cycle;
	\node at (150:1.7) {$i'$};
	\node at (-90:1.7) {$j'$};
	\node at (30:1.7) {$k'$};
	\node at (90:2.3) {$g^{-1}$};
\end{tikzpicture}
\caption{Left: A triple intersection $U_{ijk}$ hosts a GW operator for $g \in G$. We parametrized the gauge field $a$ in terms of $g_{ij} g_{jk} = g_{ik}$ and $da_{ijk} = g$.
Center: A refinement of the triangulation such that the GW operator is pulled away from the junctions. Right: A zoom on the gauge field configuration around the isolated GW operator.
\label{fig:GW}}
\end{figure}
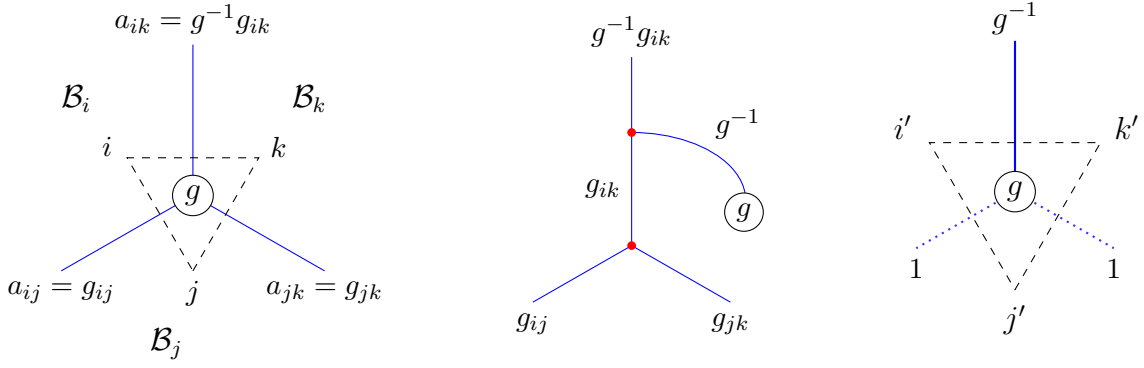

In order to discuss 1-form gauge transformations, we need to compute the square $d_a^2$ of the covariant derivative. It turns out that, to do that, we ought to be more careful and write $d_a \cB = d\cB - \sum_{U_{ij}} \delta^{(1)}(U_{ij}) \bigl( \cB_j^{(ij)} - \cB_i^{(ij)} \bigr)$ where the label $(ij)$ reminds us that we are taking the limit of $\cB_i$ or $\cB_j$ towards $U_{ij}$. Then $d_a d\cB = - \sum_{U_{ij}} \delta^{(1)}(U_{ij}) \bigl( d\cB_j^{(ij)} - d\cB_i^{(ij)} \bigr)$, and finally
\bea
\label{square of differential}
d_a^2 \, \cB &= -\sum_{U_{ijk}} \delta^{(2)}(U_{ijk}) \Bigl[ \bigl( \cB_j^{(ij)} - \cB_i^{(ij)} \bigr) + \bigl( \cB_k^{(jk)} - \cB_j^{(jk)} \bigr) - \bigl( \cB_k^{(ik)} - \cB_i^{(ik)} \bigr) \Bigr] \\
&\equiv\, - \sum_{U_{ijk}} \delta^{(2)}(U_{ijk}) \, \sigma(da_{ijk}) \, \cB \;.
\eea
In the first equality we used that $d\bigl( \delta^{(1)}(U_{ij}) \bigr) = \delta^{(2)}( \partial U_{ij})$ and that the boundary of a double intersection is a collection of triple intersections (with suitable signs due to orientations). In the second line we introduced a compact notation. Indeed, if $a$ is closed ($da = \unit$) then each $\cB_i$ can be smooth and taking the limit towards $U_{ijk}$ in each patch, the first line of (\ref{square of differential}) equals zero. If, instead, $a$ is not closed, then the $G$ bundle can have non-trivial holonomies around the triple intersections and the $\cB_i$'s cannot be smooth there. Given $(da)_{ijk} = g$, consider a gauge in which $a_{ij} = a_{jk} = 1$, $a_{ik} = g^{-1}$ (see Figure~\ref{fig:GW} right). Then the contribution to (\ref{square of differential}) from $U_{ijk}$ becomes $- \delta^{(2)}(U_{ijk}) \, \sigma(da_{ijk}) \, \cB_i^{(ik)}$. We thus write the compact formula 
\be
d_a ^2 = -\delta^{(2)}_a \, \sigma (da) \;.
\ee
In the presence of a background for $a$, 1-form gauge transformations of $\cB$ become
\be
\label{corrected gauge transformations}
\cB \to \cB + d_a \alpha \;,
\ee
and the action (\ref{covariantized action}) remains gauge invariant as long as $a$ is flat.

The theory in which $G$ is gauged involves a sum over choices of $a$ on double intersections $U_{ij}$ that satisfy the closeness condition $da = \unit$. A single symmetry defect $U(\gamma)$ in the ungauged theory is mapped to a sum over its $G$-orbit in the gauged theory. These are precisely the $[e,m]$ defect operators we introduced before.

On the other hand, we can introduce Gukow-Witten operators in the gauged theory \cite{Gukov:2006jk}. These are codimension-2 disorder operators defined by a nontrivial holonomy $g \in G$ for $a$ around a 3d submanifold $\gamma'$. In the hybrid formulation, such a GW operator displaced along a collection of triple intersections $U_{ijk}$ is defined by a sum in the path integral over cochains $a$ such that
\be
\label{GW background}
da_{ijk} = g \qquad \text{whenever $U_{ijk} \subset \gamma'$} \;,
\ee
as in Figure~\ref{fig:GW}.
More generally, a collection of GW operators is described by an exact cochain $h \in C^2(M_5, G)$, and it prescribes to sum over cochains $a$ with $da = h$ in the gauged theory. As mentioned above, requiring the $\cB_i$'s to be smooth in their own patches in a neighborhood of a triple intersection, forces them to be invariant under $g$ there.%
\footnote{If $\cB_i$ is smooth at $U_{ijk}$, then it has a well-define limit there. The limits in the three patches $U_i$, $U_j$, $U_k$ are related by $\cB_i = \theta(a_{ij}) \, \cB_j = \theta(a_{ik}) \, \cB_k$ and $\cB_j = \theta(a_{jk}) \, \cB_k$. Recalling that $G$ is Abelian, this implies $\theta(da_{ijk}) \, \cB_i = \cB_i$ and similarly for $\cB_j$ and $\cB_k$.}
This is a boundary condition naturally implemented on the GW operators, consistent with the fact that $g$-twisted sectors absorb the surfaces of $\cB$ not stabilized by $g$.

Indeed, we can identify a double intersection $U_{ij}$ with gauge field $a_{ij}$ as an alternative description for the 4d symmetry defect $V_M$ with $M = \theta(a_{ij})^\sT$. This is already apparent if we compare the relation $\cB_i = \theta(a_{ij}) \, \cB_j$ between the fields on the two sides of the intersection and (\ref{eq: LR matching}), but it also follows from the action. Let us rewrite (\ref{covariantized action}) as
\be
\label{covariantized action 2}
S = \frac{N}{4\pi} \int \langle \cB , d\cB \rangle - \frac{N}{4\pi} \sum_{U_{ij}} \int_{U_{ij}} \langle \cB, \cB_j - \cB_i \rangle \;.
\ee
The first term imposes that $\cB$ is a $\bZ_N \times \bZ_N$ gauge field.
When $\cT(M)$ is invertible, we can identify the second term with the reduced defect action (\ref{anomaly first version}). Recall from the discussion in Section~\ref{sec: continuum descr defects} that the relation between $M$ and the torsion matrix $\cT$ follows from determining the field on the defect $\cB(0) = \frac12(\cB_\text{L} + \cB_\text{R}) = - \cT\Phi$, where $\epsilon \, \Phi = \cB_\text{R} - \cB_\text{L}$, in terms of the left/right fields $\cB_\text{L/R}$. Substituting $\cB_\text{R} - \cB_\text{L} = -\epsilon \, \cT^{-1} \cB(0)$ into (\ref{covariantized action 2}), the second term becomes
\be
- \frac{N}{4\pi} \sum_{U_{ij}} \int_{U_{ij}} \cB^\sT \cT^{-1} \cB
\ee
that reproduces (\ref{anomaly first version}).

To compute how a GW operator transforms under gauge transformations (\ref{corrected gauge transformations}) we simply evaluate the variation of the action (\ref{covariantized action}) on a non-closed gauge configuration as in (\ref{GW background}):
\be
\label{gauge variation of action with a}
\delta_\alpha S = - \sum_{U_{ijk}} \frac{N}{4\pi} \int_{U_{ijk}} \bigl\langle 2\cB + d_a \alpha,\, \sigma(da_{ijk})\, \alpha \bigr\rangle \;.
\ee
In the gauge of Figure~\ref{fig:GW} right, as above, $\sigma(da_{ijk}) \, \alpha = \alpha_i^{(ik)} - \alpha_k^{(ik)} = \epsilon \, \cT^{-1} \alpha(0)$ in terms of the gauge transformation parameter on the defect. Substituting in (\ref{gauge variation of action with a}) and using that the boundary conditions fix $\cB=0$ on the GW operator, we obtain
\be
\label{gauge variation GW}
\delta_\alpha S = \sum_{U_{ijk}} \frac{N}{4\pi} \int_{U_{ijk}} \alpha^\sT \cT^{-1} d\alpha \;.
\ee
As in the description of Section~\ref{sec:TwistedSectors} in terms of symmetry defects $V_M$, also in the hybrid formulation we find that pure GW operators are not gauge invariant in this theory. We can construct gauge-invariant operators by dressing the GW operators with the twisted sectors $D[\cT]$, whose variation (\ref{eq:ausiliario}) is opposite to (\ref{gauge variation GW}).

In the case of the symmetry defect $V_C$, the field on the defect is simply $\cB(0) = 0$ and thus its gauge transformation parameter $\alpha(0)$ vanishes as well. This means that the gauge variation (\ref{gauge variation of action with a}) vanishes and the GW operator for $C$ is a well-defined gauge-invariant (invertible) topological operator in the gauged theory.

\subsection{Gapped boundaries and non-invertible fusion rules}

We consider now gapped boundaries in the gauged theory. We can use to our advantage the study and classification we already did in the ungauged theory. In order to construct a gauged boundary $|\rho^*\rangle$, we proceed in two steps. First we take a boundary $|\rho\rangle$ in the ungauged theory and make it invariant under the $G$ action:
\be
|\rho\rangle \;\;\to\;\; \frac{1}{\bigl\lvert \text{Stab}(\rho) \bigr\rvert} \, \sum_{g \in G} |\rho_g \rangle \;.
\ee 
As long as $G$ is Abelian, we can associate a stabilizer $H \subset G$ in a consistent way also to the gauged boundary $|\rho^*\rangle$, since $\text{Stab}(\rho_g) = \text{Stab}(\rho)$.
This does not specify a boundary condition completely, since it does not prescribe boundary conditions for neither the $\text{Rep}(G)$ dual symmetry lines, nor the codimension-2 defects $\fD[\cT]$. They form a canonically-conjugated pair of variables, since they braid nontrivially. Therefore, the second step is to choose boundary conditions for them. We choose to impose Dirichlet boundary conditions on $a$:%
\footnote{This is the same choice made in the holographic setup of \cite{Benini:2022hzx}.}
\be
|\text{Dir} \rangle : \quad a = 0 \;.
\ee
Then the operators $\fD[\cT]$ still exist on the gapped boundary as confined excitations.

These are not the only meaningful boundary conditions one could consider. Indeed it would be interesting to understand the effect of Dirichlet boundary conditions on the $\fD[\cT]$'s, or of mixed ones. That they might be useful to describe theories in which either charge conjugation $C$ (this has been studied, \eg, in \cite{Bourget:2018ond, Arias-Tamargo:2019jyh, Bhardwaj:2022yxj}) or the full categorical symmetry, are gauged. We hope to come back to these questions in the future.%
\footnote{In the same spirit, we could consider boundaries twisted by the dual $\check{G}$ symmetry. This amounts to choosing a representation $\alpha$ of $G$ and define, for a boundary with a trivial stabilizer,
\be
|\rho_\alpha^*\rangle =  \, \sum\nolimits_{g \in G} \chi_\alpha(g) \; |\rho_g \rangle \times |\text{Dir}\rangle \;.
\ee
These boundaries have vanishing overlap with the relative theory if we assume absolute theories in the same orbit to have the same partition function. When a stabilizer is present we can only twist by characters of $G/\text{Stab}(\rho)$, while boundaries split into copies labelled by representations of $\text{Stab}(\rho)$. We do not know how to interpret these splitted boundaries from the point of view of the 4d QFT, thus we only consider the ones labelled by the trivial representation.
}

With Dirichlet boundary conditions on $a$, we define:
\be
|\rho^* \rangle = \frac{1}{\bigl\lvert \text{Stab}(\rho) \bigr\rvert} \, \Biggl( \; \sum_{g \in G} |\rho_g \rangle \Biggr) \times |\text{Dir}\rangle \;.
\ee
The Dirichlet boundary condition on $a$ greatly simplifies the discussion. The operators $\fD[\cT]$, which away from the boundary host a $\text{Rep}(G)$ worth of lines constructed with the gauge field $a$, on the gapped boundary reduce to a direct product $\GW_{M(\cT)} \times D[\cT]$. The Gukov-Witten operators still exist on $|\rho^*\rangle$ and have group-like fusion. We will now show that the fusion of the twist operators $D_\cL[\cT]$ is the same on each gapped boundary $|\rho_g\rangle$ in the $|\rho^*\rangle$ orbit. This allows to use the results already derived for the boundary fusion.

We need to show that the various minimal theories we constructed in Section~\ref{sec: fusion gapped boundary} in order to study the fusion of twist defects, are isomorphic for boundaries in the same $G$-orbit. Let $M_g \in G \subset SL(2,\bZ_N)$ be a generator of $G$, and $M(\cT)$ be the element of $G$ associated to the twist defect $D[\cT]$ we want to study. Let $\bigl| \rho(\cL) \bigr\rangle$ be a gapped boundary defined by the Lagrangian subgroup $\cL$ with generator $l$. The Lagrangian subgroup of $|\rho_g\rangle$ is $M_g\cL$, and since $\cL^\sT \cL_\perp = 0$, we have
\be
\cL_\perp^g = M_g^{-1 \sT} \, \cL_\perp \;.
\ee
The generators $l$ and $l_\perp$ transform in a similar way. Since $G$ is Abelian and $\epsilon M_g = M_g^{-1\sT} \epsilon$, then $\cT = M_g^\sT \cT M_g$ and so both $t_l$ and $t_\perp$ are invariant along the orbit. Besides, $\Gamma M_g = M_g \Gamma$ and thus the theory $\cR_{21}$ is invariant as well. Since all relevant building blocks are isomorphic on boundaries that sit inside the same $G$-orbit, we conclude that fusion only depends on the orbit $|\rho^*\rangle$.

A new ingredient appears when fusion produces a defect $\cD[\cT_{21}]$ such that $M(\cT_{21})$ stabilizes $|\rho\rangle$. As we discussed, in these cases the minimal theory is replaced by a condensate. After gauging $G$, we are left with the GW operator $\GW_{M(\cT_{21})}$.

Using all of the above, we finally obtain the categorical fusion rules in the boundary theory specified by the gapped boundary $|\rho^*\rangle$:
\bea
\fD_{\rho^*}[\cT_2] \times \fD_{\rho^*}[\cT_1] &= \cN_{21} \; \fD_{\rho^*}[\cT_{21}] \qquad& M_{21} &\notin \text{Stab}(\rho^*) \;, \\
\fD_{\rho^*}[\cT_2] \times \fD_{\rho^*}[\cT_1] &= \cC^{\bZ_N} \; \GW_{M(\cT_{21})} \qquad& M_{21} &\in \text{Stab}(\rho^*) \;, \\
\GW_{M(\cT_2)} \times \GW_{M(\cT_1)} &= \GW_{M(\cT_{21})} \qquad& M_1 , M_2 &\in \text{Stab}(\rho^*) \;.
\eea
In the second line, the condensate is for the 1-form symmetry $\cS$ on the gapped boundary, and the DW description couples to $\tilde b_\perp$. For defects in which only one $\bZ_N$ factor is gauged, on the other hand, the fusions are as follows:%
\footnote{These can be thought of as the case of $T^k$ modulo conjugation.}
\bea
\fD_{T^k \!,\, \rho^*} \times \fD_{T^{k'} \!,\, \rho^*} &= \cA^{N,\, k^{-1} + k'^{-1}} \; \fD_{T^{k + k'} \!,\, \rho^*} \qquad& T &\notin \text{Stab}(\rho^*) \;, \\
\GW_{T^k} \times \GW_{ T^k } &= \GW_{T^{k + k'} } \qquad& T &\in \text{Stab}(\rho^*) \;.
\eea
The same can be said for conjugacy classes $g = \cH^{-1} T \cH$.

%%%%%%%%%%%%%%%%%%%%%%%%%%%%%%%%%%%%%%%%%%
%%%%%%%%%%%%%%%%%%%%%%%%%%%%%%%%%%%%%%%%%%

\section{Conclusions and future directions}
\label{sec:conclusions}

In this paper we have studied how non-invertible self-duality symmetries arise in holography, through the presence in the gravitational bulk of emergent discrete gauge fields at self-dual points on the moduli space. Although we have focused on the specific example of the 4d $\cN=4$ super-Yang-Mills theory  with gauge algebra $\su(N)$, our methods are rather general and should be applicable to a wide range of other theories, for instance to $\cN=2$ theories of class $S$.

The key role is played by a topological low-energy sector of type IIB string theory on $S^5$: a 5d Chern-Simons-like topological field theory of 2-form gauge fields --- equivalent to a $\bZ_N$ discrete 2-form gauge theory --- further orbifolded by a discrete Abelian symmetry $G$. It is essentially the symmetry TFT for $SU(N)$ $\cN=4$ SYM. This theory appears to be quite interesting and rich in its own right, both before and after gauging $G$. We have studied various aspects of the theory in this paper. Before orbifolding, we have analyzed the 4d symmetry defects associated to an $SL(2,\bZ_N)$ 0-form global symmetry of the theory, the associated twisted sectors that live at the boundary of the defects, and their fusion. We have also investigated topological (gapped) boundaries, and how the various defects reduce when they are brought there. Then, we have studied the effect of gauging a subgroup $G \subset SL(2,\bZ_N)$. In particular, the ``liberated'' twist defects $\fD_{g,\, \rho^*}$ that live on gapped boundaries turn out to be the self-duality defects of $\cN=4$ SYM. We derived their fusion rules using our formalism, confirming the results previously obtained in field theory.

We conclude listing a few open questions for future research.

\paragraph{Extension to class \matht{S}.} Our formalism can naturally be extended to study self-duality non-invertible symmetries in other theories, for instance on the conformal manifold of \mbox{$\cN=2$} theories of class $S$ (see \cite{Bashmakov:2022jtl} for related investigations). The symmetry TFT of the 6d \mbox{$\cN=(2,0)$} theory of ADE type $\fg$ is a 7d Chern-Simons theory of 3-form gauge fields with level matrix equal to the Cartan matrix $A_{ij}$ of $\fg$. After reducing on a genus-$g$ Riemann surface $\Sigma_g$, one obtains a 5d CS theory with $2g$-tuples of 2-form gauge fields $\cB_i$ and braiding:
\be
B_{v, w} = \exp\biggl(\frac{2 \pi i}{N} \sum_{i,j} A_{ij} \, \langle v^i , w^j \rangle \biggr) \;,
\ee
where $\langle v , w  \rangle$ is the homology intersection form on $\Sigma_g$. The simplest example is given by theories of type $A_{N-1}$, which give rise to $g$ copies of the CS theory we studied in this paper. Then the (projected) mapping class group $Sp(2g,\bZ_N)$ of $\Sigma_g$ acts on the TQFT as a 0-form symmetry.

\paragraph{The case of \matht{N} not prime.} For the sake of simplicity, we have restricted our analyses to the case of $N$ prime throughout our paper. This technical assumption allowed us to exploit the multiplicative group structure of $\bZ_N^*$, simplifying many formulas. When $N$ is not prime, the situation is technically more complicated, both because the number of subgroups and global structures grows with the number of prime factors in $N$, and because the fusion relations for minimal theories become more involved.

\paragraph{General formulation of the symmetry TFT.} In the last part of our work, we have resorted to a hybrid formulation of the gauged TQFT that uses both discrete and continuous gauge fields. It would be pleasant to give a completely general description in terms of the correct cohomology theory. A promising route could be to employ Deligne-Beilinson twisted cocycles.

\paragraph{Anomalies for \matht{N}-ality symmetries.} In spite of the many recent developments, a clear understanding of 't~Hooft anomalies for non-invertible symmetries in $d>2$ is still lackluster. The main obstacle is to give a concrete implementation of the associativity conditions for $n$-categories. 
The higher-dimensional TQFT approach might help to give an alternative concrete route to such questions: instead of choosing Dirichlet boundary conditions for the discrete 1-form gauge fields $a$, one might try to define Neumann boundaries instead. On these, the non-invertible defects $\fD[\cT]$ are effectively gauged and they define an absolute theory which is obtained from the ones we have studied here by gauging the non-invertible symmetry. The failure to find such a boundary would signal an 't~Hooft anomaly.

\section*{Acknowledgments}

We thank Riccardo Argurio, Michele Del Zotto, Davide Gaiotto, Justin Kaidi, Kantaro Ohmori, Sakura Schafer-Nameki, Luigi Tizzano, and Matthew Yu for helpful discussions. We are especially grateful to Davide Gaiotto for pointing out the possible role of emergent discrete gauge symmetries in the string theory setup in relation to the non-invertible defects of $\cN=4$, providing inspiration for this work.
We gratefully acknowledge support from the Simons Center for Geometry and Physics, Stony Brook University, where some of the research for this paper was performed. We thank the Perimeter Institute for Theoretical Physics and the organizers of the workshop ``Global Categorical Symmetries'' for hospitality. Research at Perimeter Institute is supported in part by the Government of Canada through the Department of Innovation, Science and Economic Development in Canada and by the Province of Ontario through the Ministry of Colleges and Universities.
The authors are partially supported by the INFN ``Iniziativa Specifica ST$\&$FI''. A.A., F.B., C.C. and G.R. are supported by the ERC-COG grant
NP-QFT No. 864583 ``Non-perturbative dynamics of quantum fields: from new deconfined phases of matter to quantum black holes'', by the MIUR-SIR grant RBSI1471GJ, and by the MIUR-PRIN contract 2015 MP2CX4.

\appendix

\section{Basic manipulations with symmetry TFT}
\label{app: symtft}

We review here some basic facts about the symmetry TFT approach to global variants of gauge theories. The idea is simple: in order to describe an $n$-dimensional gauge theory $\cT$, we introduce an auxiliary system, comprised of a $n$-dimensional \emph{relative} theory $R$ together with an $(n+1)$-dimensional non-invertible TQFT $Sym$. The relative theory contains all the information about $\cT$ which is insensitive to the global structure, such as correlators of local operators, possibly charged under flavor symmetries. Other properties of the theory, such as its 1-form symmetry, depend on the choice of the global structure and thus both the topological defects generating them and the charged objects are not part of $R$.  
The geometric setup is as follows:
$$
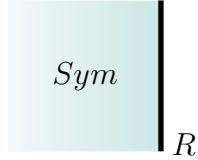
\begin{tikzpicture}
\filldraw[color=white!95!teal, left color=white!95!teal, right color=white!80!teal] (0,-1) -- (2,-1) -- (2,1) -- (0,1) -- cycle;
\draw[line width=2] (2,-1) -- (2,1);
\node at (1,0) {$Sym$};
\node[right] at (2,-0.9) {$R$};
\end{tikzpicture}
$$
From the point of view of the $(n+1)$-dimensional theory, the output of the $n$-dimensional boundary manifold is not a complex number but a vector in a finite-dimensional vector space, namely the Hilbert space of $Sym$. From the point of view of $R$, this Hilbert space is the vector space of partition functions. This does not define an absolute theory, as the bulk Hilbert space of $Sym$ is in general not one-dimensional (\ie, $Sym$ in not invertible).
This can be fixed by the choice of a topological boundary $\rho$. In general there will be multiple independent such $\rho$'s, each one specifying an absolute theory.

Since $\rho$ is topological, we take $Sym$ in a slab with $R$ and $\rho$ boundary conditions on the two sides, and collapse the picture onto $X$, thus obtaining a local (absolute) theory $A_\rho$:
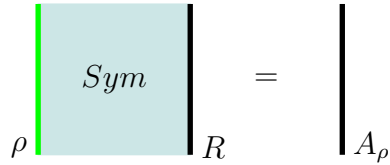
$$
\begin{tikzpicture}
\filldraw[color=white!80!teal, fill=white!80!teal] (0,-1) -- (2,-1) -- (2,1) -- (0,1) -- cycle;
\draw[line width=2] (2,-1) -- (2,1);
\draw[line width=2, color=green] (0,-1) -- (0,1);
\node at (1,0) {$Sym$};
\node[right] at (2,-0.9) {$R$};
\node[left] at (0,-0.9) {$\rho$};
\node at (3,0) {$=$};
\draw[line width=2] (4,-1) -- (4,1);
\node[right] at (4,-0.9) {$A_\rho$};
\end{tikzpicture}
$$
Using standard arguments, one can view the choice of $\rho$ as the gauging of a ``maximal" non-anomalous generalized symmetry $\cA_\rho$ inside $Sym$. On the other hand, one can expand the state on the r.h.s. as $| \cR \rangle = \sum_{\gamma} \cZ[\gamma] \, |\gamma\rangle$, with $|\gamma\rangle$ an orthonormal basis for the TQFT Hilbert space. Computing the overlap gives the partition function of the absolute theory:
\be
\cZ[A_\rho] = \langle \rho | \cR \rangle = \sum\nolimits_{\gamma} \; \langle \rho | \gamma \rangle \; \cZ[\gamma] \;.
\ee
In our case $\gamma \in H^2(X, \bZ_N \times \bZ_N)$ is a surface in the TQFT and
\be
\cZ[A_\rho] = \sum\nolimits_{\gamma \in \cL(\rho)} \cZ[\gamma] \;.
\ee
Using the symmetry TFT construction we can define various objects in the absolute theory on the slab geometry:
$$
\begin{tikzpicture}
	\filldraw[color=white!80!teal, fill=white!80!teal] (0,-1) -- (2,-1) -- (2,1) -- (0,1) -- cycle;
	\draw[line width=2] (2,-1) -- (2,1);
	\draw[line width=2, color=green] (0,-1) -- (0,1);
	\draw[fill=black!50!green] (0,0) circle (0.1);
	\node at (-0.3,0) {$\cA$};
	\node[align=center] at (1, -2) {A topological object \\[-0.7em] $\cA$ inside $A_\rho$};
\begin{scope}[shift={(6,0)}]
	\filldraw[color=white!80!teal, fill=white!80!teal] (0,-1) -- (2,-1) -- (2,1) -- (0,1) -- cycle;
	\draw[line width=2] (2,-1) -- (2,1);
	\draw[line width=2, color=green] (0,-1) -- (0,1);
	\draw[fill=black!50!blue] (2,0) circle (0.1);
	\node at (2.3,0) {$\cO$};
	\node[align = center] at (1,-2) {An $\cA$-neutral operator \\[-0.7em] $\cO$ inside $A_\rho$};
\end{scope}
\begin{scope}[shift={(12,0)}]
	\filldraw[color=white!80!teal, fill=white!80!teal] (0,-1) -- (2,-1) -- (2,1) -- (0,1) -- cycle;
	\draw[line width=2] (2,-1) -- (2,1);
	\draw[line width=2, color=green] (0,-1) -- (0,1);
	\draw (0,0) -- (2,0);
	\draw[fill=black!50!green] (0,0) circle (0.1);
	\draw[fill=black!50!blue] (2,0) circle (0.1);
	\node[align = center] at (1,-2) {An $\cA$-charged object \\[-0.7em] $\cW_\cB$ inside $A_\rho$};
	\node at (-0.3,0) {$\cB$};
	\node at (2.3,0) {$\cW$};
\end{scope}
\end{tikzpicture}
$$
The fact that $\cO$ is neutral while $\cW_\cB$ is charged under $\cA$ follows from sliding the $\cA$ operator in the pictures above before squashing the setup into the absolute theory.

\section{Properties of minimal TQFTs}
\label{app:ANp}

Three-dimensional TQFTs with discrete 1-form symmetry group $\bZ_N$ (or, more generally, products of the form $\prod_I \bZ_{N_I}$) and fixed anomaly for said 1-form symmetry admit powerful classification results in terms of ``minimal" TQFTs $\cA^{ \{N_I\} , \{p_{IJ}\} }$, as pioneered in \cite{Hsin:2018vcg}. Here we review some important consequences of the classification results for $\bZ_N$ (theories $\cA^{N,p}$) and $\bZ_N^r$ (theories $\cA^{N, \cT})$. In the main text we only use $r=1$ and $r=2$.

The possible anomalies for a $\bZ_N$ 1-form symmetry in 3d are labelled by an integer $p$ defined modulo $2N$ (or modulo $N$ on spin manifolds) and can be represented by the following 4d inflow action\cite{Hsin:2018vcg}:%
\footnote{The anomaly is generated by $-I_\cT$ as customary.}
\be
I_\cT = \frac{2 \pi p}{2N} \int_X \fP(B) \;, \qquad\qquad B \in H^2(X, \bZ_N) \;.
\ee
If we assume that there are no non-anomalous subgroups, that is $\gcd(N,p)=1$, to such anomaly we associate a minimal TQFT $\cA^{N,p}$. This theory has $N$ line operators $W_l$, $l=0, \ldots, N-1$ that form a $\bZ_N$ fusion algebra, with spins
\be
\theta_l = \exp\biggl( \frac{2 \pi i \, p}{2 N} \, l^2  \biggr) \;.
\ee
If the theory is bosonic then $\theta_{N}=1$ and $N p \in 2 \bZ$. In this paper we deal with spin theories, in which case $W_N = \psi$ can be a transparent fermion. In bosonic theories $p$ is identified with $p + 2N$, while in the spin case $p +N$ gives rise to the same spin theory. We stress that $\cA^{N,p}$ is a well defined 3d TQFT if and only if $\gcd(N,p)=1$, as otherwise the theory has transparent bosonic lines, which give a non-unitary $S$ matrix. Otherwise, the $S$-matrix is given by
\be
S_{l \,  l'} = \frac{1}{N^{1/2}} \, \exp\biggl( \frac{2 \pi i \, p}{N} \, l \, l' \biggr) \;.
\ee

An important result of \cite{Hsin:2018vcg} is that a 3d TQFT $T$ with a $\bZ_N$ 1-form symmetry with anomaly $p$ admits an expansion in the $\cA^{N,p}$ (hence the name ``minimal"):
\be
T = \cA^{N,p} \times T'  \;, \qquad\qquad T' = \frac{\cA^{N,-p} \times T}{\bZ_N} \;,
\ee 
which can be derived using the identity
\be
\bigl( \bZ_{N} \bigr)_{- N p} = \cA^{N,p} \times \cA^{N,-p} \;.
\ee
The product (or stacking) of two minimal theories $\cA^{N,p} \times \cA^{N,q}$ is also simple to compute, as long as $\gcd(p + q,N)=1$:
\be
\cA^{N,p} \times \cA^{N,q} = \cA^{N, (p^{-1} + q^{-1})^{-1}} \times \cA^{N, p + q} \;,
\ee
where inverses are taken in $\bZ_N$. Minimal theories have a large degree of redundance, indeed let $r$ be coprime with $N$, then
\be
\cA^{N,p} \sim \cA^{N, r^2 p}
\ee
as MTCs. The transformation is equivalent to choosing a different generator for $\bZ_N$. Note that this implies that $\cA^{N,p} \sim \cA^{N, p^{-1}}$. 
Using these conventions, we can set the line $W_{l=1}$ as the generator of $\bZ_N$. It then follows from the $S$-matrix that the generator has charge $p$ under $\bZ_N$. Alternatively, we could use the line $W_{l=p^{-1}}$ as the fundamental line. This line has unit charge under $\bZ_N$. The change of variables affects the inflow action, which is then labelled by $p^{-1}$ instead.
In the main text we choose to work under this choice of generator.

The construction can be generalized to multiple $\bZ_N$ factors. For simplicity we treat the case in which $N$ is prime, as in the main text. A theory $\cA^{N, \cT}$ is then described by a symmetric $r \times r$ matrix $\cT$, whose lines have spins
\be
\theta_l = \exp \left(\frac{2 \pi i}{2N} \, l^\sT \, \cT \, l\right) = \exp\left( \frac{2 \pi i}{2 N} \left[\sum_{i} \cT_{ii} l_i^2 + 2 \sum_{i>j} \cT_{ij} l_i l_j \right]\right) \, .
\ee 
A bosonic theory requires $N \cT_{ii} \in 2 \bZ$ and $ N \cT_{ij} \in \bZ$, while a spin theory can have $\cT_{ii} \in \bZ$ and $\cT_{ij} \in \bZ$. The condition of having a well defined $S$-matrix requires that $\cT$ is an invertible matrix over $\bZ_N^r$, this is the natural generalization of the $\gcd$ condition for $r=1$.
To this TQFT we can associate an anomaly theory as in the previous case:
\be
I_\cT = \frac{2 \pi}{2 N} \fP^\cT(B) \, , \ \ \fP^\cT(B) =  \sum_i^r \cT_{ii} \fP(B_i) + \sum_{i>j}^r 2 \cT_{ij} B_i \cup B_j  \, , \ \ \ B \in H^2(X, \bZ_N^r) \, .
\ee
As in the previous case there is a large degree of redundancy in these theories. Let $\cN$ be an invertible matrix over $\bZ_N$, then:
\be
\cA^{N, \; \cN^\sT \cT \cN} = \cA^{N, \; \cT} \, ,
\ee
as $\cN$ just implements a redefinition of the generators. Since $\cT$ is a nondegenerate symmetric quadratic form it can be diagonalized with coefficients in $\bZ_N$ by a suitable $\cN$: $\cT = \diag(t_i, \; ... , \; t_r)$ with $\gcd(t_i, N)=1$. The only relevant information about the theory (without specifying the coupling to the two-form gauge field $B$) are thus the quadratic residue classes for the $t_i$.
As for the one-dimensional case, in the main text we use a slightly different convention in which the fundamental lines have charge one. To go back to the standard convention one has to substitute $\cT$ by $\cT^{-1}$ in the formulas.

An important novelty with respect to the one-dimensional case is that generalized minimal theories can have anomaly free subgroups. For spin theories these are generated by vectors $l$ such that:
\be
l^\sT \cT l = 0 \mod N \, . \label{eq:appgaugiable}
\ee
Let us take $r=2$ and $N$ prime. For $r=2$ $\cA^{N,\cT}$ theories also contain twisted $\bZ_N$ DW theories $\text{DW}(\alpha)$ but with a twist matrix which is a multiple of $N$. If $N$ is prime every $l$ will generate a Lagrangian subgroup, so a solution to \eqref{eq:appgaugiable} implies that the theory is DW for a certain choice of torsion.
This is important in the main text, as it implements the correct fusion laws for twisted sectors on invariant boundaries.

Notice that the factorization theorem for $r=1$ still applies. This means that, given a $\bZ_N$ subgroup with nontrivial anomaly $p$, we can write:
\be
\cA^{N, \cT} = \cA^{N, p} \times \cA^{N , \cT'} \, ,
\ee
And $\cT'$ is an $r-1 \times r-1$ matrix. We use this decomposition property multiple times throughout our work.

\section{The case of charge conjugation}
\label{app:changeconj}

Here we expand on the case of charge conjugation $C$, which is the only 0-form symmetry defect with vanishing torsion. Since the 4d defect theory (\ref{eq: defect lagrangian full gauging}) or (\ref{new defect Lagrangian}) with $\cT = 0$ is a non-invertible TQFT, the twisted sector of the charge-conjugation defect does not host a well defined MTC of line operators \cite{Hsin:2018vcg}. Consider the case of two defects $V[\cT_1]$ and $V[\cT_2]$ whose fusion is $C$. If $\cT_1$ and $\cT_2$ are such that $\cT_{21}=0$ (they fuse onto $C$), then
\be
\cT_2 = - \frac{\epsilon}{2} \, \cT_1^{-1} \, \frac{\epsilon}{2} \;.
\ee
Using our formalism, we indeed find that the braiding in $\cA^{N,-\cT_1} \times_\cB \cA^{N,-\cT_2}$ is degenerate, because the braiding matrix $\cK_{21}$ in (\ref{spin in product theory}) has $\det \cK_{21} = 0$.%
\footnote{One uses that if $K= \smat{A& B \\ C & D}$ and $A$ is invertible, then $\det(K)=\det(A) \, \det(D - C A^{-1} B)$. Another simple way to see this is to perform the redefinition $n_1 \to \cT_1^{-1} \frac{\epsilon}{2} n_1$, then the matrix becomes: $\cK_{2,1} = \mat{\cT_2 & \cT_2 \\ \cT_2 & \cT_2}$, which indeed has half rank.}

\paragraph{Bulk fusion.}
Lines in the kernel of $K$ couple to $\cB$, but have vanishing spin and do not braid with anything else and thus do not form a well defined MTC. They are the naive restriction of the lines $\Psi$ of the $C$-twisted sector when we decouple the lines charged under $\Phi$.

To understand the fusion we must also take into account nonlocal lines. The $D[\cT_1]\times D[\cT_2]$ system is described by a braiding matrix:
\be
K^{-1} = \frac{1}{N} \mat{0 & \unit & 0 & 0 \\ \unit & \cT_1 & 0 &   \frac{\epsilon}{2} \\ 0 & 0 & 0 & \unit \\ 0 &- \frac{\epsilon}{2} & \unit & \cT_2} \, ,
\ee
with a basis made up of $\left( \Gamma_1, \Psi_1, \Gamma_2 , \Psi_2\right)$. Thus we label a line by its charges $\left(n_1, m_1, n_2, m_2\right)$ under the above generators. The vectors $(\cT_1 m_1,-m_1,0,0)$ and $(0,0,\cT_2 m_2,-m_2)$ are charged only under $\cB$ transformations, while the vectors $(n_1,0,0,0)$ and $(0,0,n_2,0)$ are charged under $\Phi_1$ and $\Phi_2$ respectively. In the variables $\Phi$, $\Phi_1$ the lines charged only under $\Phi$ are $(n_1,0,n_1,0)$.

We want to decompose this system. First, lines of the form:
\be
\widetilde{L} = \mat{ \cT_1n \\ -n \\ -\cT_2n \\ n } \, ,
\ee
are neutral w.r.t. all gauge transformations of the 4d bulk and form an $\cA^{N, - (\cT_1 + \cT_2)}$ decoupled theory. Lines which do not braid with them must satisfy the condition 
\be
n_1-n_2+\frac{\epsilon}{2}(m_1+m_2)=0.
\ee
We choose the basis:
\be
L_{n_1} =  \mat{n_1 \\ 0 \\ n_1 \\ 0}  \, , \ \ L_{n_2} =  \mat{ \frac{\epsilon}{2}n_2\\ n_2 \\ \frac{\epsilon}{2}n_2 \\ -n_2} \, , \ \ L_{n_3} =  \mat{  \frac{\epsilon}{2}( \Gamma-1)n_3 \\ \Gamma n_3 \\   \frac{\epsilon}{2}\Gamma n_3\\(1 - \Gamma)n_3} 
\ee
with $\Gamma = \left(\cT_1 + \cT_2 \right)^{-1}(\cT_2 + \frac{\epsilon}{2}) $. Notice that the relevant definitions can be read off from our Lagrangian computations in Section \ref{sec: Lag descr D[T]}.  The line $L_1$ is charged only under $\Phi$, $L_{n_2}$ under $\Phi$ and $\Phi_1$ while $L_{n_3}$ only under $\cB$ and $\Phi$. Between these lines, $L_{n_2}$ have nontrivial spin :
\be
\theta_{n_2} = \exp\left(\frac{ \pi i }{ N} n_2^\sT (\cT_1 + \cT_2) n_2\right) \, ,
\ee
and does not braid with both $L_1$ and $L_3$. Therefore it forms a decoupled $\cA^{N, \cT_1 + \cT_2}(\widetilde{\Phi})$ factor where $\widetilde{\Phi} = (T_1+T_2)(\Phi_1+\Gamma\Phi)$. Since $\theta_{n_1} = 1 $, $\theta_{n_3}= \exp\left(\frac{\pi i}{N}n_3^\sT \cT_{2,1}  n_3\right)$, $B_{n_1 n_3} = \exp\left(\frac{2 \pi i }{N} n_1^\sT n_3\right)$ they form a $(\bZ_{N\times N})_{N \cT_{2,1}} (\cB,\Phi)$ theory, which is the twisted sector for $V[\cT_{2,1}]$:
\be \label{eq: fusion defects}
D[\cT_1] \times D[\cT_2] = \left[\cA^{N, \cT_1 + \cT_2}(\widetilde{\Phi}) \times \cA^{N, - \cT_1 - \cT_2} \right] D[\cT_{21}] \, .
\ee
Notice that:
\be
\cA^{N, \cT_1 + \cT_2} \times \cA^{N, - \cT_1 - \cT_2} = (\bZ_{N } \times \bZ_N)_{N(\cT_1 + \cT_2)}
\ee
and also that the decoupled coefficient $\cA^{N, -\cT_1 -\cT_2}$ is the same decoupled TQFT as in the normal bulk fusion, which is the correct leftover coefficient once $\Phi_1$ is integrated out.
This formula generalizes smoothly to the case of charge conjugation for which $\cT_{2,1}=0$.

\paragraph{Boundary fusion}
Now we can understand fusions on a gapped boundary $\cL$. All gapped boundaries are $C$-invariant, thus the twisted sector will host a genuine $\GW$ operator in the gauged theory, plus a condensate coming from the fusion.

First we must discuss what happens to the full defect $D[\cT]$ when it approaches the gapped boundary. In the 4d zero form symmetry defect $V[\cT]$ we have a coupling $\cB^\sT \; \Phi$. As we move to the boundary this becomes $ \cB^{\sT}\;\Phi \rightarrow \tilde{b}_\perp \; l_\perp^\sT \Phi$. We expand
\bea
&\Phi = l \phi_l + \frac{u \phi_\perp }{l_{\perp}^{\sT}u}\; \Rightarrow \; l_{\perp}^{\sT} \Phi = \Phi_{\perp} \, ,\\
& \cB = l_{\perp} \tilde{b}_{\perp} + \frac{u_{\perp}b_l}{u_{\perp}^{\sT}l} \; \Rightarrow \; l^{\sT}\cB = b_l \, ,
\eea
where $u$ is the generator of $\cS$ defined in Section \ref{sec: CS theory} and $u_\perp = \langle \, , u \rangle$ is such that $l_\perp^\sT u = l^\sT u_\perp \not= 0$ \footnote{When $\cT \not = 0$ we can choose $u = \cT^{-1}l_{\perp}$ and we find $b_{\perp} \equiv l_{\perp}^{\sT} \cT^{-1} \cB = t_{\perp} \tilde{b}_{\perp}$.}. Labelling a line $L_{n,m}$ in the twisted sector by its charges $(n , \, m)$ under $\Gamma$ and $\Psi$ we find that the lines
\be
\mat{u_\perp \\ 0} \, , \mat{- \cT l \\ l} \, 
\ee
are charged only under $\phi_l$ and $b_l$ respectively. They form a DW theory with braiding matrix
\be
K^{-1} = \mat{0 & l^\sT u_\perp \\ l^\sT u_\perp & - t_l}\,.
\ee
After a rescaling of the electric generator this becomes a $(\bZ_N)_{N t_l}(\phi_l, b_l)$ DW theory.
Remaining lines need to have trivial braiding with these generators. They have a basis given by
\be
\mat{0 \\ u} \, , \ \ \mat{l_\perp \\ 0} \, ,
\ee 
with braiding matrix
\be
\tilde{K}^{-1} = \mat{0 & l_\perp^\sT u \\ l_\perp^\sT u & u^\sT \cT u}\,.
\ee
To get a more familiar result notice that $u= \cT^{-1} l_\perp$ is a good choice for $u$ as long as the boundary is not invariant. In these variables the lines
\be
\mat{l_\perp \\ - \cT^{-1} l_\perp} \, ,
\ee
have spin $\exp\left(-\frac{\pi i}{N} t_\perp \right)$ and couple only to $b_\perp$ with unit charge. They thus correctly reproduce the sub-theory $\cA^{N, -t_\perp}(b_\perp)$. We conclude that this procedure is consistent with the one used in section \ref{sec: fusion gapped boundary} where the field $\Phi$ was integrated out. However this procedure is more general and in particular it can be extended to the case of charge conjugation. We have shown that in general
\be
D[\cT] = (\bZ_N)_{-Nt_l} (b_l, \phi_l) \times (\bZ_N)_{N u^{\sT}\cT u}(\tilde{b}_{\perp},\phi_{\perp}) \,. 
\label{eq:boundary defect with phi}
\ee

We want now to discuss the fusions of two twist defects on the gapped boundary. A simple way to derive the boundary fusion is to start from the formula \eqref{eq: fusion defects}, impose boundary conditions which set the decoupled DW theories to one on the boundary (which is a consistent boundary condition) and divide
\be
(\bZ_{N } \times \bZ_N)_{0} (\cB,\Phi) = (\bZ_N)_0 ( \phi_l,b_l) \times (\bZ_N)_0(\tilde{b}_\perp,\phi_{\perp}) \, .
\ee
The first term is generated by lines:
\be
\mat{u_\perp \\ 0} \, , \ \ \mat{0 \\ l} \, ,
\ee
While the second by lines:
\be
\mat{0 \\ u} \, , \ \ \mat{l_\perp \\ 0} \, .
\ee
Notice that the braiding are non-degenerate owning to $l_\perp^\sT u = l^\sT u_\perp \neq 0$.
The first term is also a decoupled DW theory which can be set to one on the boundary, while the second term is a condensate for the $\bZ_N$ surviving there.
One would then conclude
\be
D_\cL[\cT_1] \times D_\cL[\cT_2] = (\bZ_N)_0(\tilde{b}_\perp,\phi_{\perp}) \, D_\cL[0] \, ,
\ee
for a trivial $D_\cL[0]$.

\bibliographystyle{ytphys}
\baselineskip=0.92\baselineskip
\bibliography{TopGravity}

\end{document}